\documentclass{aa}
\usepackage{graphicx}
\usepackage{txfonts}
\usepackage{lscape}

\begin{document}

\title{Activity trends in young solar-type stars
\thanks{Based on observations made as part of the automated astronomy program at Tennessee State University and with the Nordic Optical Telescope, operated on the island of La Palma jointly by Denmark, Finland, Iceland, Norway, and Sweden, in the Spanish Observatorio del Roque de los Muchachos of the Instituto de Astrofisica de Canarias.}
\fnmsep
\thanks{The analysed photometry and numerical results of the analysis are published electronically at the CDS and available via anonymous ftp to cdsarc.u-strasbg.fr (130.79.128.5) or via http://cdsarc.u-strasbg.fr/viz-bin/qcat?J/A+A/yyy/Axxx}}
\author{J. Lehtinen \inst{1} \and L. Jetsu \inst{1} \and T. Hackman \inst{1} \and P. Kajatkari \inst{1} \and G.W. Henry \inst{2}}
\institute{Department of Physics, PO Box 64, 00014 University of Helsinki, 00014 Helsinki, Finland \\ \email{jyri.j.lehtinen@helsinki.fi}
\and Center of Excellence in Information Systems, Tennessee State University, 3500 John A. Merritt Blvd., Box 9501, Nashville,
TN 37209, USA}
\date{Received date / Accepted date}

\abstract{}
{We study a sample of 21 young and active solar-type stars with spectral types ranging from late F to mid K and characterize the behaviour of their activity.}
{We apply the continuous period search (CPS) time series analysis method on Johnson B- and V-band photometry of the sample stars, collected over a period of 16 to 27 years. Using the CPS method, we estimate the surface differential rotation and determine the existence and behaviour of active longitudes and activity cycles on the stars. We supplement the time series results by calculating new $\log{R'_{\rm HK}} = \log{F'_{HK}/\sigma T_{\rm eff}^4}$ emission indices for the stars from high resolution spectroscopy.}
{The measurements of the photometric rotation period variations reveal a positive correlation between the relative differential rotation coefficient and the rotation period as $k \propto P_{\rm rot}^{1.36}$, but do not reveal any dependence of the differential rotation on the effective temperature of the stars. Secondary period searches reveal activity cycles in 18 of the stars and temporary or persistent active longitudes in 11 of them. The activity cycles fall into specific activity branches when examined in the $\log{P_{\rm rot}/P_{\rm cyc}}$ vs. $\log{{\rm Ro}^{-1}}$, where ${\rm Ro}^{-1} = 2\Omega\tau_c$, or $\log{P_{\rm rot}/P_{\rm cyc}}$ vs. $\log{R'_{\rm HK}}$ diagram. We find a new split into sub-branches within this diagram, indicating multiple simultaneously present cycle modes. Active longitudes appear to be present only on the more active stars. There is a sharp break at approximately $\log{R'_{\rm HK}}=-4.46$ separating the less active stars with long-term axisymmetric spot distributions from the more active ones with non-axisymmetric configurations. In seven out of eleven of our stars with clearly detected long-term non-axisymmetric spot activity the estimated active longitude periods are significantly shorter than the mean photometric rotation periods. This systematic trend can be interpreted either as a sign of the active longitudes being sustained from a deeper level in the stellar interior than the individual spots or as azimuthal dynamo waves exhibiting prograde propagation.}
{}


\maketitle

\section{Introduction}

\begin{table*}
\caption{Basic observational properties of the sample stars, the comparison (Cmp) and check (Chk) stars used for the photometry, and the time span of the photometric record.}
\center
\begin{tabular}{llccccllc}
\hline
\hline
Star & HD/SAO & $V$\tablefootmark{1} & $B-V$\tablefootmark{1} & $d$ [pc]\tablefootmark{1} & Sp. & Cmp & Chk & photometry \\
\hline
\object{PW And}    & \object{HD 1405}   & 8.86 & 0.95 & 21.9 & K2V\tablefootmark{2}    & \object{HD 1406}   & \object{HD 1439}   & Aug 1988--Jun 2014 \\
\object{EX Cet}    & \object{HD 10008}  & 7.66 & 0.80 & 23.6 & G9V\tablefootmark{3}    & \object{HD 10116}  & \object{HD 9139}   & Dec 1998--Jan 2014 \\
\object{V774 Tau}  & \object{HD 26923}  & 6.32 & 0.57 & 21.2 & G0V\tablefootmark{4}    & \object{HD 27497}  & \object{HD 26292}  & Dec 1998--Feb 2014 \\
\object{V834 Tau}  & \object{HD 29697}  & 8.09 & 1.09 & 13.5 & K4V\tablefootmark{3}    & \object{HD 284676} & \object{HD 29169}  & Sep 1993--Mar 2014 \\
\object{V1386 Ori} & \object{HD 41593}  & 6.76 & 0.81 & 15.5 & G9V\tablefootmark{3}    & \object{HD 41304}  & \object{HD 42784}  & Dec 1998--Mar 2014 \\
\object{V352 CMa}  & \object{HD 43162}  & 6.37 & 0.71 & 16.7 & G6.5V\tablefootmark{4}  & \object{HD 43879}  & \object{HD 43429}  & Dec 1998--Feb 2014 \\
\object{V377 Gem}  & \object{HD 63433}  & 6.90 & 0.68 & 21.8 & G5V\tablefootmark{3}    & \object{HD 64467}  & \object{HD 63432}  & Dec 1998--Apr 2014 \\
\object{V478 Hya}  & \object{HD 70573}  & 8.69 & 0.62 & 88.5 & G6V\tablefootmark{2}    & \object{HD 71136}  & \object{HD 70458}  & Apr 1993--Apr 2014 \\
\ldots             & \object{HD 72760}  & 7.32 & 0.79 & 21.8 & K0-V\tablefootmark{3}   & \object{HD 71640}  & \object{HD 72660}  & Dec 1998--Apr 2014 \\
\object{V401 Hya}  & \object{HD 73350}  & 6.74 & 0.66 & 23.6 & G5V\tablefootmark{3}    & \object{HD 73400}  & \object{HD 72412}  & Dec 1998--Apr 2014 \\
\object{DX Leo}    & \object{HD 82443}  & 7.05 & 0.78 & 17.7 & K1V\tablefootmark{4}    & \object{HD 83098}  & \object{HD 83821}  & Feb 1992--May 2014 \\
\object{LQ Hya}    & \object{HD 82558}  & 7.82 & 0.93 & 18.3 & K0V\tablefootmark{2}    & \object{HD 82477}  & \object{HD 82428}  & Nov 1987--Apr 2014 \\
\object{NQ UMa}    & \object{HD 116956} & 7.29 & 0.80 & 21.9 & G9V\tablefootmark{3}    & \object{HD 114446} & \object{HD 119992} & Dec 1998--Jun 2014 \\
\object{KU Lib}    & \object{HD 128987} & 7.24 & 0.71 & 23.6 & G8V\tablefootmark{4}    & \object{HD 127170} & \object{HD 126679} & Dec 1998--Jun 2014 \\
\object{HP Boo}    & \object{HD 130948} & 5.86 & 0.58 & 17.9 & F9IV-V\tablefootmark{5} & \object{HD 128402} & \object{HD 127739} & Dec 1998--Jun 2014 \\
\object{V379 Ser}  & \object{HD 135599} & 6.92 & 0.83 & 15.6 & K0V\tablefootmark{6}    & \object{HD 136118} & \object{HD 137006} & Dec 1998--Jun 2014 \\
\object{V382 Ser}  & \object{HD 141272} & 7.44 & 0.80 & 21.3 & G9V\tablefootmark{3}    & \object{HD 141103} & \object{HD 139137} & Dec 1998--Jun 2014 \\
\object{V889 Her}  & \object{HD 171488} & 7.39 & 0.62 & 37.2 & G2V\tablefootmark{2}    & \object{HD 171286} & \object{HD 170829} & Apr 1994--Jun 2014 \\
\object{MV Dra}    & \object{HD 180161} & 7.04 & 0.80 & 20.0 & G8V\tablefootmark{2}    & \object{HD 182735} & \object{HD 177249} & Feb 1999--Jun 2014 \\
\object{V453 And}  & \object{HD 220182} & 7.36 & 0.80 & 21.9 & G9V\tablefootmark{3}    & \object{HD 219224} & \object{HD 221661} & Dec 1998--Jun 2014 \\
\object{V383 Lac}  & \object{SAO 51891} & 8.57 & 0.85 & 23.7 & K1V\tablefootmark{2}    & \object{HD 212072} & \object{HD 212712} & May 1994--Jun 1996 \\
                   &                    &      &      &      &                         &                    &                    & Oct 2011--Jun 2014 \\
\hline
\end{tabular}
\tablebib{
\tablefoottext{1}{\cite{esa1997hipparcos}}
\tablefoottext{2}{\cite{montes2001late}}
\tablefoottext{3}{\cite{gray2003contributions}}
\tablefoottext{4}{\cite{gray2006contributions}}
\tablefoottext{5}{\cite{gray2001physical}}
\tablefoottext{6}{\cite{gaidos2000spectroscopy}}
}
\label{startab}
\end{table*}

Extended time series observations are among the most useful resources for studying the phenomena related to magnetic activity of stars. While the general activity level of a star can quickly be estimated from a single observation of its chromospheric or coronal emission level, most of the relevant phenomena are time dependent and require us to follow the stars over a longer period of time. Importantly, the variations of the observed activity indicators are often periodic in their nature ranging from the rotation signal to decadal cyclic activity variations. Thus, time series analysis is a crucial tool for studying stellar activity.

Although rotation itself is not an activity phenomenon, it plays a crucial role in stellar dynamos. In particular, differential rotation is an important parameter and determines whether the stellar dynamo more resembles an $\alpha\Omega$ or an $\alpha^2$ dynamo \citep{charbonneau2010dynamo}. Quenching of differential rotation is predicted for rapidly rotating stars transitioning into $\alpha^2$ dynamos \citep{kitchatinov1999differential}.

The magnitude of stellar surface differential rotation has been investigated with a number of different observational methods. Most of them rely on the idea that dark starspots or bright chromospheric active regions forming on different latitudes will display different rotation periods and produce multiple varying periodic signals in the observed time series \citep{hall1991learning}. A massive study based on this principle was recently undertaken by \cite{reinhold2013rotation} and \cite{reinhold2015rotation} using Kepler data. Previous studies have revealed a positive correlation on the one hand between the relative differential rotation and the rotation period \citep{hall1991learning} and on the other between the absolute differential rotation and the effective temperature \citep{barnes2005dependence, colliercameron2007differential}.

The rotational signal also reveals the longitudes or rotational phases of the major spots or chromospheric active regions \citep{hall2009activity}. In many stars these concentrate on active longitudes where most of the activity appears on one or two narrow longitudes, which can stay intact for decades \citep{jetsu1996active,lehtinen2011continuous}. Long-lived stable active longitudes can be interpreted as signs of non-axisymmetric dynamo modes present in the stars \citep{radler1986investigations, moss1995nonaxisymmetric}.

On longer time scales, time series observations using both photometry and chromospheric line emission have been used in the search of activity cycles \citep{baliunas1995chromosphere, messina2002magnetic, olah2009multiple}. This field of study is biased because the existing observational records are still quite short compared to the longest activity cycles that have been found. Moreover, the cyclic activity variations are typically quasiperiodic rather than stationary processes, which further decreases the efficiency of the period search methods. Nevertheless, it has been possible to relate the estimated cycle periods to other stellar parameters. This has revealed a sequence of activity branches, suggesting different dynamo modes being excited on different stars \citep[e.g.][]{saar1999time}.

In this study we perform an analysis of 21 young nearby solar-type stars using ground-based monitoring photometry gathered over a period of 16 to 27 years. Most of the stars have been selected from the sample studied by \cite{gaidos1998nearby}, but the observing programme also includes six additional stars with notably high levels of activity. 

The stars in our sample are listed in Table \ref{startab} along with their basic observational properties. The Johnson V band magnitudes and $B-V$ colours are taken from the Hipparcos and Tycho catalogues \citep{esa1997hipparcos}, and represent the mean magnitudes and colour indices from the observations over the full Hipparcos mission. The distances $d$ are likewise derived from the Hipparcos and Tycho trigonometric parallaxes. The spectral types reported for the stars are taken from a number of sources given individually for each star.

\begin{figure*}
\centering
\includegraphics[width=0.75\linewidth]{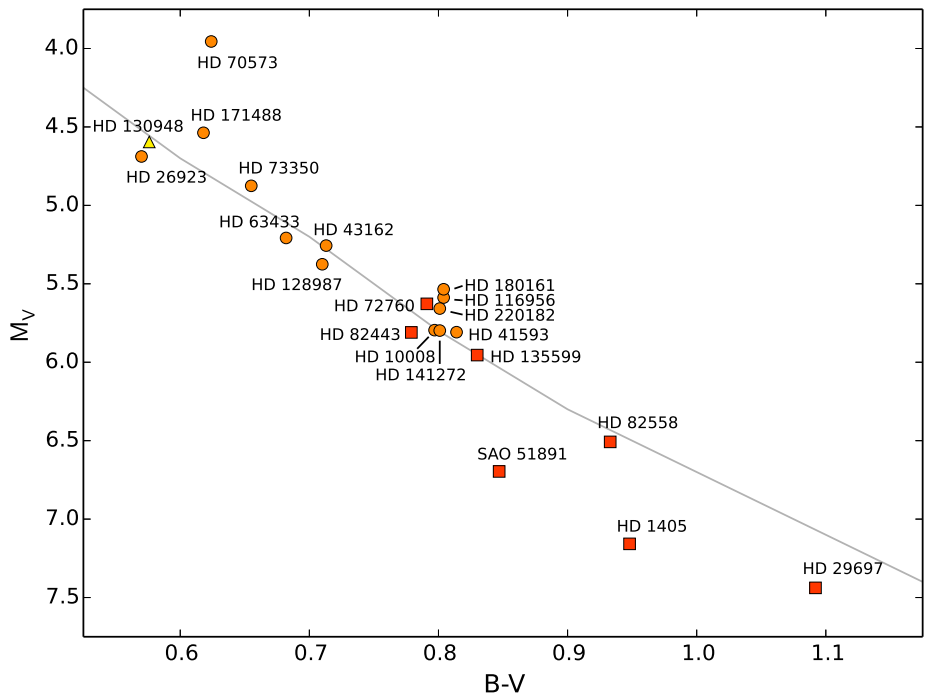}
\caption{Colour magnitude diagram of our sample stars. The plot symbols denote spectral types: yellow triangle for F-type, light orange circle for G-type, and dark orange square for K-type stars. The ZAMS is indicated by the grey line according to \cite{cox2000allen}.}
\label{fighr}
\end{figure*}

The colour magnitude diagram of the stars is shown in Fig. \ref{fighr} based on the Hipparcos and Tycho data and assuming negligible interstellar extinction. The plot also indicates the zero age main sequence (ZAMS) according to \cite{cox2000allen}. It shows that the stars all lie on the main sequence although there is an apparent offset to larger absolute magnitudes for the coolest K-type stars.

\section{Observations}

\subsection{Photometry}

Our study is based on photometry obtained with the T3 0.4~m Automatic Photoelectric Telescope (APT) at the Fairborn Observatory in Arizona, which has been monitoring our programme stars since late 1987. In this paper we include all of the standard Johnson B- and V-band photometry collected with the telescope up to June 2014. The time span of the observations is given in Table \ref{startab} and ranges from 16 to 27 years depending on the star. The only exception is a 15 year gap in the observations of SAO 51891.

The observations from the stars consist of differential photometry, where the variable target stars (Var) are compared to constant, usually F-type, comparison stars (Cmp) as the difference $\rm Var-Cmp$. In addition the constancy of the comparison stars is simultaneously monitored by observing separate constant check stars (Chk) as the difference $\rm Chk-Cmp$.

We estimate the typical error of the target star photometry to be between 0.003 and 0.004 magnitudes based on monitoring constant stars with the same setup \citep{henry1995development}. Errors of the check star observations can be expected to be somewhat larger since fewer individual integrations are used to determine their values. For a brief description of the operation of the APT and reduction of the data, see \cite{fekel2005chromospherically} and references therein.

\subsection{Spectroscopy}

To determine the chromospheric activity level of the sample stars we have observed their visible spectra with the high resolution fibre-fed echelle spectrograph FIES at the Nordic Optical Telescope \citep{telting2014fies}. This instrument provides full spectral coverage within the wavelength interval 3640 -- 7360~\AA \ in 79 overlapping orders. The spectroscopic observations were obtained in 2012 and 2014. We performed the observations in the high resolution mode, which gives a spectral resolution of $R=67000$. The observations were reduced using the FIEStool pipeline.

\section{Time series analysis of the photometry}
\label{sectcps}

For the time series analysis of our photometry we used the continuous period search method (hereafter CPS) formulated by \cite{lehtinen2011continuous}. The method models the light curve data with non-linear harmonic fits,
\begin{equation}
\hat{y}(t_i) = M + \sum_{k=1}^K [B_k\cos(k2\pi ft_i) + C_k\sin(k2\pi ft_i)],
\end{equation}
to short datasets selected from the full time series data by applying a sliding window of predetermined length. The order $K$ of the harmonic fits is adaptive and determined from the data using the Bayesian information criterion. In this study we allowed fits of the orders $K=0$ (i.e. a constant brightness model), $K=1$, and $K=2$. For each dataset we calculated the mean magnitude $M$, the full light curve amplitude $A$, the period $P=f^{-1}$, and the epochs of the primary and secondary light curve minima $t_{\rm min,1}$ and $t_{\rm min,2}$. Naturally $t_{\rm min,2}$ can only exist for $K=2$ models and for $K=0$ models only $M$ is defined. Each of the fits was checked for reliability by comparing the distribution of the fit residuals and the error distributions of the model parameters against Gaussian distributions. If any of them was found to be significantly non-Gaussian, the whole dataset was labelled as unreliable. To find the initial search range for the light curve periods, we first applied the three stage period analysis method \cite[hereafter TSPA]{jetsu1999three} with a wide frequency range before proceeding with the CPS analysis.

The lengths of the individual datasets are defined by a maximum time span $\Delta T_{\rm max}$. If a dataset starts with a data point at $t_0$, all following data points within $[t_0,t_0+\Delta T_{\rm max}]$ are included in the dataset. Because of the sliding window approach, most of the datasets have data points in common with the adjacent datasets, which will introduce inherent correlation in the model parameters. To overcome this correlation, we chose a set of independent non-overlapping datasets as the basis of our further statistical analysis. We also discarded all datasets having $n_{\rm data}<12$ data points because their fits are likely to have low quality.

\begin{figure}
\centering
\includegraphics[width=\linewidth]{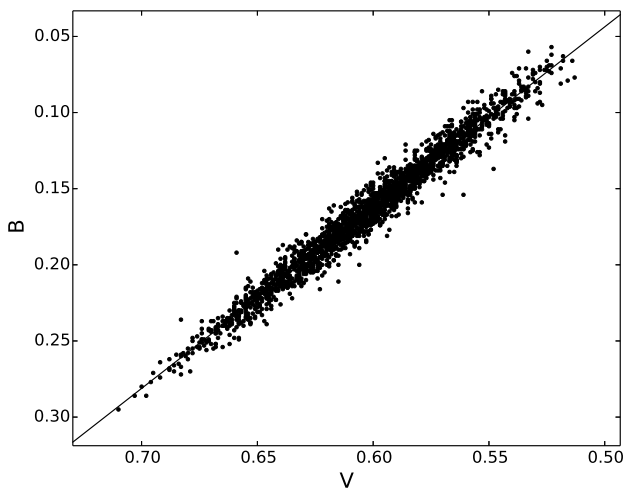}
\caption{B-band vs. V-band differential photometry for HD 171488 with a fit showing the linear correlation between the two colour bands.}
\label{figcorr}
\end{figure}

As opposed to previous studies using the CPS method \citep{lehtinen2011continuous, lehtinen2012spot, hackman2011spot, hackman2013flipflops, kajatkari2014spot, kajatkari2015periodicity}, we chose to use both the B and V bands in this study to get more data points for the periodic fits and to increase their precision. As can be seen in Fig. \ref{figcorr} for HD~171488, there is a linear correlation between coeval photometry in the two photometric bands. This applies generally to the photometry of all our stars. The linear relation means that the two bands contain essentially the same information. It is thus reasonable to simply scale one band on top of the other and use the resulting combined time series as a single photometric band. We used the V-band as the basis and found the empirical scaling relation $B=c_1V-c_0$ between the two sets of differential photometry separately for each star. The proportionality coefficients $c_1$ are listed in Table \ref{modeltab}. In each case we found $c_1>1$ with the mean value $\langle c_1 \rangle=1.29$, meaning that the B-band light curves always have larger amplitudes than the V-band curves. This behaviour is consistent with modulation caused by spots that are cooler than the surrounding photosphere.

The use of combined B- and V-band photometry has increased the number of reliable period detections for all stars. For some of the lowest amplitude stars in our sample we were able to detect periodicity in 10--20\% more datasets in the combined data than when using the V-band only. With the largest amplitude stars the use of combined B- and V-band data removed all $K=0$ models, allowing period detection throughout the data. Using the combined photometry also increased the overall ability of CPS to find reliable model fits for the datasets. As a median, the number of reliable fits was 8\% higher for the combined data than for the V-band only.

The crucial parameter to be set for the CPS analysis is the upper limit of the dataset length $\Delta T_{\rm max}$. To find the most reasonable value, we performed the CPS analysis for our sample stars with values $\Delta T_{\rm max}=20$~d, 30~d, and 45~d. We found that the dispersion of the period estimates, $\Delta P_{\rm w}$ (see Sect. \ref{sectrot}, Eqs. \ref{pweq} and \ref{dpweq}), from the individual datasets was relatively small for the two longer values, but for $\Delta T_{\rm max}=20$~d there was more scatter. This indicates that at $\Delta T_{\rm max}<30$~d the dataset length and the number of data points contained within these datasets have become inadequate and it is no longer possible to produce repeatable period estimates for many of our sample stars. If we intend to use the estimated period fluctuations as a measure of differential rotation for the stars, we should use $\Delta T_{\rm max}\ge30$~d. With these dataset lengths the estimated period fluctuations appear to be approaching a lower limit likely governed by physical processes on the observed stars.

We can also define a time scale of change for the light curves, $T_{\rm C}$, and see how this changes with varying $\Delta T_{\rm max}$. This is the time from the start of a dataset during which a model fit can adequately describe the following data \citep{lehtinen2011continuous}. As expected, we found that as $\Delta T_{\rm max}$ increased so did $T_{\rm C}$. This shows that as the datasets get longer the light curve of the observed star has more time to evolve and smear out finer details. When details are lost, the light curve fits tend to get simpler and consequently better describe future data.

We settled on using $\Delta T_{\rm max}=30$~d for all of the sample stars. This is a good compromise between getting reasonably precise period estimates and restricting the light curve evolution within the datasets. It is also similar to the dataset lengths we have used previously in other studies using the CPS \citep[e.g.][]{lehtinen2011continuous, lehtinen2012spot, kajatkari2015periodicity}.

\begin{figure}
\centering
\includegraphics[width=\linewidth]{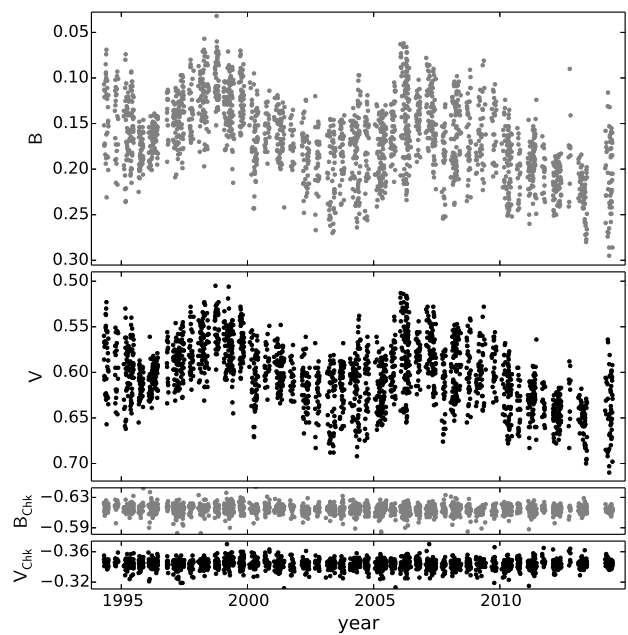}
\caption{Differential B- and V-band photometry from HD 171488 (\textit{top two panels}) and the constant check star HD 170829 used for it (\textit{bottom two panels}). The same magnitude scale is used for all the panels.}
\label{fig171488phot}
\end{figure}

Fig. \ref{fig171488phot} shows as an example what the differential photometry looks like for HD~171488. This is a star for which we see a clear periodic signal. As can be seen from the raw photometry, the envelope of the light curve of HD 171488 indicating the light curve amplitude due to rotational modulation is clearly much wider than the scatter seen in the photometry of the constant check star HD 170829. Consequently, the CPS was able to find good periodic fits both for the V-band alone and for the combined B- and V-bands. There were, however, slight improvements associated with using the combined bands; the number of datasets with reliable fits increased from 517 to 585 and the number of those where a period could be detected from 513 to 585.

When using the combined bands, the variations of the estimated periods showed an expected drop with increasing $\Delta T_{\rm max}$. For HD~171488 using $\Delta T_{\rm max}=20$~d, 30~d, and 45~d we found $\Delta P_{\rm w}=0.048$~d, 0.012~d, and 0.009~d respectively. From $\Delta T_{\rm max}=20$~d to $\Delta T_{\rm max}=30$ d there is a large drop in $\Delta P_{\rm w}$ indicating increased precision in period determination but increasing the dataset length to $\Delta T_{\rm max}=45$~d left the level of precision practically unchanged. This is an encouraging result if we intend to measure physical period variations on the stars.

For the same three values of $\Delta T_{\rm max}$, we found the estimated time scale of change of the light curve to be $T_{\rm C}=22.7$~d, 28.3~d, and 36.9~d. There is a clear increase in $T_{\rm C}$ when moving from $\Delta T_{\rm max}=20$~d to $\Delta T_{\rm max}=45$~d, indicating the loss of finer detail in the light curve when using longer datasets. This shows that using longer datasets will lead to some level of information loss but also that the calculated time scale of change $T_{\rm C}$ seems to depend mostly on the chosen analysis parameters and not the observed stars themselves.

\begin{figure}
\centering
\includegraphics[width=\linewidth]{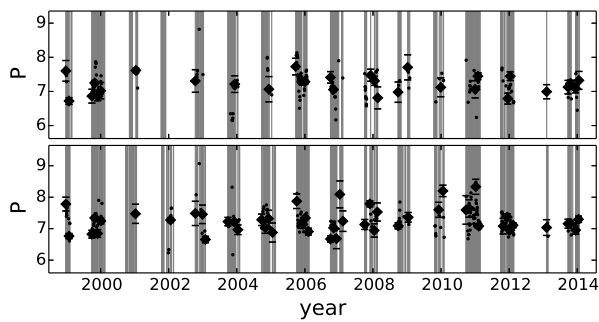}
\caption{Comparison of period estimates for HD 43162 from plain V-band photometry (\textit{top}) and combined B- and V-band photometry (\textit{bottom}). The $P$ estimates from the independent datasets are denoted by the diamonds with error bars and from the rest of the datasets as points. The grey shading indicates the time span of all reliable model fits.}
\label{fig43162p}
\end{figure}

Another star, HD~43162, with a lower light curve amplitude shows clearly the advantages of using the combined photometric bands. In our previous study of the V-band photometry of this star \citep{kajatkari2015periodicity} we were able to detect periodicity in 192 of the 370 datasets with reliable model fits. Here, with two more years of data, we found periodicity in 229 of the 407 reliable datasets from the V-band photometry. This increased to 326 out of a total of 425 reliable datasets when using the combined bands. Graphically this is visible in Fig. \ref{fig43162p} where both the number of period estimates and the total number of reliable model fits, indicated by the grey shaded area, increased when combining the two photometric bands. Clearly the availability of more data points improved our capability to detect periodicity from the very noisy lowest amplitude light curves.

\begin{table}
\caption{Properties of the light curve fits. The proportionality coefficient $c_1$ between the B- and V-band photometry, the numbers of independent reliable $K=0$ and $K>0$ order fits, and the mean V-band amplitude of the $K>0$ order fits in the CPS results.}
\center
\begin{tabular}{lcccc}
\hline
\hline
Star & $c_1$ & $n_{K=0}$ & $n_{K>0}$ & $\langle A \rangle$ \\
\hline
HD 1405   & 1.18 & 0  & 77  & 0.102 \\
HD 10008  & 1.52 & 30 & 16  & 0.010 \\
HD 26923  & 1.28 & 48 & 15  & 0.013 \\
HD 29697  & 1.09 & 0  & 52  & 0.096 \\
HD 41593  & 1.25 & 1  & 70  & 0.026 \\
HD 43162  & 1.63 & 14 & 48  & 0.019 \\
HD 63433  & 1.27 & 7  & 68  & 0.022 \\
HD 70573  & 1.30 & 3  & 100 & 0.043 \\
HD 72760  & 1.23 & 23 & 52  & 0.014 \\
HD 73350  & 1.37 & 21 & 40  & 0.017 \\
HD 82443  & 1.22 & 1  & 130 & 0.063 \\
HD 82558  & 1.20 & 0  & 117 & 0.093 \\
HD 116956 & 1.27 & 0  & 81  & 0.033 \\
HD 128987 & 1.31 & 36 & 35  & 0.014 \\
HD 130948 & 1.39 & 8  & 66  & 0.018 \\
HD 135599 & 1.25 & 24 & 39  & 0.018 \\
HD 141272 & 1.25 & 0  & 63  & 0.031 \\
HD 171488 & 1.23 & 0  & 82  & 0.067 \\
HD 180161 & 1.38 & 15 & 47  & 0.013 \\
HD 220182 & 1.22 & 3  & 55  & 0.032 \\
SAO 51891 & 1.18 & 0  & 15  & 0.086 \\
\hline
\end{tabular}
\label{modeltab}
\end{table}

\section{Data analysis and results}

\begin{table*}
\caption{Results relating to the weighted mean photometric rotation periods $P_{\rm w}$, the active longitude periods $P_{\rm al}$, and the cycle periods $P_{\rm cyc}$.}
\center
\begin{tabular}{lcccccccccc}
\hline
\hline
Star & $P_{\rm w}$ [d]\tablefootmark{a} & $Z$ & $P_{\rm al}$ [d]\tablefootmark{a} & $Q_{\rm K}$ & $P_{\rm cyc}$ [yr] & cycle type\tablefootmark{b} & $FAP$ & $FAP$ grade\tablefootmark{c} \\
\hline
HD 1405   & 1.75622(62) & 0.019 & 1.752212(71)  & $3.82\cdot10^{-3}$  & 8.0        & $\rm MAM_-M_+$ & $3.0\cdot10^{-7}$ & good \\
HD 10008  & 6.78(13)    & 0.47  & 6.80          & 0.81                & 10.9       & $\rm MM_-M_+$  & $1.1\cdot10^{-3}$ & fair, long \\
HD 26923  & 11.08(20)   & 0.42  & 10.66         & 0.99                & 7.0        & $\rm MM_-M_+$  & $3.8\cdot10^{-2}$ & poor \\
HD 29697  & 3.9651(59)  & 0.065 & 3.9433(28)    & $9.24\cdot10^{-5}$  & 7.3        & $\rm A$        & $3.8\cdot10^{-2}$ & poor \\
HD 41593  & 8.135(30)   & 0.19  & 8.0417(23)    & $3.45\cdot10^{-3}$  & 3.3        & $\rm M_-$      & $8.2\cdot10^{-3}$ & fair \\
HD 43162  & 7.168(38)   & 0.22  & 7.1323(10)    & $1.88\cdot10^{-3}$  & 8.1        & $\rm A$        & $5.7\cdot10^{-2}$ & poor \\
HD 63433  & 6.462(40)   & 0.31  & 6.46414(98)   & $4.20\cdot10^{-2}$  & 2.7        & $\rm MM_+$     & $1.2\cdot10^{-3}$ & fair \\
          &             &       &               &                     & 8.0        & $\rm AM_-$     & $2.5\cdot10^{-2}$ & poor, long \\
HD 70573  & 3.3143(34)  & 0.061 & 3.29824(36)   & $2.69\cdot10^{-5}$  & 6.9        & $\rm MM_-M_+$  & $5.1\cdot10^{-6}$ & good \\
HD 72760  & 9.57(11)    & 0.50  & 9.68          & 0.70                & \ldots     & \ldots         & \ldots            & \ldots \\
HD 73350  & 12.14(13)   & 0.40  & 12.59         & 0.58                & 3.5        & $\rm MM_-M_+$  & $3.6\cdot10^{-5}$ & fair \\
HD 82443  & 5.4244(43)  & 0.054 & 5.41471(22)   & $2.50\cdot10^{-8}$  & 4.1        & $\rm MAM_-M_+$ & $9.2\cdot10^{-5}$ & fair \\
          &             &       &               &                     & 20.0       & $\rm M$        & $3.1\cdot10^{-6}$ & good, long \\
HD 82558  & 1.60435(42) & 0.017 & 1.6037330(68) & $1.44\cdot10^{-2}$  & 14.5--18.0 & $\rm MAM_-M_+$ & $<10^{-12}$       & excellent, long \\
HD 116956 & 7.860(16)   & 0.11  & 7.84203(72)   & $2.06\cdot10^{-11}$ & 2.9        & $\rm MM_-$     & $4.0\cdot10^{-5}$ & fair \\
          &             &       &               &                     & 14.7       & $\rm A$        & $5.5\cdot10^{-7}$ & good, long \\
HD 128987 & 9.80(12)    & 0.44  & 9.63          & 0.25                & 5.4        & $\rm MM_-M_+$  & $4.7\cdot10^{-3}$ & fair \\
HD 130948 & 7.849(26)   & 0.16  & 7.89          & 0.85                & 3.9        & $\rm MM_+$     & $3.3\cdot10^{-2}$ & poor \\
HD 135599 & 5.529(68)   & 0.46  & 5.13          & 0.99                & 14.6       & $\rm MM_+$     & $2.9\cdot10^{-9}$ & good, long \\
HD 141272 & 13.843(84)  & 0.29  & 13.80         & 0.27                & 6.4        & $\rm M$        & $9.7\cdot10^{-2}$ & poor \\
HD 171488 & 1.3454(13)  & 0.054 & 1.336923(41)  & $3.98\cdot10^{-5}$  & 9.5        & $\rm MM_-M_+$  & $2.1\cdot10^{-9}$ & good \\
HD 180161 & 9.91(11)    & 0.46  & 9.41          & 0.92                & \ldots     & \ldots         & \ldots            & \ldots \\
HD 220182 & 7.678(23)   & 0.13  & 7.62002(21)   & $2.37\cdot10^{-4}$  & 13.7       & $\rm MM_+$     & $2.9\cdot10^{-4}$ & fair, long \\
SAO 51891 & 2.4179(41)  & 0.039 & 2.40          & 0.98                & \ldots     & \ldots         & \ldots            & \ldots \\
\hline
\end{tabular}
\tablefoot{
\tablefoottext{a}{Numbers in parentheses show the error in the last digits of the period estimates.}
\tablefoottext{b}{Cycle type: "M" cycle found from $M$ results, "A" cycle found from $A$ results, "$\rm M_-$" cycle found from $M-A/2$ results, "$\rm M_+$" cycle found from $M+A/2$ results}
\tablefoottext{c}{$FAP$ grade: "excellent" $FAP \le 10^{-9}$, "good" $10^{-9} < FAP \le 10^{-5}$, "fair" $10^{-5} < FAP \le 10^{-2}$, "poor" $10^{-2} < FAP \le 10^{-1}$, "long" cycle longer than half the length of the photometric record}}
\label{pertab}
\end{table*}

Here we describe the methodology used for analysing the photometric and spectroscopic observations. We summarize the basic characteristics of the CPS fits in Table \ref{modeltab} by giving for each star the number of independent reliable datasets with non-periodic $K=0$ and periodic $K>0$ order fits as well as the mean V-band model amplitudes from the $K>0$ fits. In Table \ref{pertab} we then present the main results from the period analysis. Results from the spectroscopy are presented in Table \ref{rhktab}.

\subsection{Rotation}
\label{sectrot}

The photometric periods derived from the independent datasets for each star show various levels of repeatability from star to star.  These period variations may be the result of sparse data, low amplitude, or low $S/N$ observations \citep{lehtinen2011continuous}. On the other hand, they may be signs of surface differential rotation or active region growth and decay occurring simultaneously at various latitudes and longitudes on the star. For example, a large active region may be forming in one location while another is decaying at a second location well separated in longitude. The resulting shifting phases of light curve minima may result in measured photometric periods that do not correspond to the true rotation period on any latitude. We have been careful to minimize these problems by suitably defining the independent datasets as described in Section \ref{sectcps} above, in particular by limiting the duration of any dataset to 30~d. Therefore, we will assume, as in \cite{hall1991learning}, that the observed range of periods for a particular star is a measure of the differential rotation with stellar latitude.

We characterize the mean rotation periods of our stars by the weighted mean
\begin{equation}
P_{\rm w} = \frac{\sum w_iP_i}{\sum w_i}
\label{pweq}
\end{equation}
and the weighted standard deviation
\begin{equation}
\Delta P_{\rm w} = \sqrt{\frac{\sum w_i(P_i-P_{\rm w})^2}{\sum w_i}}
\label{dpweq}
\end{equation}
of the independent period estimates, where the weights are the inverse square errors of the individual periods, $w_i=\sigma_{{\rm P},i}^{-2}$. The mean rotation periods $P_{\rm w}$ of the stars are listed in Table \ref{pertab} where their errors are given as standard errors of the mean, $\sigma_{\rm P, w}=\Delta P_{\rm w}n_{K>0}^{-1/2}$, calculated from the weighted standard deviation and the total number $n_{K>0}$ of independent datasets with a periodic model. As the differential rotation estimate we use the relative $\pm3\sigma$ range of the period fluctuations \citep{jetsu1993decade}
\begin{equation}
Z = \frac{6\Delta P_{\rm w}}{P_{\rm w}}.
\end{equation}

As mentioned above, there are certain caveats to directly interpreting the $Z$ value as a measure of differential rotation. If we can be certain that the entire range of periods is due to differential rotation, the $Z$ value should approximate the relative difference between the fastest and slowest rotating spot areas on the star, approximately corresponding to the 99\% interval of the estimated period values. Moreover, if these extreme period values correspond to the rotation periods at the equator and the poles of the star, there would be an unambiguous connection between $Z$ and the differential rotation coefficient $k=\Delta\Omega/\Omega_{\rm eq}$. In practice we know from solar observations and Doppler imaging of other stars that this is generally not the case, and the observed spot distribution is confined to a narrower latitude band. The observed range of rotation periods from the spots is thus expected to be smaller than the full range, thus predicting $Z<k_{\rm true}$. On the other hand, observational errors, sparse data coverage, and the need to use short datasets to get local period estimates will cause additional uncertainty in the period values and cause somewhat increased values of $Z$. In general, the best scaling of the period variations into a differential rotation estimate, minimizing both over- and underestimation of $k$, is not an obvious choice. We recommend that our $Z$ values be primarily used to determine the functional relation of the differential rotation to other astrophysical quantities or as an indicator of whether the differential rotation of a star is weak or strong. We note that the exact numerical factor chosen for the calculation of $Z$ does not affect the results presented in Sect. \ref{diffsect}.

Because of the above mentioned uncertainties, we report the raw $Z$ values in Table \ref{pertab}. There is a tendency for the large $Z$ values to be found in stars with poorer observational coverage and fewer available independent period estimates. However, at the same time, the star with the poorest observational coverage, SAO 51891, has one of the smallest values of $Z$. In Sect. \ref{diffsect} we further investigate the effect of the spurious period fluctuations on the reliability of differential rotation estimation from the $Z$ values.

\subsection{Active longitudes}

The longitudinal distribution of major spot areas is tracked by the light curve minimum epochs $t_{\rm min}$. Using the rotation period $P$, we can transform these into minimum phases $\phi_{\rm min} = (t_{\rm min}-t_0)/P\ \rm mod\ 1$, where $t_0$ is an arbitrary epoch chosen to define the zero phase. Each dark spot on a star will contribute a depression in the observed brightness and for equally strong spots with a phase separation larger than $\Delta\phi\approx0.33$ these can be observed as separate light curve minima \citep{lehtinen2011continuous}.

If there are active longitudes present on a spotted star, we expect to observe a long-term phase coherence of $t_{\rm min}$ with an active longitude period $P_{\rm al}$ \citep{jetsu1996active}. This period can be found for example by using the Kuiper test \citep{kuiper1960tests} with a range of folding period values. We applied the unweighted Kuiper test as formulated by \cite{jetsu1996searching} on the primary and secondary light curve minimum epochs $t_{\rm min,1}$ and $t_{\rm min,2}$ from the CPS results. For each detected $P_{\rm al}$ we calculated the error estimate using bootstrap.

The active longitude periods and their critical level values $Q_{\rm K}$ for the Kuiper test, i.e. the $p$-values against the null hypothesis of uniform distribution of $t_{\rm min}$, are both listed in Table \ref{pertab}. We found active longitudes for nine of our sample stars with $Q_{\rm K}<0.01$ and two more with $Q_{\rm K}<0.1$. These stars show a wide range of different behaviours from the very weak and only occasionally appearing phase coherence on HD~63433 to the stable and persistent active longitudes on HD~116956. In the case of stars for which no $P_{\rm al}$ could be found with $Q_{\rm K}<0.1$, we display the best candidate period from the Kuiper periodogram in the vicinity of their $P_{\rm w}$. We have not calculated error estimates for these periods since they are very uncertain.

As is visible from the light curve minimum phases plotted for the individual stars in Figs. \ref{figres1}--\ref{figres6} using their $P_{\rm al}$, the estimated active longitude periods are generally only average values over the full observation records. Most stars with active longitudes show either some jumps in the minimum phases or more gradual phase migration. The best defined migration patterns are seen on HD~70573, HD~82443 and HD~220182. These are discussed further under the individual stars in Sect. \ref{individualsect}.

\subsection{Activity cycles}

From even a cursory inspection of our plots below, it is evident that the active stars show long term variability that can be linked to variations in the activity level and often appears to follow cyclic patterns. Like \cite{rodono2000magnetic}, we applied the period detection method of \cite{horne1986prescriotion} (hereafter the HB method) to four parameters from the CPS results: the mean magnitude $M$, the light curve amplitude $A$, and the combined values $M-A/2$ and $M+A/2$. These parameters describe respectively the axisymmetric and non-axisymmetric parts of the spot distribution and the minimum and maximum spotted area on the star. We excluded SAO 51891 from this cycle search since its photometric record is dominated by the long gap of 15 years in the observations and the remaining data covers a time span that is too short to warrant reliable cycle detection.

We grade these cycles by the false alarm probabilities ($FAP$) given by the period detection method following the scheme of \cite{baliunas1995chromosphere}: ``excellent'' for $FAP \le 10^{-9}$, ``good'' for $10^{-9} < FAP \le 10^{-5}$, ``fair'' for $10^{-5} < FAP \le 10^{-2}$ and ``poor'' for $10^{-2} < FAP \le 10^{-1}$. Our secondary grade label ``long'' differs from their scheme. It is given to cycle periods that have lengths larger than half the length of the whole photometric record and are thus less certain than the shorter cycle periods. We do not state the error estimates for the cycle periods given by the HB method since the true uncertainties of the cycles are dominated by their non-periodic behaviour and are impossible to estimate from the currently available data.

We found cyclic behaviour in 18 out of 20 stars and evidence of two separate cycles in three of the stars. However, many of these cycle periods either have a large $FAP$ or are of the same order as the length of the photometric record. In all except five cases the cycle periods are found in the $M$ data. Typically the same periods are also present in the $M-A/2$ or $M+A/2$ data, often with a somewhat larger $FAP$ value. Overall, the $A$ data shows much less evidence of cyclic behaviour than the other parameters $M$, $M-A/2$, and $M+A/2$. A cycle in $A$ can be found only for seven stars. For three of them this cycle is unique to the $A$ data.

In the case of HD~82558 there is a strong signature of a long cycle in all four parameters. They fall into a range from 14.5~yr to 18.0~yr, suggesting that they correspond to the same physical cycle, but they do not give a clear indication of what the exact length of this cycle is. In this case we have to conclude that the cycle of HD~82558 is too long to allow more than an estimate of the period.

\subsection{Chromospheric indices}

\begin{table}
\caption{Chromospheric emission indices $\log{R'_{\rm HK}}$ and activity classes from this work and the chromospheric emission indices $\log{R'_{\rm HK, calib}}$ from the literature sources used for calibration.}
\center
\begin{tabular}{lccc}
\hline
\hline
Star & $\log{R'_{\rm HK}}$ & activity & $\log{R'_{\rm HK, calib}}$ \\
     &                     & class\tablefootmark{a} & \\
\hline
HD 1405   & -4.217 & A  & \ldots \\
HD 10008  & -4.480 & A  & -4.530\tablefootmark{1} \\
HD 26923  & -4.618 & MA & -4.555\tablefootmark{1} \\
          &        &    & -4.521\tablefootmark{2} \\
HD 29697  & -4.036 & VA & -4.066\tablefootmark{1} \\
HD 41593  & -4.427 & A  & -4.456\tablefootmark{1} \\
HD 43162  & -4.425 & A  & -4.480\tablefootmark{2} \\
HD 63433  & -4.452 & A  & -4.424\tablefootmark{1} \\
HD 70573  & -4.488 & A  & -4.10\tablefootmark{3} \\
HD 72760  & -4.609 & MA & -4.454\tablefootmark{1} \\
          &        &    & -4.57\tablefootmark{3} \\
HD 73350  & -4.700 & MA & -4.657\tablefootmark{1} \\
HD 82443  & -4.286 & A  & -4.264\tablefootmark{1} \\
          &        &    & -4.234\tablefootmark{2} \\
          &        &    & -4.30\tablefootmark{3} \\
HD 82558  & -4.079 & VA & -4.03\tablefootmark{3} \\
HD 116956 & -4.366 & A  & -4.447\tablefootmark{1} \\
HD 128987 & -4.505 & MA & -4.439\tablefootmark{2} \\
HD 130948 & -4.533 & MA & \ldots \\
HD 135599 & -4.462 & A  & -4.663\tablefootmark{1} \\
HD 141272 & -4.566 & MA & -4.452\tablefootmark{1} \\
HD 171488 & -4.175 & VA & -4.21\tablefootmark{3} \\
HD 180161 & -4.541 & MA & \ldots \\
HD 220182 & -4.388 & A  & -4.503\tablefootmark{1} \\
SAO 51891 & -4.327 & A  & -4.05\tablefootmark{3} \\
\hline
\end{tabular}
\tablefoot{
\tablefoottext{a}{Activity classes: ``VA'' very active stars with $\log{R'_{\rm HK}}>-4.20$, ``A'' active stars with $-4.20\ge\log{R'_{\rm HK}}>-4.50$, ``MA'' moderately active stars with $-4.50\ge\log{R'_{\rm HK}}>-4.75$.}
}
\tablebib{
\tablefoottext{1}{\cite{gray2003contributions}}
\tablefoottext{2}{\cite{gray2006contributions}}
\tablefoottext{3}{\cite{white2007high}}
}
\label{rhktab}
\end{table}

We quantified the chromospheric activity of our sample stars by measuring the Mount Wilson S-index \citep{vaughan1978flux}
\begin{equation}
S = \alpha \frac{H+K}{R+V}
\end{equation}
and transforming it into the fractional emission flux $\log{R'_{\rm HK}}$ at the \ion{Ca}{ii} H\&K lines. The process involves measuring the emission flux through two triangular bands $H$ and $K$ with FWHM of 1.09 \AA \ centred at the two line cores and normalizing it to the flux at two flat continuum bands $V$ and $R$ with a full width of 20 \AA \ and centred around 3901 \AA \ and 4001 \AA. The normalizing constant $\alpha$ is needed to adjust the measured values to the original Mount Wilson HKP-1 and HKP-2 spectrometers. We chose to calibrate our measurements against the values of \cite{gray2003contributions, gray2006contributions} since they provide uniform measurements for most of our sample stars. To improve the calibration at the strong emission end, we also used values of \cite{white2007high}. For normalized FIES spectra we obtained $\alpha=19.76$.

The S-indices are transformed into fractional fluxes $R_{\rm HK}=F_{\rm HK}/\sigma T_{\rm eff}^4$ in the line cores with respect to the black-body luminosity of the stars using the conversion formula \citep{middelkoop1982magnetic}
\begin{equation}
R_{\rm HK} = 1.34\cdot10^{-4}C_{\rm cf}S
\end{equation}
where the colour dependent conversion factor
\begin{equation}
\log{C_{\rm cf}} = 0.25(B-V)^3 - 1.33(B-V)^2 + 0.43(B-V) + 0.24
\end{equation}
is applicable to main-sequence stars with $0.3 \le B-V \le 1.6$ \citep{rutten1984magnetic}. This value still contains a photospheric contribution which is described as
\begin{equation}
\log{R_{\rm phot}} = -4.898 + 1.918(B-V)^2 - 2.893(B-V)^3
\end{equation}
for stars with $B-V \ge 0.44$ \citep{noyes1984magnetic}. This is subtracted from the $R_{\rm HK}$ value to get the corrected value
\begin{equation}
R'_{\rm HK} = R_{\rm HK} - R_{\rm phot}.
\end{equation}

The final logarithmic $\log{R'_{\rm HK}}$ values are given in Table \ref{rhktab} together with values from the sources used for calibrating the measured S-indices. All of our stars are either ``active'' or ``very active'' according to the classification of \cite{henry1996survey}, i.e. their $\log{R'_{\rm HK}}>-4.75$. To help differentiate our stars according to their chromospheric activity level, we divide the ``active'' class of \cite{henry1996survey} into two subclasses (``active'' and ``moderately active'') at its mean $\log{R'_{\rm HK}}$. The resulting activity classes are ``very active'' for $\log{R'_{\rm HK}}>-4.20$, ``active'' for $-4.20\ge\log{R'_{\rm HK}}>-4.50$ and ``moderately active'' for $-4.50\ge\log{R'_{\rm HK}}>-4.75$.

\section{Individual stars}
\label{individualsect}

\begin{longtab}
\begin{landscape}
\begin{longtable}{lccccc|lccccc}
\caption{Rotation periods, activity indices, age estimates and identifications of kinematic group membership for the sample stars from previous studies.} \\
\hline
\hline
Star & $P_{\rm rot} [d]$ & $\log{R'_{\rm HK}}$ & $\log{R_{\rm X}}$ & age\tablefootmark{a} [Myr] & group\tablefootmark{b} &
Star & $P_{\rm rot} [d]$ & $\log{R'_{\rm HK}}$ & $\log{R_{\rm X}}$ & age\tablefootmark{a} [Myr] & group\tablefootmark{b} \\
\hline
\endfirsthead
\caption{continued.}\\
\hline
\hline
Star & $P_{\rm rot} [d]$ & $\log{R'_{\rm HK}}$ & $\log{R_{\rm X}}$ & age\tablefootmark{a} [Myr] & group\tablefootmark{b} &
Star & $P_{\rm rot} [d]$ & $\log{R'_{\rm HK}}$ & $\log{R_{\rm X}}$ & age\tablefootmark{a} [Myr] & group\tablefootmark{b} \\
\hline
\endhead
\hline
\endfoot
HD 1405  & 1.745\tablefootmark{1}   & -3.85\tablefootmark{26}   & \ldots                  & $T_{\rm iso}$: 20\tablefootmark{2}      & AB\tablefootmark{3,48,49,50} & HD 82558  & 1.6009\tablefootmark{11}  & -3.66\tablefootmark{29}  & -3.03\tablefootmark{29} & $T_{\rm X}$: 80\tablefootmark{29}    & IC\tablefootmark{51} \\
         & 1.76159\tablefootmark{2} &                           &                         & $T_{\rm iso}$: 30--80\tablefootmark{45} & LA\tablefootmark{46}         &           & 1.6603\tablefootmark{14}  & -4.03\tablefootmark{37}  & -3.06\tablefootmark{43} &                                      & \\
         & 1.76\tablefootmark{3}    &                           &                         &                                         &                              &           & 1.5978\tablefootmark{15}  & -3.63\tablefootmark{39}  & -3.17\tablefootmark{44} &                                      & \\
HD 10008 & 7.15\tablefootmark{4}    & -4.2\tablefootmark{4}     & -4.48\tablefootmark{6}  & $T_{\rm gyro}$: 211\tablefootmark{4}    & HLA\tablefootmark{49}        &           & 1.60114\tablefootmark{16} &                          &                         &                                      & \\
         &                          & -4.38\tablefootmark{6}    & -4.43\tablefootmark{29} & $T_{\rm HK}$: 360\tablefootmark{28}     & LA\tablefootmark{29,46}      &           & 1.6007\tablefootmark{17}  &                          &                         &                                      & \\
         &                          & -4.530\tablefootmark{27}  &                         & $T_{\rm HK}$: 130\tablefootmark{29}     & TWA\tablefootmark{51}        &           & 1.6043\tablefootmark{18}  &                          &                         &                                      & \\
         &                          & -4.427\tablefootmark{28}  &                         & $T_{\rm X}$: 440\tablefootmark{29}      &                              &           & 1.60514\tablefootmark{19} &                          &                         &                                      & \\
         &                          & -4.29\tablefootmark{29}   &                         &                                         &                              & HD 116956 & 7.80\tablefootmark{6}     & -4.22\tablefootmark{41}  & -4.48\tablefootmark{6}  & $T_{\rm HK}$: 260\tablefootmark{28}  & LA\tablefootmark{6,29,46} \\
         &                          & -4.45\tablefootmark{30}   &                         &                                         &                              &           & 7.8288\tablefootmark{20}  & -4.447\tablefootmark{27} & -4.36\tablefootmark{29} & $T_{\rm HK}$: 70\tablefootmark{41}   & TWA\tablefootmark{51} \\
         &                          & -4.41\tablefootmark{31}   &                         &                                         &                              &           &                           & -4.385\tablefootmark{28} & -4.35\tablefootmark{35} & $T_{\rm X}$: 360\tablefootmark{46}   & \\
HD 26923 & \ldots                   & -4.52\tablefootmark{6}    & -4.64\tablefootmark{6}  & \ldots                                  & UMa\tablefootmark{6,46,51}   &           &                           & -4.22\tablefootmark{35}  &                         &                                      & \\
         &                          & -4.503\tablefootmark{9}   & -4.41\tablefootmark{35} &                                         &                              & HD 128987 & 9.35\tablefootmark{6}     & -4.451\tablefootmark{28} & -4.78\tablefootmark{6}  & $T_{\rm HK}$: 430\tablefootmark{28}  & IC\tablefootmark{29} \\
         &                          & -4.555\tablefootmark{27}  &                         &                                         &                              &           &                           & -4.51\tablefootmark{29}  & -4.86\tablefootmark{29} & $T_{\rm HK}$: 650\tablefootmark{29}  & \\
         &                          & -4.496\tablefootmark{32}  &                         &                                         &                              &           &                           & -4.45\tablefootmark{30}  &                         & $T_{\rm X}$: 680\tablefootmark{29}   & \\
         &                          & -4.49\tablefootmark{33}   &                         &                                         &                              &           &                           & -4.439\tablefootmark{34} &                         &                                      & \\
         &                          & -4.521\tablefootmark{34}  &                         &                                         &                              & HD 130948 & 7.85\tablefootmark{6}     & -4.45\tablefootmark{6}   & -4.69\tablefootmark{6}  & $T_{\rm HK}$: 380\tablefootmark{28}  & \ldots \\
         &                          & -4.39\tablefootmark{35}   &                         &                                         &                              &           &                           & -4.434\tablefootmark{28} & -4.65\tablefootmark{29} & $T_{\rm HK}$: 190\tablefootmark{29}  & \\
HD 29697 & 3.936\tablefootmark{5}   & -4.066\tablefootmark{27}  & -3.25\tablefootmark{35} & \ldots                                  & UMa\tablefootmark{46}        &           &                           & -4.34\tablefootmark{29}  & -4.50\tablefootmark{35} & $T_{\rm HK}$: 870\tablefootmark{36}  & \\
         &                          & -4.06\tablefootmark{35}   &                         &                                         &                              &           &                           & -4.64\tablefootmark{35}  &                         & $T_{\rm X}$: 330\tablefootmark{29}   & \\
HD 41593 & 7.97\tablefootmark{6}    & -4.36\tablefootmark{6}    & -4.59\tablefootmark{6}  & $T_{\rm HK}$: 320\tablefootmark{28,36}  & UMa\tablefootmark{6,46,51}   &           &                           & -4.50\tablefootmark{36}  &                         &                                      & \\
         &                          & -4.456\tablefootmark{27}  &                         &                                         &                              &           &                           & -4.45\tablefootmark{42}  &                         &                                      & \\
         &                          & -4.409\tablefootmark{28}  &                         &                                         &                              & HD 135599 & 5.97\tablefootmark{6}     & -4.663\tablefootmark{27} & -4.94\tablefootmark{6}  & $T_{\rm HK}$: 200\tablefootmark{29}  & UMa\tablefootmark{6} \\
HD 43162 & 7.24\tablefootmark{7}    & -4.40\tablefootmark{6}    & -4.31\tablefootmark{6}  & $T_{\rm HK}$: 460\tablefootmark{28}     & IC\tablefootmark{29,51}      &           &                           & -4.35\tablefootmark{29}  & -5.11\tablefootmark{29} & $T_{\rm HK}$: 950\tablefootmark{36}  & \\
         &                          & -4.460\tablefootmark{28}  & -4.29\tablefootmark{29} & $T_{\rm HK}$: 280\tablefootmark{29}     &                              &           &                           & -4.52\tablefootmark{35}  &                         & $T_{\rm X}$: 1340\tablefootmark{29}  & \\
         &                          & -4.39\tablefootmark{29}   &                         & $T_{\rm HK}$: 370\tablefootmark{36}     &                              &           &                           & -4.52\tablefootmark{36}  &                         &                                      & \\
         &                          & -4.480\tablefootmark{34}  &                         & $T_{\rm X}$: 280\tablefootmark{29}      &                              & HD 141272 & 14.045\tablefootmark{4}   & -4.2\tablefootmark{4}    & -4.62\tablefootmark{6}  & $T_{\rm gyro}$: 773\tablefootmark{4} & HLA\tablefootmark{52} \\
         &                          & -4.40\tablefootmark{36}   &                         &                                         &                              &           & 14.01\tablefootmark{6}    & -4.452\tablefootmark{27} & -4.40\tablefootmark{29} & $T_{\rm HK}$: 420\tablefootmark{28}  & LA\tablefootmark{29,46} \\
HD 63433 & 6.46\tablefootmark{6}    & -4.34\tablefootmark{6}    & -4.55\tablefootmark{6}  & $T_{\rm HK}$: 170\tablefootmark{28}     & UMa\tablefootmark{6,46,51}   &           &                           & -4.447\tablefootmark{28} & -4.40\tablefootmark{35} & $T_{\rm HK}$: 200\tablefootmark{29}  & \\
         &                          & -4.424\tablefootmark{27}  & -4.52\tablefootmark{35} & $T_{\rm HK}$: 350\tablefootmark{36}     &                              &           &                           & -4.35\tablefootmark{29}  &                         & $T_{\rm HK}$: 340\tablefootmark{36}  & \\
         &                          & -4.331\tablefootmark{28}  &                         &                                         &                              &           &                           & -4.39\tablefootmark{35}  &                         & $T_{\rm X}$: 430\tablefootmark{29}   & \\
         &                          & -4.39\tablefootmark{35}   &                         &                                         &                              &           &                           & -4.39\tablefootmark{36}  &                         &                                      & \\
         &                          & -4.39\tablefootmark{36}   &                         &                                         &                              &           &                           & -4.49\tablefootmark{39}  &                         &                                      & \\
HD 70573 & 3.296\tablefootmark{5}   & -4.187\tablefootmark{28}  & \ldots                  & $T_{\rm HK}$: 50\tablefootmark{28}      & HLA\tablefootmark{49}        & HD 171488 & 1.338\tablefootmark{5}    & -4.21\tablefootmark{37}  & -3.61\tablefootmark{43} & $T_{\rm iso}$: 30\tablefootmark{21}  & LA\tablefootmark{46} \\
         &                          & -4.10\tablefootmark{37}   &                         &                                         & LA\tablefootmark{46}         &           & 1.3371\tablefootmark{21}  &                          &                         & $T_{\rm iso}$: 50\tablefootmark{47}  & \\
HD 72760 & \ldots                   & -4.39\tablefootmark{6}    & -4.78\tablefootmark{6}  & $T_{\rm HK}$: 200\tablefootmark{28}     & Hya\tablefootmark{29,46}     &           & 1.313\tablefootmark{22}   &                          &                         & $T_{\rm Li}$: 50\tablefootmark{21}   & \\
         &                          & -4.454\tablefootmark{27}  & -4.76\tablefootmark{29} & $T_{\rm HK}$: 440\tablefootmark{29}     &                              &           & 1.337\tablefootmark{23}   &                          &                         &                                      & \\
         &                          & -4.350\tablefootmark{28}  &                         & $T_{\rm HK}$: 310\tablefootmark{36}     &                              &           & 1.3372\tablefootmark{24}  &                          &                         &                                      & \\
         &                          & -4.454\tablefootmark{29}  &                         & $T_{\rm X}$: 690\tablefootmark{29}      &                              & HD 180161 & 5.49\tablefootmark{4}     & -4.6\tablefootmark{4}    & -4.67\tablefootmark{6}  & $T_{\rm gyro}$: 127\tablefootmark{4} & Hya\tablefootmark{6,29,46} \\
         &                          & -4.38\tablefootmark{36}   &                         &                                         &                              &           & 9.7\tablefootmark{6}      & -4.44\tablefootmark{6}   & -4.68\tablefootmark{29} & $T_{\rm HK}$: 530\tablefootmark{28}  & \\
         &                          & -4.57\tablefootmark{37}   &                         &                                         &                              &           &                           & -4.481\tablefootmark{28} & -4.66\tablefootmark{35} & $T_{\rm HK}$: 690\tablefootmark{29}  & \\
         &                          & -4.21\tablefootmark{38}   &                         &                                         &                              &           &                           & -4.52\tablefootmark{36}  &                         & $T_{\rm HK}$: 950\tablefootmark{36}  & \\
HD 72760 &                          & -4.28\tablefootmark{39}   &                         &                                         &                              & HD 180161 &                           & -4.52\tablefootmark{29}  &                         & $T_{\rm X}$: 560\tablefootmark{29}   & \\
HD 73350 & 6.14\tablefootmark{6}    & -4.49\tablefootmark{6}    & -4.85\tablefootmark{6}  & $T_{\rm HK}$: 1170\tablefootmark{28}    & Hya\tablefootmark{6,29}      &           &                           & -4.52\tablefootmark{35}  &                         &                                      & \\
         & 12.3\tablefootmark{8}    & -4.48\tablefootmark{8}    & -4.78\tablefootmark{29} & $T_{\rm HK}$: 300\tablefootmark{29}     &                              & HD 220182 & 7.489\tablefootmark{4}    & -4.2\tablefootmark{4}    & -4.59\tablefootmark{6}  & $T_{\rm gyro}$: 230\tablefootmark{4} & \ldots \\
         &                          & -4.657\tablefootmark{27}  & -4.80\tablefootmark{43} & $T_{\rm HK}$: 830\tablefootmark{36}     &                              &           & 7.66\tablefootmark{6}     & -4.34\tablefootmark{6}   & -4.61\tablefootmark{29} & $T_{\rm HK}$: 100\tablefootmark{28}  & \\
         &                          & -4.605\tablefootmark{28}  &                         & $T_{\rm X}$: 460\tablefootmark{29}      &                              &           &                           & -4.503\tablefootmark{27} & -4.58\tablefootmark{35} & $T_{\rm HK}$: 40\tablefootmark{29}   & \\
         &                          & -4.40\tablefootmark{29}   &                         &                                         &                              &           &                           & -4.269\tablefootmark{28} &                         & $T_{\rm HK}$: 260\tablefootmark{36}  & \\
         &                          & -4.49\tablefootmark{36}   &                         &                                         &                              &           &                           & -4.15\tablefootmark{29}  &                         & $T_{\rm X}$: 550\tablefootmark{29}   & \\
         &                          & -4.49\tablefootmark{40}   &                         &                                         &                              &           &                           & -4.37\tablefootmark{35}  &                         &                                      & \\
HD 82443 & 5.409\tablefootmark{4}   & -4.22\tablefootmark{6}    & -4.01\tablefootmark{6}  & $T_{\rm gyro}$: 130\tablefootmark{4}    & LA\tablefootmark{29,46}      &           &                           & -4.37\tablefootmark{36}  &                         &                                      & \\
         & 5.43\tablefootmark{5}    & -4.211\tablefootmark{9}   & -3.93\tablefootmark{29} & $T_{\rm HK}$: 20\tablefootmark{29}      & Ple\tablefootmark{6}         & SAO 51891 & 2.42\tablefootmark{5}     & -4.05\tablefootmark{37}  & \ldots                  & \ldots                               & LA\tablefootmark{46} \\
         & 5.42\tablefootmark{6}    & -4.0\tablefootmark{11}    & -3.94\tablefootmark{35} & $T_{\rm X}$: 220\tablefootmark{29}      & THA\tablefootmark{51}        &           & 2.62\tablefootmark{25}    &                          &                         &                                      & \\
         & 6\tablefootmark{9}       & -4.264\tablefootmark{27}  &                         &                                         &  \\
         & 5.40\tablefootmark{10}   & -4.08\tablefootmark{29}   &                         &                                         &  \\
         & 5.405\tablefootmark{11}  & -4.234\tablefootmark{34}  &                         &                                         &  \\
         & 5.377\tablefootmark{12}  & -4.08\tablefootmark{35}   &                         &                                         &  \\
         & 5.424\tablefootmark{13}  & -4.30\tablefootmark{37}   &                         &                                         &  \\
         &                          & -4.15\tablefootmark{39}   &                         &                                         &  \\
\end{longtable}\tablefoot{
\tablefoottext{a}{Age estimates are based on: $T_{\rm iso}$ isochrone fitting, $T_{\rm Li}$ lithium abundance, $T_{\rm gyro}$ gyrochronology, $T_{\rm HK}$ \ion{Ca}{ii} H\&K emission, $T_{\rm X}$ X-ray emission.}
\tablefoottext{b}{Labels for kinematic groups: "AB" AB Dor Moving Group, "HLA" Her-Lyr Association, "Hya" Hyades Supercluster, "IC" IC 2391 Supercluster, "LA" Local Association, "Ple" Pleiades Moving Group, "UMa" UMa Moving Group, "THA" Tuc-Hor Association, "TWA" TW Hya Association}
}
\tablebib{
\tablefoottext{1}{\cite{hooten1990photometry}}
\tablefoottext{2}{\cite{strassmeier2006first}}
\tablefoottext{3}{\cite{messina2011race}}
\tablefoottext{4}{\cite{strassmeier2000vienna}}
\tablefoottext{5}{\cite{henry1995automated}}
\tablefoottext{6}{\cite{gaidos2000spectroscopy}}
\tablefoottext{7}{\cite{kajatkari2015periodicity}}
\tablefoottext{8}{\cite{petit2008toroidal}}
\tablefoottext{9}{\cite{baliunas1996magnetic}}
\tablefoottext{10}{\cite{messina1996starspot}}
\tablefoottext{11}{\cite{strassmeier1997starspot}}
\tablefoottext{12}{\cite{messina1999activity}}
\tablefoottext{13}{\cite{messina2002magnetic}}
\tablefoottext{14}{\cite{fekel1986chromospherically}}
\tablefoottext{15}{\cite{strassmeier1988photometric}}
\tablefoottext{16}{\cite{jetsu1993decade}}
\tablefoottext{17}{\cite{kovari2004doppler}}
\tablefoottext{18}{\cite{lehtinen2012spot}}
\tablefoottext{19}{\cite{olspert2015multiperiodicity}}
\tablefoottext{20}{\cite{lehtinen2011continuous}}
\tablefoottext{21}{\cite{strassmeier2003doppler}}
\tablefoottext{22}{\cite{marsden2006surface}}
\tablefoottext{23}{\cite{jarvinen2008magnetic}}
\tablefoottext{24}{\cite{huber2009long}}
\tablefoottext{25}{\cite{biazzo2009young}}
\tablefoottext{26}{\cite{lopezsantiago2010high}}
\tablefoottext{27}{\cite{gray2003contributions}}
\tablefoottext{28}{\cite{isaacson2010chromospheric}}
\tablefoottext{29}{\cite{maldonado2010spectroscopy}}
\tablefoottext{30}{\cite{arriagada2011chromospheric}}
\tablefoottext{31}{\cite{jenkins2011chromospheric}}
\tablefoottext{32}{\cite{noyes1984magnetic}}
\tablefoottext{33}{\cite{soderblom1993dwarf}}
\tablefoottext{34}{\cite{gray2006contributions}}
\tablefoottext{35}{\cite{mishenina2012activity}}
\tablefoottext{36}{\cite{wright2004chromospheric}}
\tablefoottext{37}{\cite{white2007high}}
\tablefoottext{38}{\cite{busa2007caii}}
\tablefoottext{39}{\cite{martinezarnaiz2010chromospheric}}
\tablefoottext{40}{\cite{plavchan2009new}}
\tablefoottext{41}{\cite{montes2001chromospheric}}
\tablefoottext{42}{\cite{henry1996survey}}
\tablefoottext{43}{\cite{vidotto2014stellar}}
\tablefoottext{44}{\cite{hempelmann1995coronal}}
\tablefoottext{45}{\cite{lopezsantiago2003rotational}}
\tablefoottext{46}{\cite{montes2001late}}
\tablefoottext{47}{\cite{frasca2010photometric}}
\tablefoottext{48}{\cite{zuckerman2004young}}
\tablefoottext{49}{\cite{lopezsantiago2006nearest}}
\tablefoottext{50}{\cite{mccarthy2012sizes}}
\tablefoottext{51}{\cite{nakajima2012potential}}
\tablefoottext{52}{\cite{fuhrmann2004nearby}}
}
\label{littab}
\end{landscape}
\end{longtab}

\begin{table}
\caption{Age estimates adopted for the kinematic groups.}
\center
\begin{tabular}{lll}
\hline
\hline
Group                 & abbr. & age [Myr] \\
\hline
AB Dor Moving Group   & AB    & 50--120\tablefootmark{1} \\
Her-Lyr Association   & HLA   & 120--280\tablefootmark{2} \\
Hyades Supercluster   & Hya   & 600\tablefootmark{3} \\
IC 2391 Supercluster  & IC    & 35--55\tablefootmark{3} \\
Local Association     & LA    & 20--150\tablefootmark{3} \\
Pleiades Moving Group & Ple   & 100\tablefootmark{4} \\
UMa Moving Group      & UMa   & 300\tablefootmark{3} \\
Tuc-Hor Association   & THA   & 20--40\tablefootmark{2} \\
TW Hya Association    & TWA   & 4--12\tablefootmark{2} \\
\hline
\end{tabular}
\tablebib{
\tablefoottext{1}{\cite{malo2013bayesian}}
\tablefoottext{2}{\cite{evans2012mapping}}
\tablefoottext{3}{\cite{montes2001late}}
\tablefoottext{4}{\cite{zuckerman2004young}}
}
\label{agetab}
\end{table}

In this section we discuss the noteworthy results for each of the sample stars individually and compare our results with past research. The main characterizing parameters from the literature are the rotation period estimate, the chromospheric \ion{Ca}{ii} H\&K and the coronal X-ray emission indices $\log{R'_{\rm HK}}$ and $\log{R_{\rm X}}$, the reported age estimates, and the identifications of kinematic group membership. These parameters are listed in Table 5 along with references to their sources. Ages adopted for the various kinematic groups are given in Table \ref{agetab}. This table also lists the abbreviations used for the groups.

\begin{figure*}
\centering
\begin{tabular}{cc}
\includegraphics[width=0.485\linewidth]{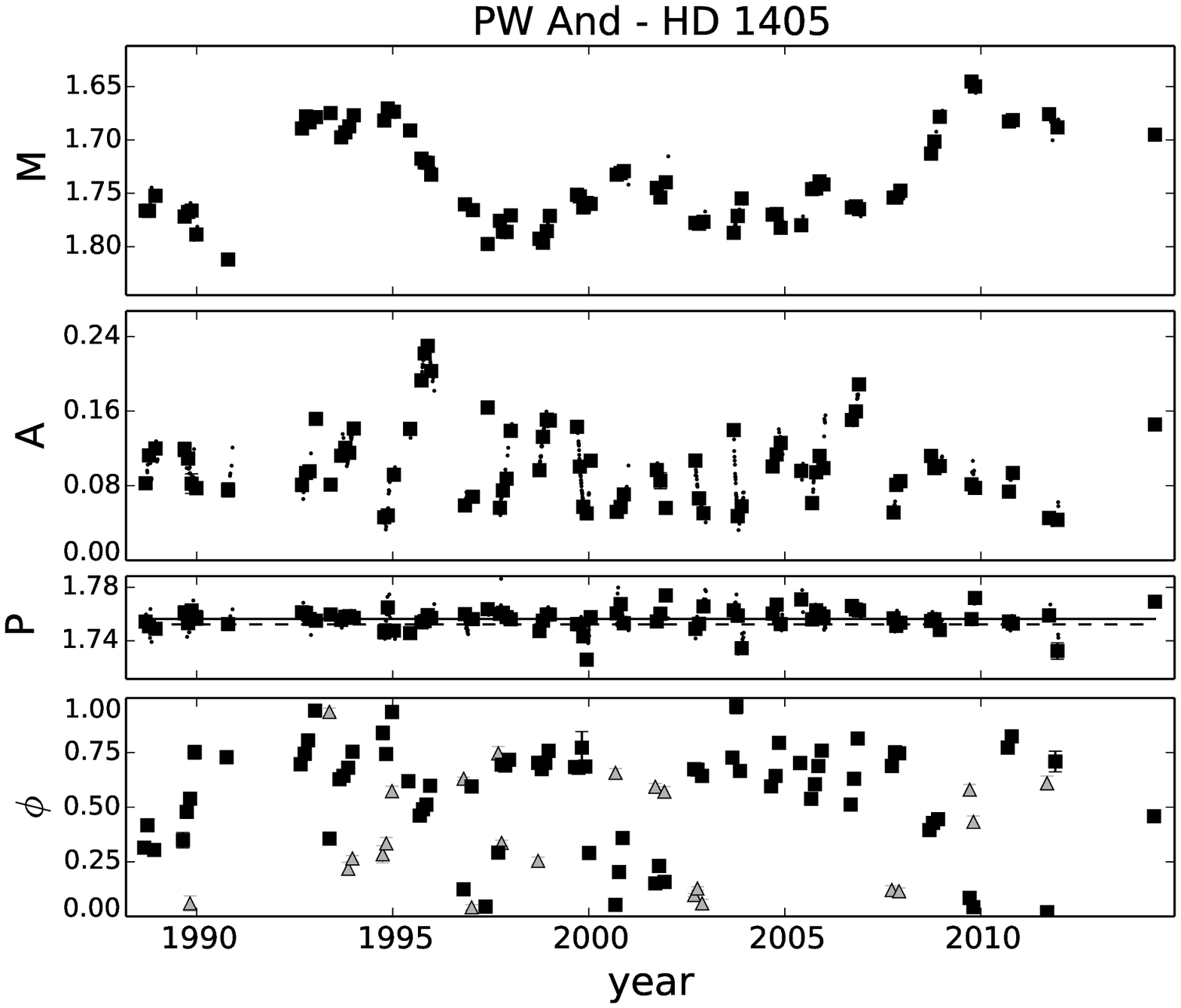} &
\includegraphics[width=0.485\linewidth]{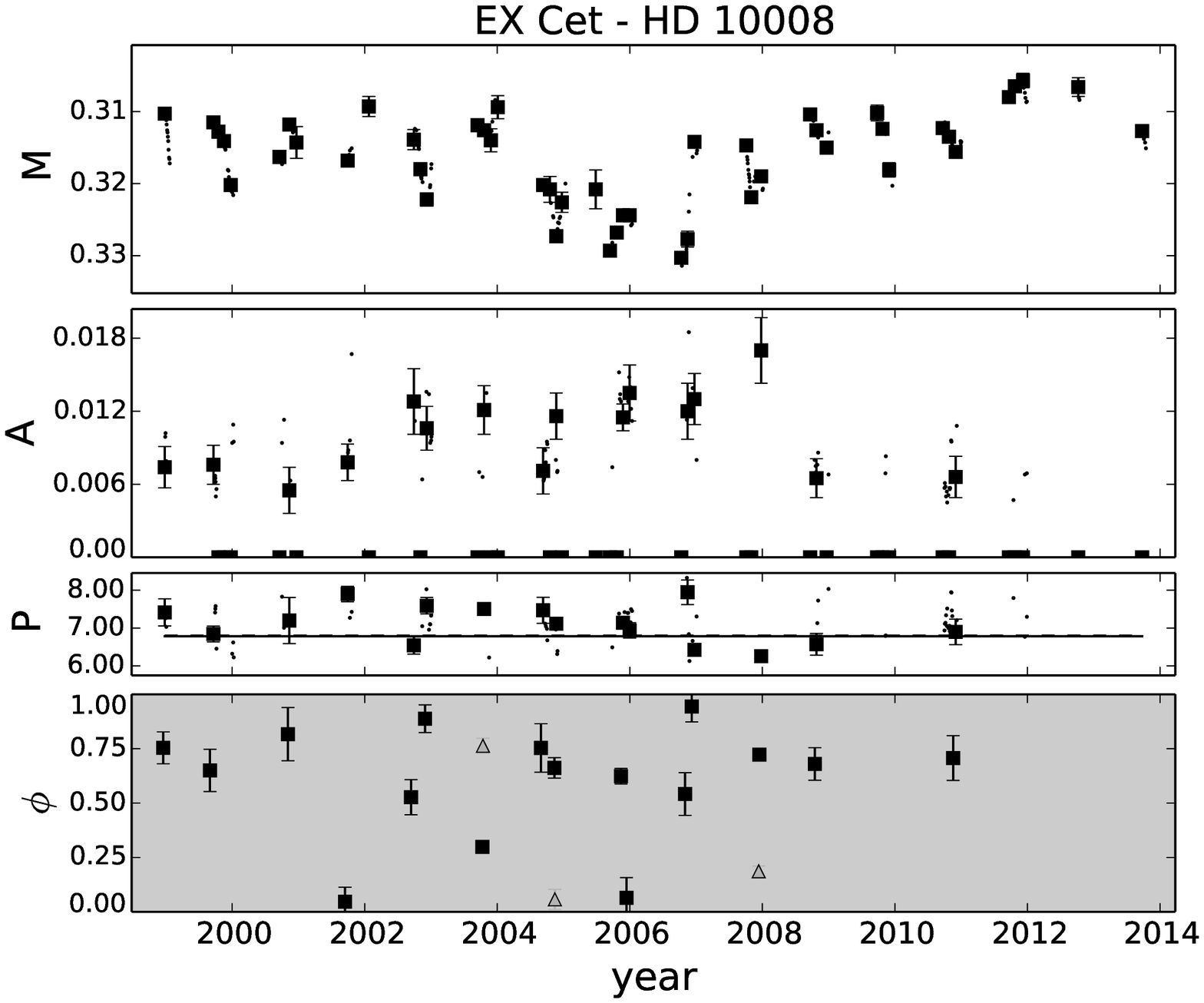} \\
\includegraphics[width=0.485\linewidth]{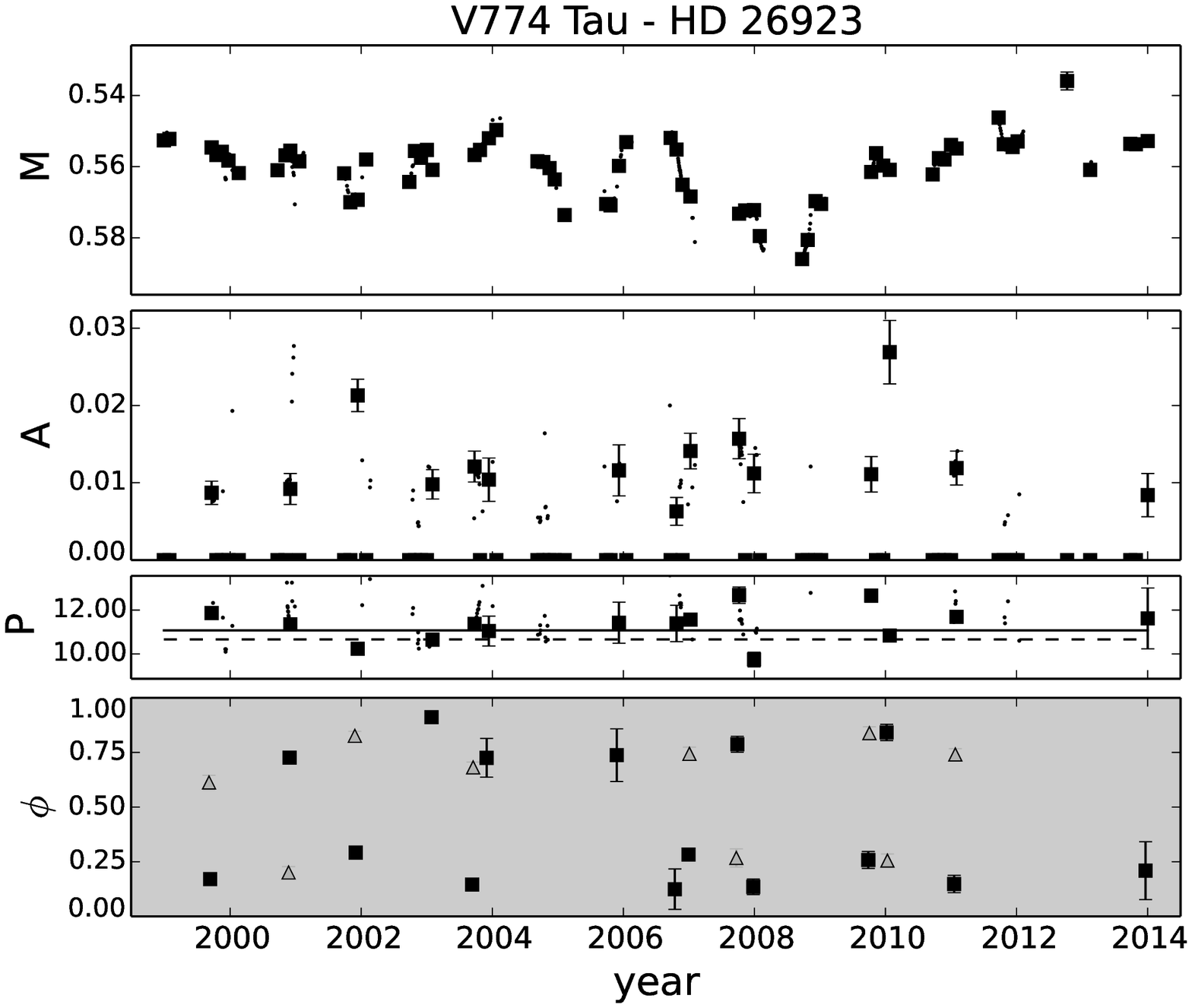} &
\includegraphics[width=0.485\linewidth]{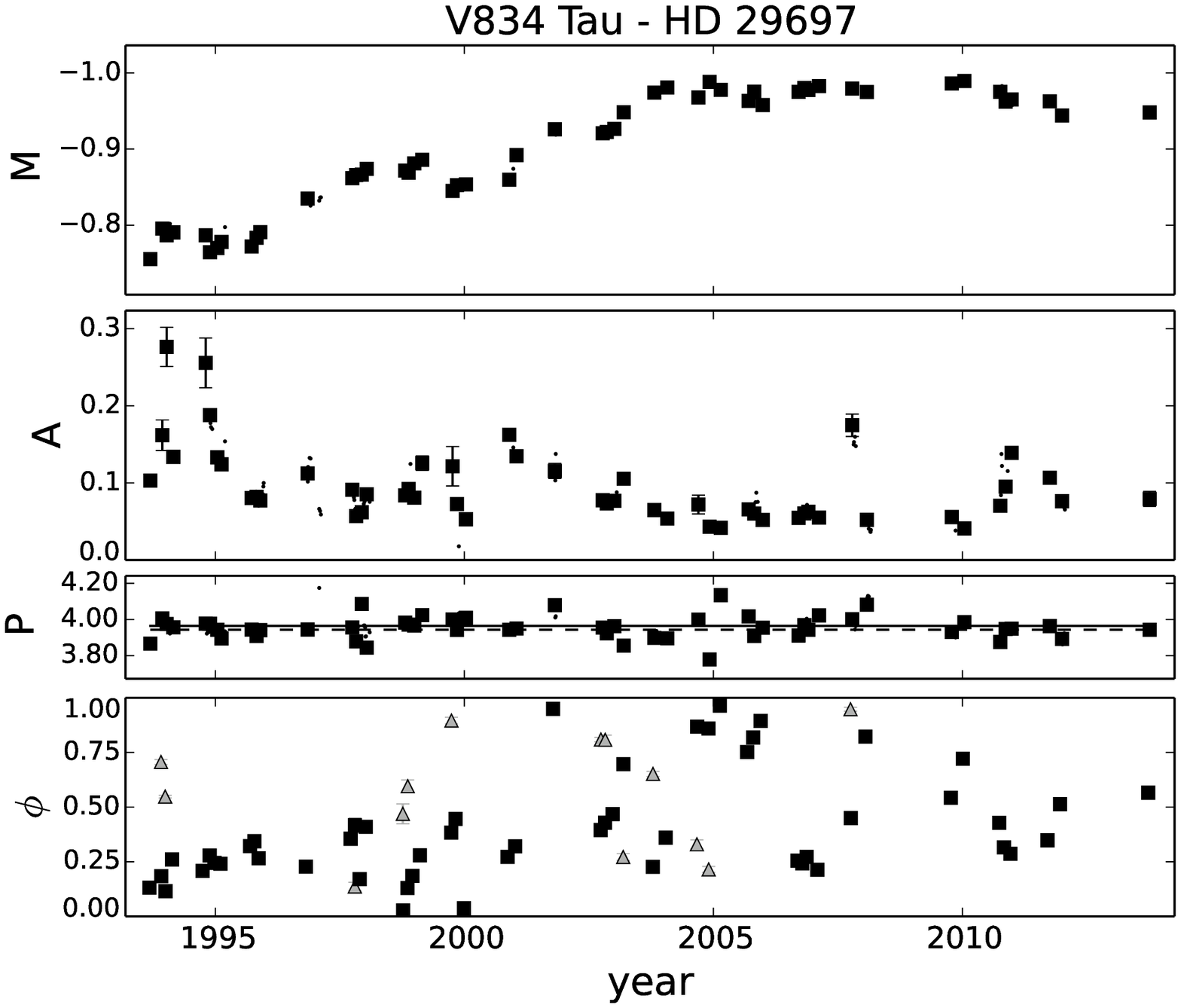}
\end{tabular}
\caption{CPS results of $M$, $A$, $P$ and $\phi_{\rm min}$ for HD 1405, HD 10008, HD 26923, and HD 29697. For $M$, $A$, and $P$ the results from the independent datasets are shown with the black squares with error bars while results from the rest of the datasets are shown with points. Datasets with a constant brightness model ($K=0$) are shown with $A$ set to 0. The light curve minimum phases $\phi_{\rm min}$ phased with $P_{\rm al}$ are shown in the fourth panel with black squares denoting the primary minima and grey triangles the secondary minima. This panel is shaded grey if the candidate $P_{\rm al}$ does not have a significant Kuiper statistic. The mean photometric period $P_{\rm w}$ is shown in the third panel as a solid line and the active longitude period $P_{\rm al}$ with the dashed line. We note that the error bars are often smaller than the plot symbols.}
\label{figres1}
\end{figure*}

The results from the CPS analysis are presented graphically for each star in Figs. \ref{figres1}--\ref{figres6}. These plots show the $M$, $A$, and $P$ results from the independent datasets for each star, as well as the primary and secondary minimum epochs $t_{\rm min}$ folded into minimum phases $\phi_{\rm min}$ with the best periods found by the Kuiper method. In the cases where the folding period does not have a significant Kuiper statistic and cannot be identified as an active longitude period, we have coloured the fourth panel in the plots grey.

If not mentioned otherwise, the stars lack observed stellar companions. For some of the stars distant companions have been found and we note the basic characteristics of each of them. In all of the cases the observed companions are on wide orbits and the primary components in our sample are thus effectively single stars.

\subsection{PW And -- HD 1405}

\object{HD 1405} (\object{PW And}) is an ``active'' rapidly rotating ($P_{\rm w}=1.7562$~d) K2V star. It is among the most active stars in our sample, with the emission index $\log{R'_{\rm HK}}=-4.217$. It is known to have strong rotationally modulated chromospheric emission \citep{montes2001chromospheric, lopezsantiago2006nearest, zhang2015chromospheric}, and the emission index $\log{R'_{\rm HK}}=-3.85$ reported by \cite{lopezsantiago2010high} puts it firmly into the ``very active'' class. The star is very young; isochrone fitting and identification as an AB or LA member place its age between 20~Myr and 150~Myr. The high lithium abundance is consistent with a Pleiades-type age \citep{montes2001chromospheric}. \cite{strassmeier2006first} presented a Doppler imaging temperature map indicating that the spot activity is concentrated at latitudes below $+40^{\circ}$.

We found both a well-defined 8.0~yr activity cycle and active longitudes on HD~1405. As expected, the activity variations are not stationary and although the cycle has repeated itself multiple times during the observational record, none of the consecutive cycles have been identical.

The active longitudes show more stability and the main activity area has stayed for the most of the time at a single rotational phase in the reference frame of $P_{\rm al}=1.75221$~d. On some occasions, most notably before 1990 and between 2000 and 2002, the main activity area was located for a while on the opposite side of the star with a roughly $\Delta\phi=0.5$ phase separation from the typical active longitude phase. The occasional phase jumping of the main active area is analogous to the "flip-flop" events found by \cite{jetsu1993spot} for \object{FK Com}, where the light curve minima alternated between two active longitudes on opposite sides of the star. During 1989, we saw one flip-flop event underway; during the course of the second half of the year the light curve minima quickly but gradually migrated from one active longitude to the other.

\subsection{EX Cet -- HD 10008}

\object{HD 10008} (\object{EX Cet}, \object{HIP 7576}) is an ``active'' G9V star with age estimates ranging from below 10~Myr up to 440~Myr. Most of the age estimates, including the identification as a HLA or LA member, are consistent with an age above 100~Myr. On the other hand, \cite{nakajima2012potential} identified the star as a TWA member implying an age below 10~Myr. The age discrepancy makes the TWA membership seem unlikely.

HD~10008 has a low amplitude light curve. We were able to detect periodicity in only 16 of the 46 independent datasets. The mean V-band light curve amplitude in the periodic $K>0$ order fits is $0\fm010$. We found evidence of a 10.9~yr activity cycle in the photometry. There is an apparent anticorrelation between $M$ and $A$ such that the light curve amplitude is highest when the mean magnitude is at its faintest. This indicates that the spot activity becomes increasingly non-axisymmetric on this star as the activity level rises. Nevertheless, the HB method was unable to find a cycle period from the $A$ results.

\subsection{V774 Tau -- HD 26923}

\object{HD 26923} (\object{V774 Tau}, \object{HIP 19859}) is a ``moderately active'' G0V star that forms a loose binary system with the G8V star \object{HD~26913} at an angular separation of 1\arcmin \ on the sky. The projected orbital separation between the components is 2800~AU \citep{abt1988maximum}. Kinematically it is identified as an UMa member.

HD~26923 is another low amplitude star. We detected periodicity in 15 of the 63 independent datasets with a mean V-band light curve amplitude of $0\fm013$. There is no previous measurement available for the rotation period of the star and our period value $P_{\rm w}=11.1$~d is the first one published. Some kind of meandering phase grouping may be perceived in the minimum epochs with a folding period of $P=10.7$~d. However, the Kuiper statistic suggests that this period is highly insignificant and does not allow the features to be reliably interpreted as active longitudes.

We found a ``poor'' 7.0~yr cycle and looking at the modelled M and A results in Fig. \ref{figres1}, the activity variations are clearly quite erratic. Between 2004 and 2009 there seems to have been another quasiperiodic structure present in the $M$ results with a period of about 2~yr. This structure is not, however, present in the rest of the data and was not detected by the HB method. \cite{baliunas1995chromosphere} reported that the \ion{Ca}{ii} H\&K emission level of the star is variable but did not find any cycle period from their observations.

\subsection{V834 Tau -- HD 29697}

\object{HD 29697} (\object{V834 Tau}, \object{HIP 21818}) is a ``very active'' K4V star identified as an UMa member. \cite{henry1995automated} found that its lithium abundance is consistent with the age of the Pleiades.

We found signs of a ``poor'' 7.3~yr cycle in the $A$ results of the star. The $M$ results show a strong brightening trend reaching its tip around 2010 and turning into a slow decline thereafter. This might be a sign of a long cycle with a length of approximately 30~yr or more, but claiming such a cycle at this point is highly premature. What is striking is the large $0\fm23$ amplitude seen in the seasonal $M$ values. This is more than double the mean peak to peak value $0\fm096$ of the light curve amplitude due to rotational modulation. \cite{lopezsantiago2010high} reported a projected rotation velocity of $v \sin i = 10.2$~km~s$^{-1}$, which implies a high inclination of the rotation axis. Thus the large seasonal variation in the mean magnitude is best explained by axisymmetric spot structures, e.g. a large concentration of spots around the pole during the brightness minimum.

There is one apparent active longitude on the star with $P_{\rm al}=3.943$~d (compare with $P_{\rm w}=3.965$~d) that was particularly well defined before 1998. During the following decade the phase coherence of the light curve was much poorer but after 2010 the active area has again been tightly confined around nearly the same rotational phase as before 1998.

\begin{figure*}
\centering
\begin{tabular}{cc}
\includegraphics[width=0.485\linewidth]{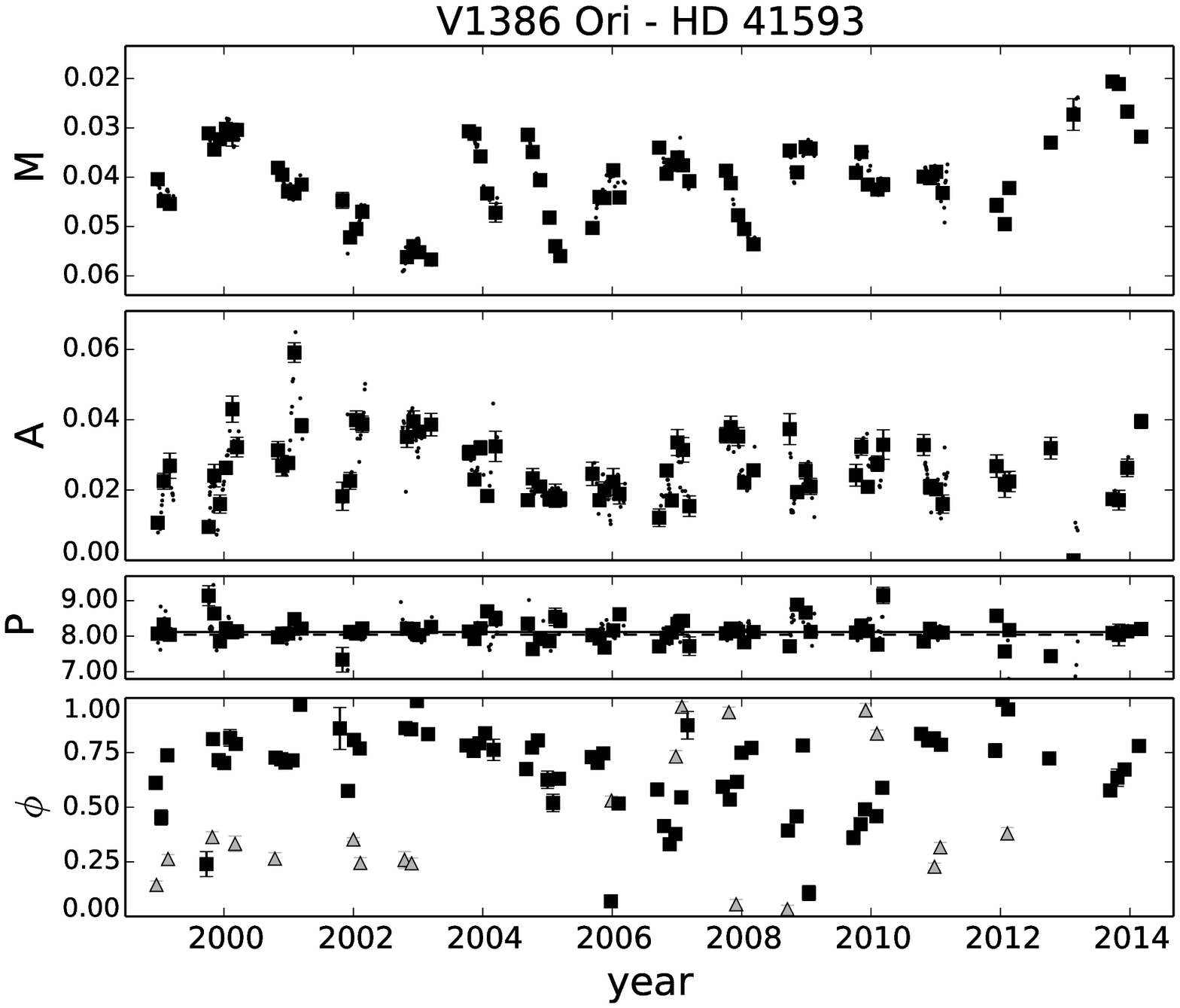} &
\includegraphics[width=0.485\linewidth]{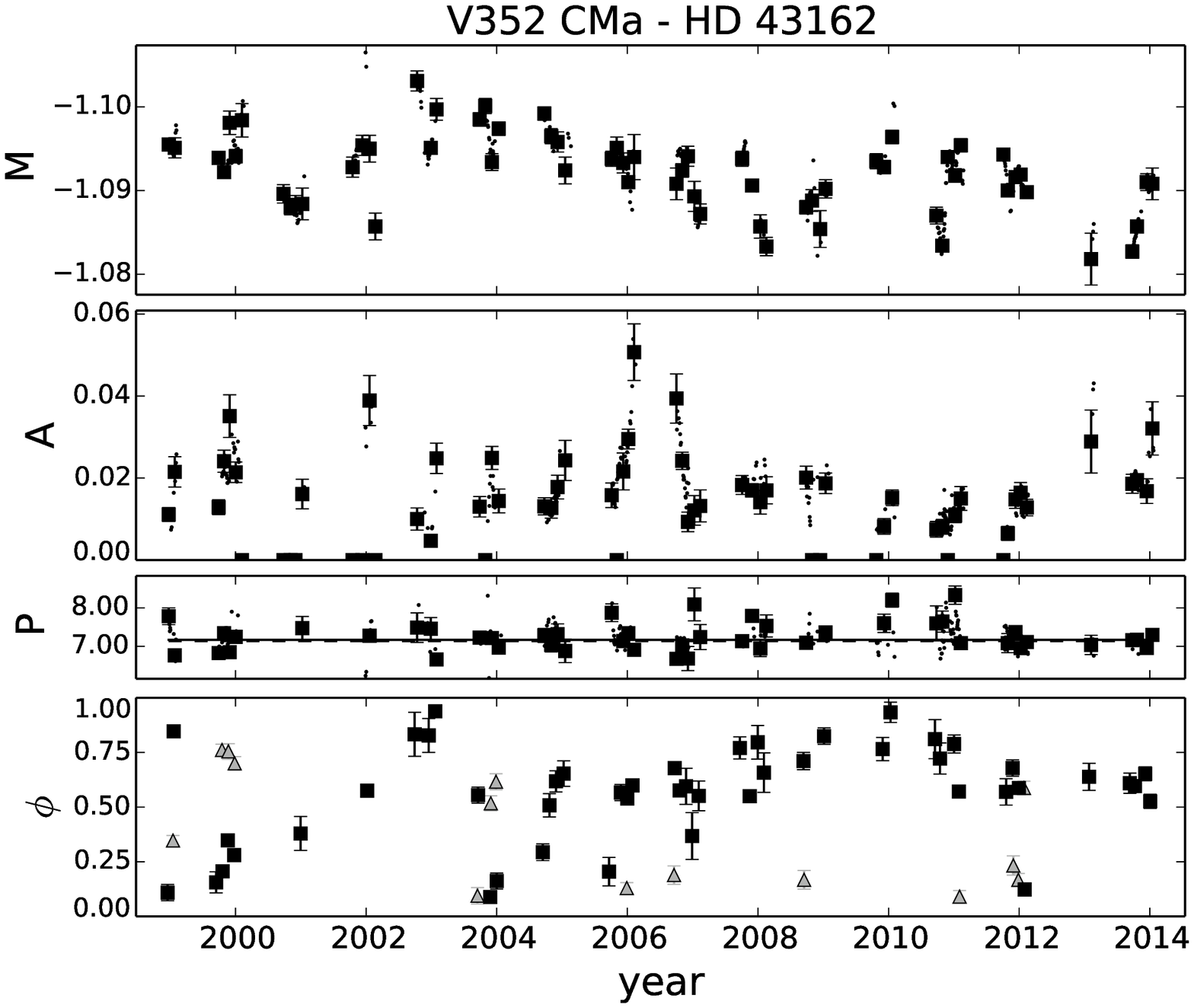} \\
\includegraphics[width=0.485\linewidth]{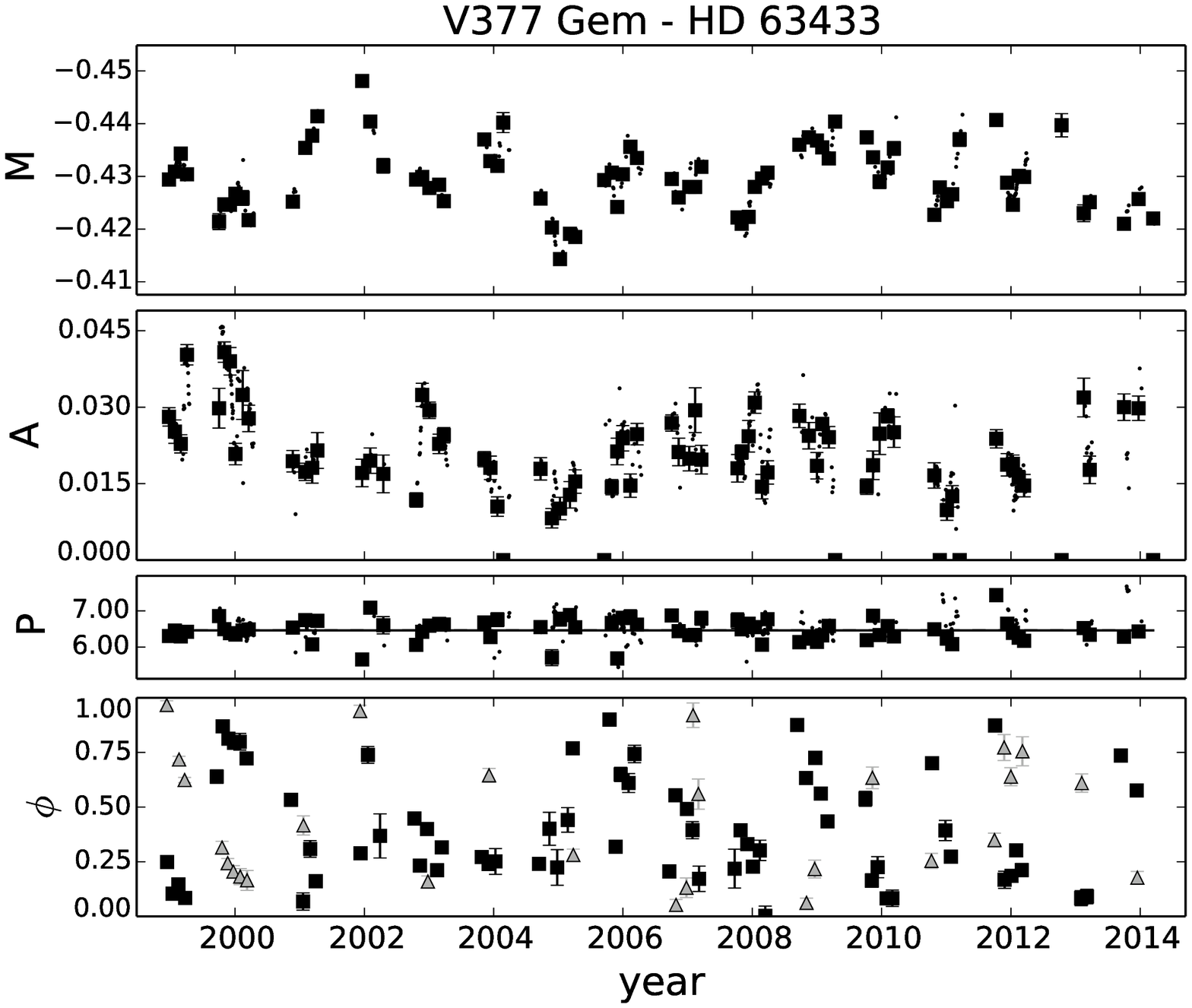} &
\includegraphics[width=0.485\linewidth]{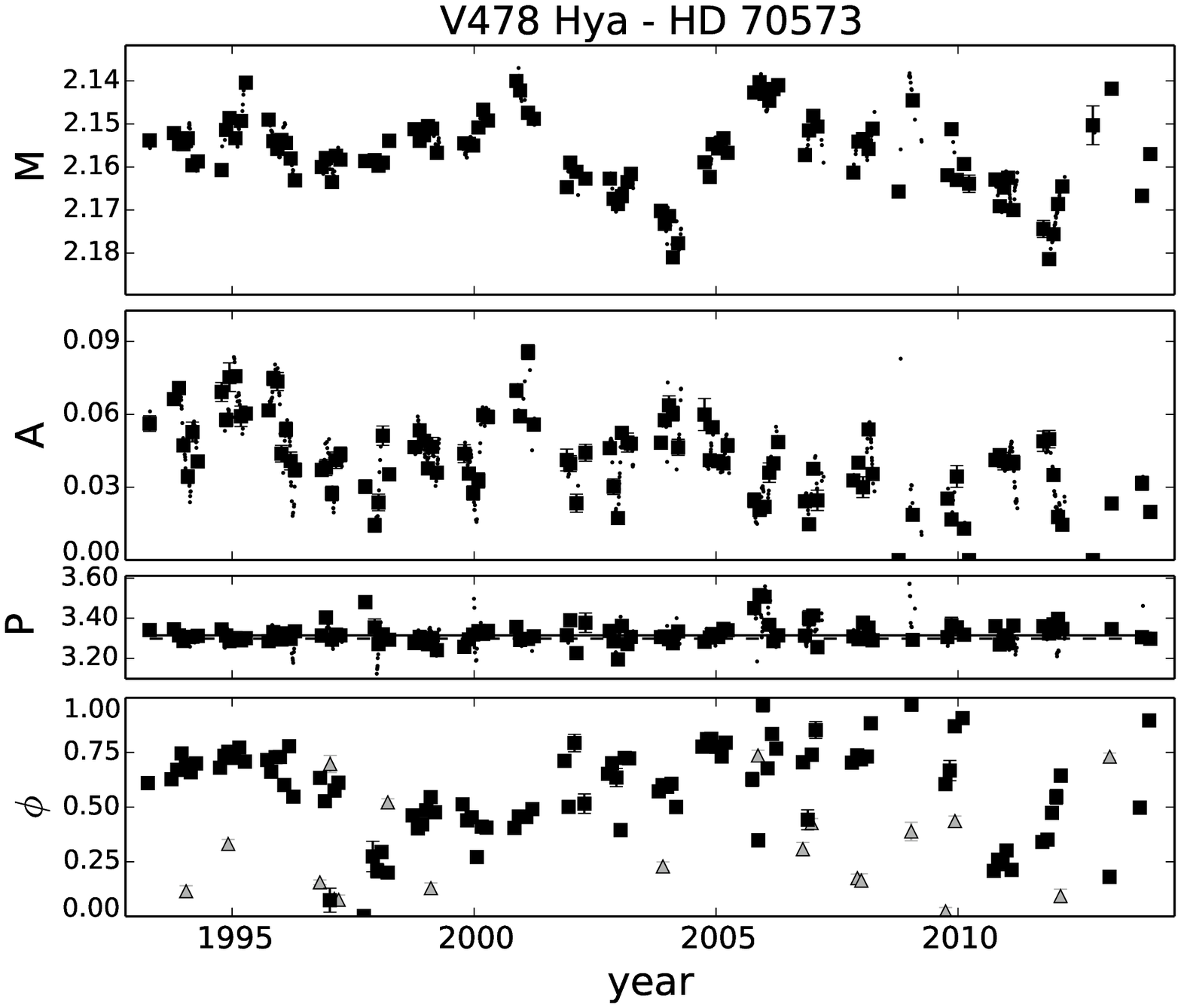}
\end{tabular}
\caption{Same as Fig. \ref{figres1} for HD 41593, HD 43162, HD 63433, and HD 70573.}
\label{figres2}
\end{figure*}

\subsection{V1386 Ori -- HD 41593}

\object{HD 41593} (\object{V1386 Ori}, \object{HIP 28954}) is an ``active'' G9V star identified as an UMa member. Our results reveal a short 3.3~yr cycle from its $M-A/2$ results. There is clear variability seen directly in both the $M$ and $A$ results but the HB method was unable to find any significant cycle in either one of them. The active longitudes of the star with the period $P_{\rm al}=8.042$~d (compare with $P_{\rm w}=8.14$~d) are more easily discerned and were especially clear before 2006. Between 2006 and 2010 the phase coherence of the observed light curve minima was lost, but recently the main active longitude has reappeared at the same rotational phase where it was located before 2006.

\subsection{V352 CMa -- HD 43162}

\object{HD 43162} (\object{V352 CMa}, \object{HIP 29568}) is the ``active'' G6.5V primary component of a loose triple system. The other two components in the system are both M dwarfs whose projected distances from the primary are 410~AU \citep{chini2014new} and 2740~AU \citep{raghavan2010survey}. Age estimates of the star based on coronal and chromospheric activity levels indicate that it is between 280~Myr and 460~Myr old. However, kinematically it is identified as an IC member, implying that it has an age of only a few tens of million years.

We found a ``poor'' 8.1~yr cycle in the $A$ results of HD~43162. This is reminiscent to the 11.7~yr cycle found by \cite{kajatkari2015periodicity} in their $M$ results based on much of the same data as used in the current study. However, we could not find a period from our series of $M$ results with $FAP<0.1$ although there are quasiperiodic variations. Compared to \cite{kajatkari2015periodicity}, we used two more years of photometry in our study. The inclusion of extra data might have thus introduced enough instability in the variations to increase the $FAP$ above the rejection limit.

We found active longitudes on the star, as did \cite{kajatkari2015periodicity}; the most significant period $P_{\rm al}=7.158$~d found by \cite{kajatkari2015periodicity} is much closer to the photometric rotation period $P_{\rm w}=7.17$~d than our active longitude period $P_{\rm al}=7.132$~d, and also has a higher $Q_{\rm K}$ value. The active longitude structure connected to our $P_{\rm al}$ has been quite stable through the observation record with only minor back and forth phase migration.

\cite{kajatkari2015periodicity} also estimated the light curve period variation of the star using the same methods that we used. Their $3\sigma$ range of fluctuation was $Z=0.19$, which is similar to our value $Z=0.22$.

\subsection{V377 Gem -- HD 63433}

\object{HD 63433} (\object{V377 Gem}, \object{HIP 38228}) is an ``active''  G5V star identified as an UMa member. We found two different cycles for it, a short 2.7~yr cycle in the $M$ and $M+A/2$ and a ``long'' and ``poor'' 8.0~yr cycle in the $A$ and $M-A/2$ data. The active longitudes with the period $P_{\rm al}=6.4641$~d (compare with $P_{\rm w}=6.46$~d) are marginally detected. These are mostly visible in the data between 2002 and 2005 and possibly also after 2012. During the first two observing seasons there was also a well-defined but short-lived drift structure seen in the light curve minima. This appears in the Kuiper test as a short periodicity of 6.4114~d but with $Q_{\rm K}=0.16$, which is not significant. If this drift pattern is also connected to active longitudes, their period variations have been quite large over time and the change between the two measured periods has been quite abrupt.

\subsection{V478 Hya -- HD 70573}

\object{HD 70573} (\object{V478 Hya}) is an ``active'' G6V star identified as a HLA or LA member. \cite{isaacson2010chromospheric} estimated a very young age of 50~Myr for it on the basis of chromospheric emission. The young age is supported by the fact that in our colour magnitude diagram (Fig. \ref{fighr}) the star is located quite far above the ZAMS. There is a 6.1 Jupiter mass planetary companion orbiting the star at a 1.76~AU semimajor axis orbit \citep{setiawan2007evidence}.

There is a discrepancy between the values of our chromospheric emission index at $\log{R'_{\rm HK}}=-4.488$ and the previously reported values at $\log{R'_{\rm HK}}=-4.10$ \citep{white2007high} and $\log{R'_{\rm HK}}=-4.187$ \citep{isaacson2010chromospheric}, which would classify the star as ``very active''. Our spectrum from February 2012 seems to have caught the star at a state of lower-than-average activity. However, at the same time the star was at a brightness minimum implying high spottedness. It is possible that our spectrum was simply timed so that all the major plage areas were located behind the limb of the star and invisible to Earth.

There is a well-defined 6.9~yr activity cycle apparent in the photometry of HD~70573 that can be quite clearly traced in the $M$ results. We did not find cyclicity directly in the $A$ results, but variability with a slightly shorter time scale can also be seen in them. There has been one strong active longitude visible in the data throughout the observation record. This active longitude has been constantly migrating with respect to the average $P_{\rm al}=3.2982$~d. From the migration rates it is possible to estimate the seasonal coherence periods roughly as $P_{\rm migr}=3.2966$~d between 1996 and 2000 and $P_{\rm migr}=3.2997$~d both before and after this. Between 2010 and 2011 there was also a sudden jump in the active longitude phase some $\Delta\phi=0.25$ forward. Each of these was shorter than the mean photometric rotation period $P_{\rm w}=3.314$~d.

\begin{figure*}
\centering
\begin{tabular}{cc}
\includegraphics[width=0.485\linewidth]{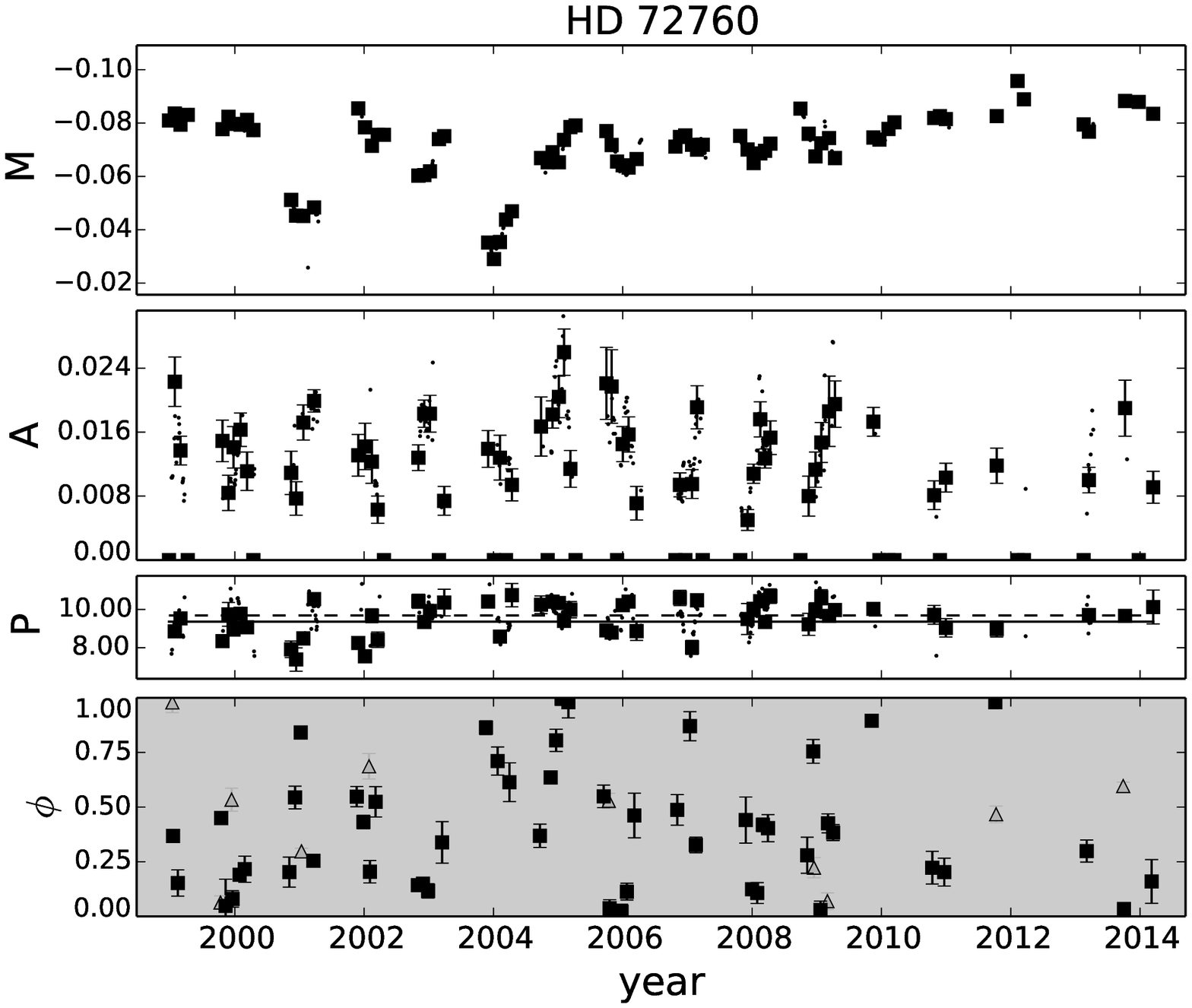} &
\includegraphics[width=0.485\linewidth]{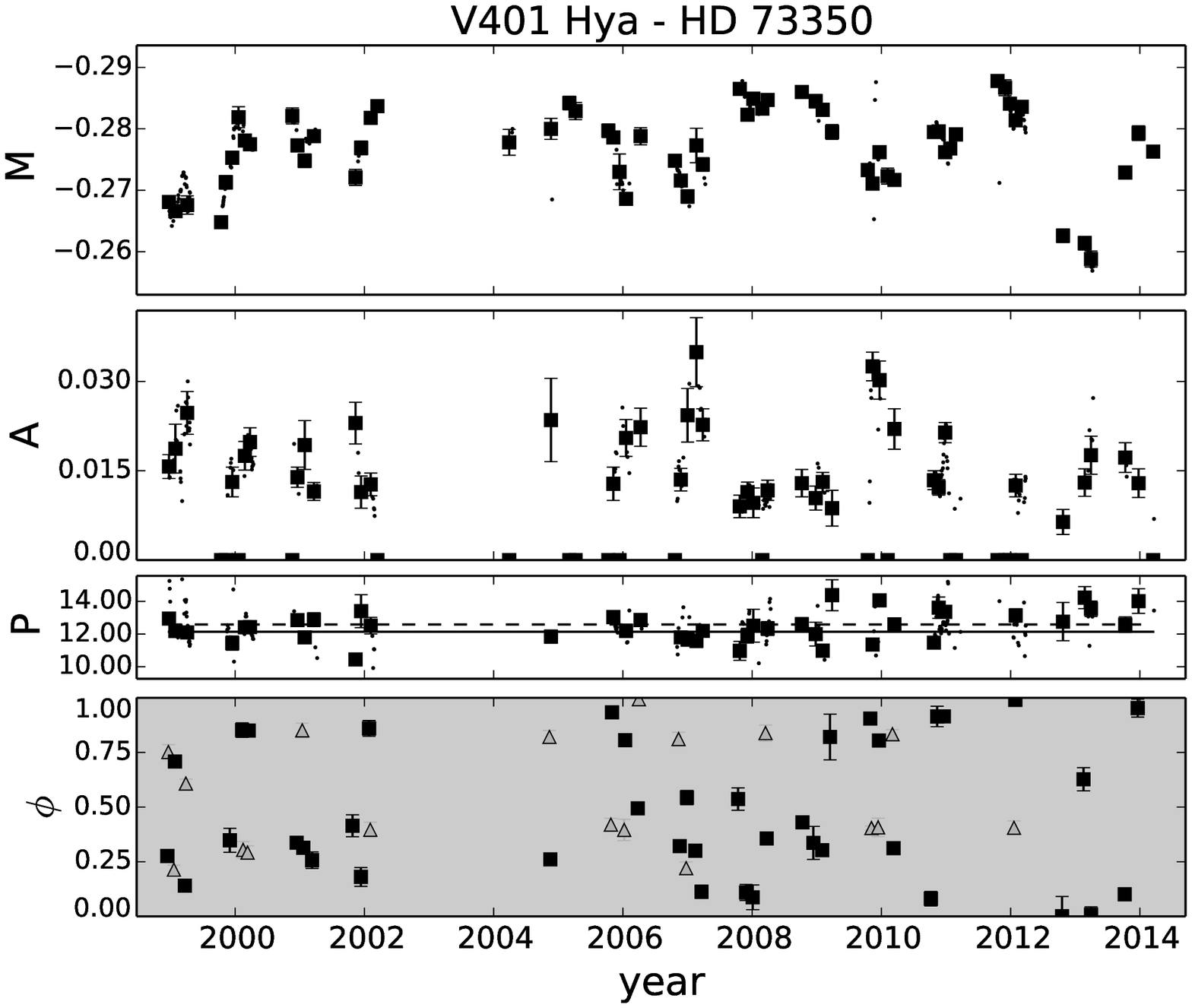} \\
\includegraphics[width=0.485\linewidth]{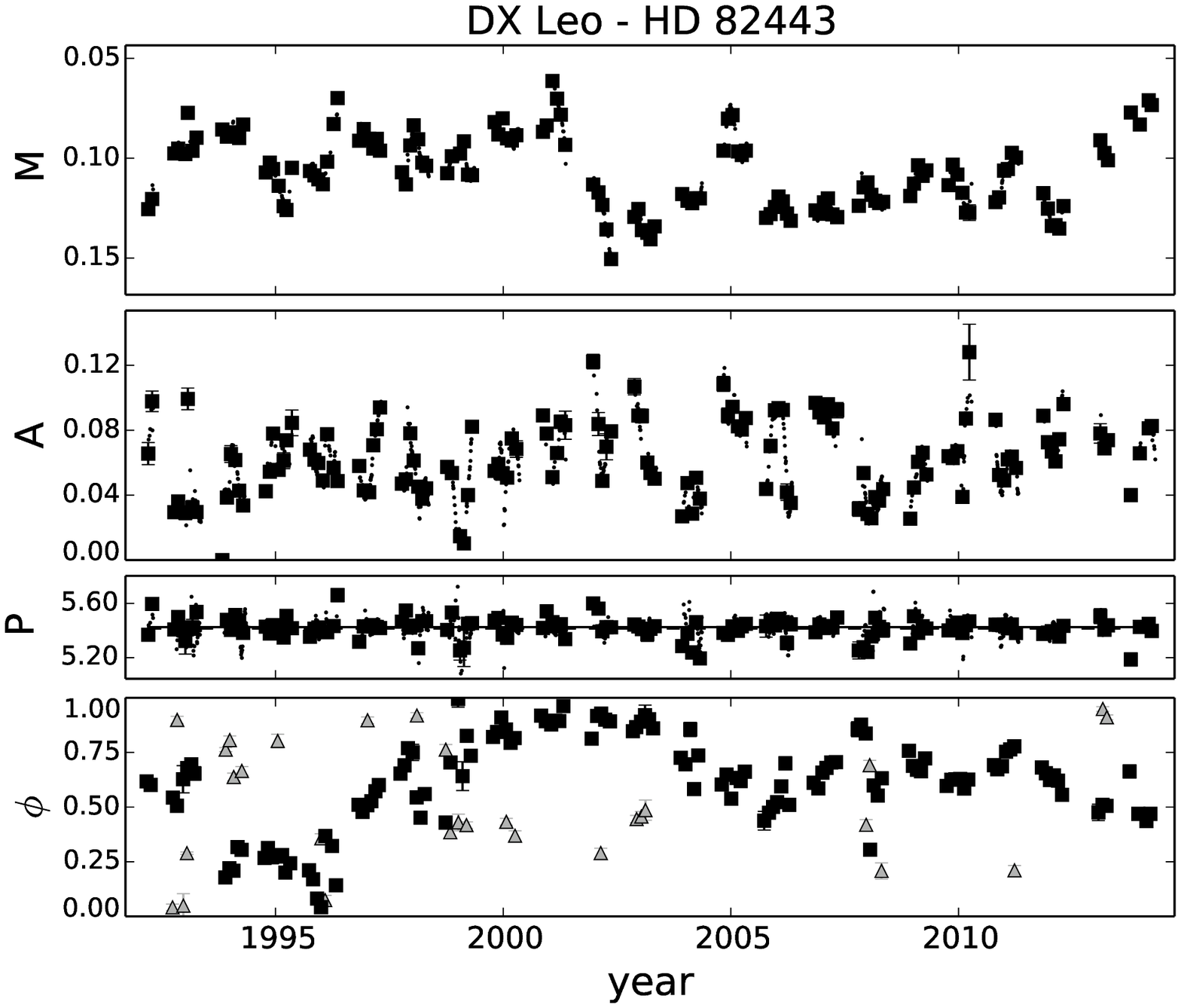} &
\includegraphics[width=0.485\linewidth]{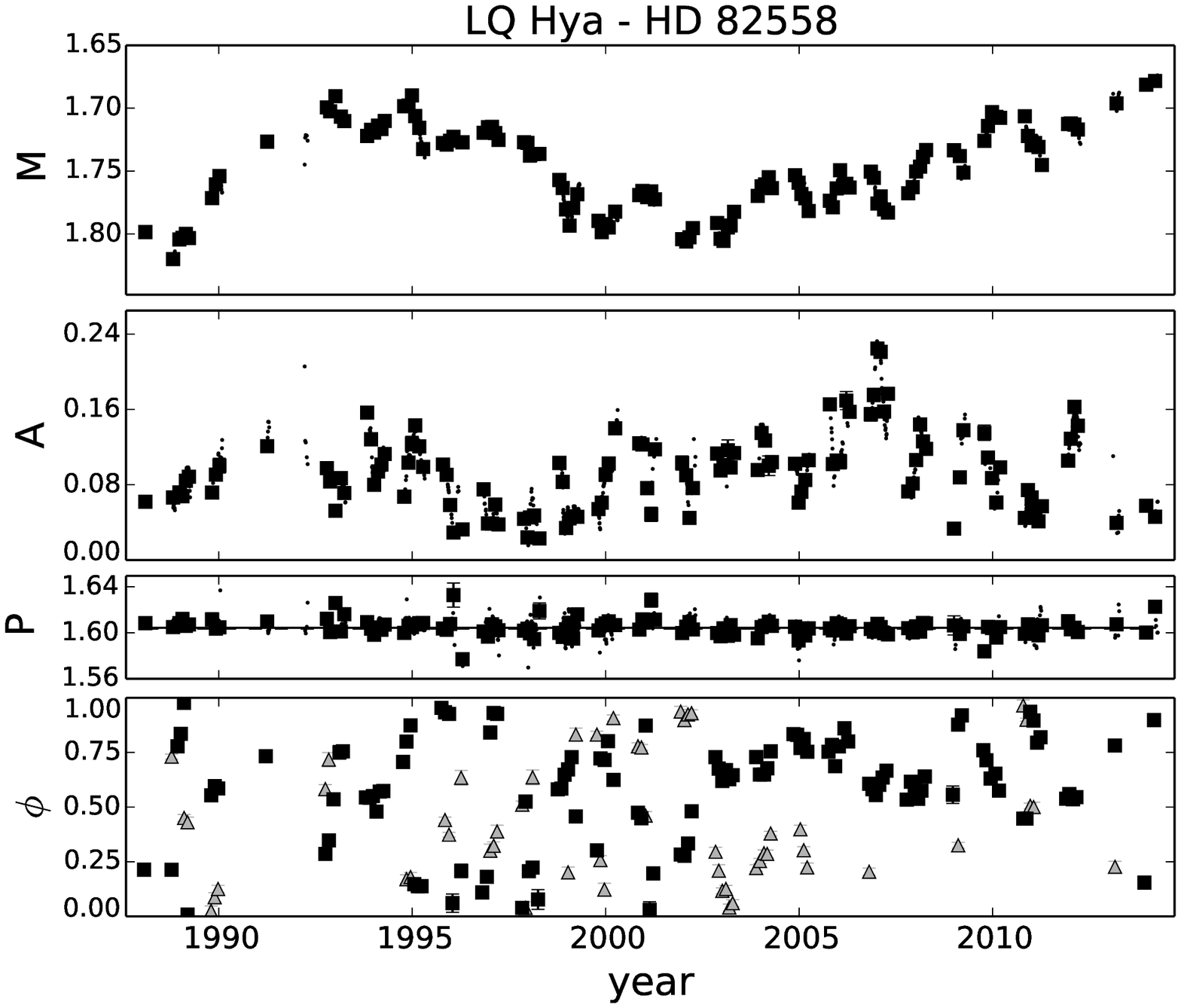}
\end{tabular}
\caption{Same as Fig. \ref{figres1} for HD 72760, HD 73350, HD 82443, and HD 82558.}
\label{figres3}
\end{figure*}

\subsection{HD 72760}

\object{HD 72760} (\object{HIP 42074}) is a ``moderately active'' K0V star identified as a Hya member. There is no previous rotation period available for it, so our period value $P_{\rm w}=9.6$~d is the first one published. We found no evidence of any activity cycles or active longitudes on the star. In the $M$ results there are two short depressions, but otherwise both the light curve mean and amplitude have stayed at stable levels.

\subsection{V401 Hya -- HD 73350}

\object{HD 73350} (\object{V401 Hya}, \object{HIP 42333}) is a ``moderately active'' G5V star identified as a Hya member. It had previously been thought to form a loose binary with the A0 type star \object{HD 73351} at an angular separation of 1\arcmin \ on the sky, but later research identified the companion candidate as a more distant field star \citep{raghavan2010survey}.

There are two previous and discrepant estimates available for the rotation period of HD~73350, $P_{\rm rot}=6.14$~d \citep{gaidos2000spectroscopy} and $P_{\rm rot}=12.3$~d \citep{petit2008toroidal}. Using these periods and the known projected rotation velocity $v\sin{i}=4.0$~km~s$^{-1}$ \citep{petit2008toroidal}, the minimum predicted radius of the star can be calculated as either $R\sin{i}=0.48\ R_{\sun}$ for the shorter period or $R\sin{i}=0.97\ R_{\sun}$ for the longer period. Considering that the spectral type is close to solar makes the longer period more likely. More conclusive proof of the longer period value comes from our preliminary TSPA analysis of the photometry which revealed the roughly 12~d period as the primary periodicity in the data and the roughly 6~d period as its first overtone, which results from the secondary minima often present in the light curve. Our final mean photometric rotation period from the CPS was $P_{\rm w}=12.1$~d. It is remarkable that \cite{petit2008toroidal} were able to find this period from only 13 nights of data by using the period as a fitting parameter in their Zeeman Doppler imaging.

\cite{petit2008toroidal} also tried to estimate the absolute value of the surface differential rotation by similar means but obtained a value with large error bars, $\Delta\Omega=0.2\pm0.2$~rad~d$^{-1}$. If we interpret the range of our rotation period fluctuations, $Z=0.40$, as a direct measure of the differential rotation coefficient $k$, we obtain $\Delta\Omega=0.21$~rad~d$^{-1}$ from our measurements. Although this is a high value, it fits the mean estimate of \cite{petit2008toroidal} very closely.

We found a short 3.5~yr cycle in the mean magnitudes $M$. The cycle looks fairly stable and although we did not find it in the $A$ results, there is similar behaviour visible in them as well. The light curve amplitude is typically higher during the times when the mean brightness is at its minima. A candidate folding period at $P=12.59$~d can be found by the Kuiper method and corresponds to a phase structure seen before the year 2002. However, this period has a high $Q_K$ value and cannot be identified with stable well-defined active longitudes.

\subsection{DX Leo -- HD 82443}

\object{HD 82443} (\object{DX Leo}, \object{HIP 46843}) is a thoroughly studied ``active'' K1V star. It forms a loose binary system with a M4.5 type dwarf at a distance of 1716~AU \citep{poveda1994statistical}. Kinematically it has been grouped as a THA, Ple, or LA member and its lithium abundance is consistent with the Pleiades age \citep{montes2001chromospheric}.

A short activity cycle has been found by many authors for the star. \cite{baliunas1995chromosphere} reported a 2.8~yr cycle in the chromospheric emission while \cite{messina1999activity} reported a cycle of 3.89~yr and \cite{messina2002magnetic} of 3.21~yr, both from photometry. Our results for the variation of $M$ and $A$ show at first glance quite erratic behaviour, but a ``fair'' 4.1~yr cycle is still firmly present in both of them in addition to the $M-A/2$ and $M+A/2$ results. Furthermore, we were able to find a new ``long'' 20.0~yr cycle from the $M$ values.

The active longitude behaviour of HD~82443 is both well developed and complex. First, between the 1993 and 1994 observing seasons there was a flip--flop event on the star. While the event was underway, both of the active longitudes were present for a while but one of them gradually weakened and eventually disappeared entirely, while the other one took over as the new primary active longitude. Later around 1996, this new active longitude started to migrate with respect to the average active longitude period $P_{\rm al}=5.4147$~d with a period of $P_{\rm migr}=5.4243$~d estimated from the migration rate. This migration pattern eventually stopped and turned into an opposite migration between 2003 and 2005 with a period of $P_{\rm migr}=5.4029$~d. Each of these periods is shorter than or equal to the mean photometric period $P_{\rm w}=5.424$~d. Since 2005 the star has shown one stable active longitude with some migration back and forth but mostly staying close to a single phase in the average $P_{\rm al}=5.4147$~d rotational frame of reference.

\cite{messina1999activity} estimated the surface differential rotation of the star from observed variations in the photometric period as $\Delta P/P\ge0.04$ and \cite{messina2003magnetic} similarly as $\Delta P/P=0.024$. These estimations are in line with our small value $Z=0.054$ calculated using similar means. \cite{messina2003magnetic} also correlated their estimated period values with the phase of the activity cycle and claimed that the differential rotation would be antisolar. No trace of such behaviour can be confirmed by our results and it remains impossible to specify the sign of the surface differential rotation.

\subsection{LQ Hya -- HD 82558}

\object{HD 82558} (\object{LQ Hya}, \object{HIP 46816}) is another ``very active'' K0V star with a rich reference history. It exhibits strong chromospheric emission modulated by rotation \citep{fekel1986chromosphere, strassmeier1993surface, montes2001chromospheric, alekseev2002starspot} in such a way that stronger emission is concentrated on the same phases as the dark photospheric spots \citep{frasca2008chromospheric, cao2014chromospheric, flores2015chromospheric}. It has also been observed to have strong flares in the optical and X-ray wavelengths \citep{montes1999optical, covino2001quiescent}. The star is very young as demonstrated by the identification as an IC member. The high lithium abundance also points to a Pleiades-type age \citep{fekel1986chromosphere}.

Complex surface magnetic fields were directly observed on the star by \cite{donati1997spectropolarimetric}. Based on Doppler imaging and Zeeman Doppler imaging, the spot activity appears to be concentrated in two areas, one at low latitudes and the other near the pole \citep{strassmeier1993surface, rice1998doppler, donati1999magnetic, donati2003dynamo, kovari2004doppler, cole2015doppler}. Comparison of spot longitudes by \cite{cole2015doppler} between Doppler imaging results and the photometric results by \cite{lehtinen2012spot} and \cite{olspert2015multiperiodicity} revealed good agreement between the two observational methods.

Activity cycles have been widely reported for HD 82558 with lengths mostly grouping into three distinct ranges. Cycle lengths reported previously from the photometry are: 6.24~yr by \cite{jetsu1993decade}; 6.8~yr and 11.4~yr by \cite{olah2000multiperiodic}; 3.4~yr by \cite{olah2002starspot}; 7.7~yr and 15~yr by \cite{berdyugina2002magnetic}; 3.2~yr, 6.2~yr, and 11.4~yr by \cite{messina2002magnetic}; 3.7~yr, 6.9~yr, and 13.8~yr by \cite{kovari2004doppler}; 2.5~yr, 3.6~yr, and a cycle increasing in length from 7~yr to 12.4~yr by \cite{olah2009multiple}; and 13~yr by \cite{lehtinen2012spot}. \cite{berdyugina2002magnetic} also reported a 5.2~yr cycle from the flip--flop behaviour of the active longitudes, though this does not seem likely in the light of our results.

We were able to find a highly significant ``long'' cycle in our data but its exact length varied quite a bit depending on which set of photometric results we looked at. The cycle lengths retrieved by the HB method were 17.4~yr for the $M$ results, 14.5~yr for the $A$ results, 15.8~yr for the $M-A/2$ results and 18.0 for the $M+A/2$ results. These doubtless represent the same underlying cycle but its long duration and seemingly non-stationary behaviour mean that there is considerable instability in the time series analysis. Our results also point to a longer cycle than any of the past results. While the analysis of \cite{jetsu1993decade} revealed a good 6.24~yr fit to the light curve mean magnitude between 1984 and 1992, the long-term variations of our $M$ and $A$ results have shown increasingly sluggish behaviour after that. Since 2000 the average mean brightness of the star has been continuously increasing with no sign of turning back. In this light the result of \cite{olah2009multiple} of a cycle with an increasing length seems like an apt description of what is happening on the star. We also note that our $M$ results show small amplitude oscillations with a time scale of 2 to 3 years similar to of the shortest cycles reported by other authors. The HB method was, however, unable to detect these shorter cycles from our data.

HD~82558 shows only occasional active longitude behaviour. Similarly to our previous study using only the V-band photometry \citep{lehtinen2012spot}, we were able to find coherent active longitudes between 2003 and 2008 but for the rest of the time no other active longitude patterns could be found. A short-lived pattern seen for some years before 1995 at a period of $P=1.68929$~d or $P=1.61208$~d could be found from our previous data but is absent from our current results. The inclusion of more data also meant that the significance of the detected period at $P_{\rm al}=1.603733$~d (compare with $P_{\rm w}=1.6044$~d) dropped quite a bit from our previous study. The lack of stable active longitudes contrasts with the results of \cite{jetsu1993decade} and \cite{berdyugina2002magnetic} but agrees with \cite{donati2003dynamo} and \cite{olspert2015multiperiodicity} who find no persistent longitudinal activity concentrations. Furthermore, \cite{olspert2015multiperiodicity}, who analysed the same V-band data we did, found that no consistent light curve period could be identified from the data using correlation times longer than approximately 230~d.

Differential rotation of HD 82558 has been estimated from the detected photometric period variations as $Z=0.015$ \citep{jetsu1993decade} and $Z=0.020$ \citep{lehtinen2012spot} using the same methodology we usre in this paper and as $\Delta P/P=0.013$ \citep{messina2003magnetic} and $\Delta P/P=0.025$ \citep{you2007photometric} from the total observed period range. These are all in line with our current estimate of $Z=0.017$. Other methods have produced smaller differential rotation estimates. \cite{berdyugina2002magnetic} found the value $k=0.002$ from tracing the migration rates of their detected active longitudes and \cite{kovari2004doppler} got a value of $k=0.0056$ by cross-correlating their Doppler imaging temperature maps. \cite{donati2003temporal} used surface differential rotation as an optimized parameter in their Zeeman Doppler imaging and found values of $0.0037 \le k \le 0.049$ for the Stokes I inversion and $-0.013 \le k \le 0.051$ for the Stokes V inversion. They concluded that this range of values possibly represents physical changes in the differential rotation rate. \cite{messina2003magnetic} correlated their local period estimates with the phase of their 6.2~yr cycle and claimed antisolar differential rotation. As in the case of HD~82443, this behaviour is not seen in our results. In any case, it appears safe to say that the differential rotation of HD~82558 is small and the star exhibits close to rigid body rotation.

\begin{figure*}
\centering
\begin{tabular}{cc}
\includegraphics[width=0.485\linewidth]{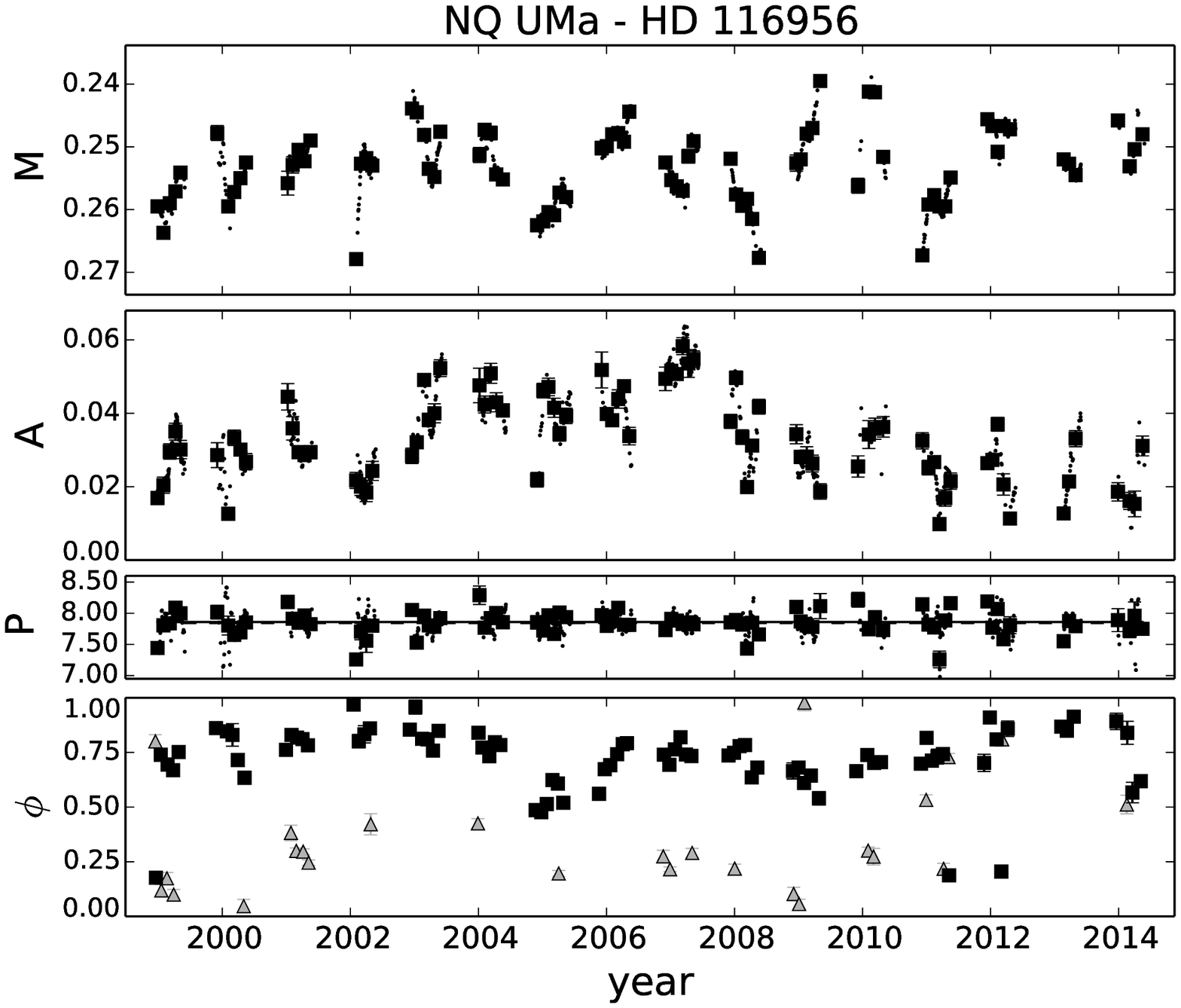} &
\includegraphics[width=0.485\linewidth]{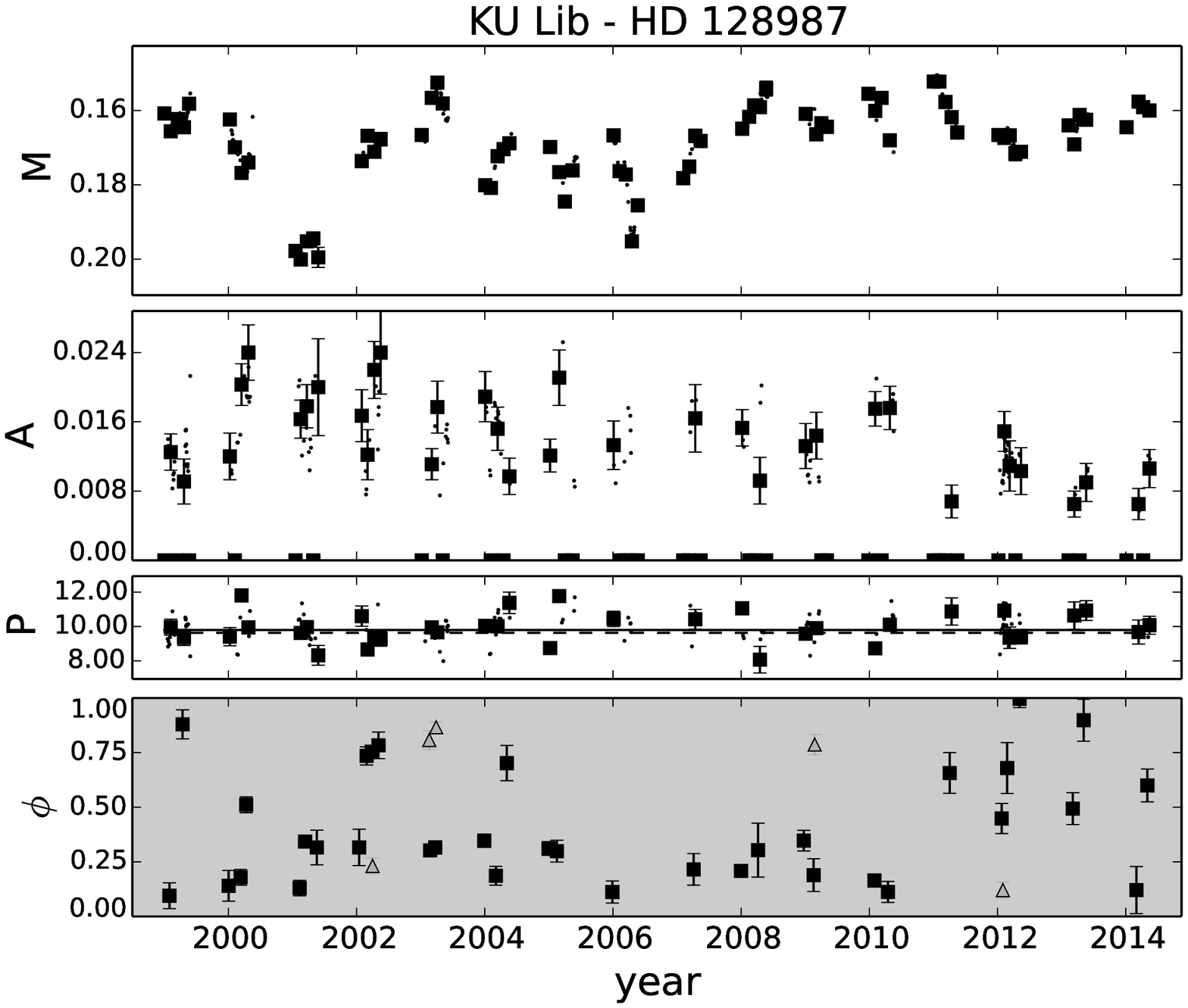} \\
\includegraphics[width=0.485\linewidth]{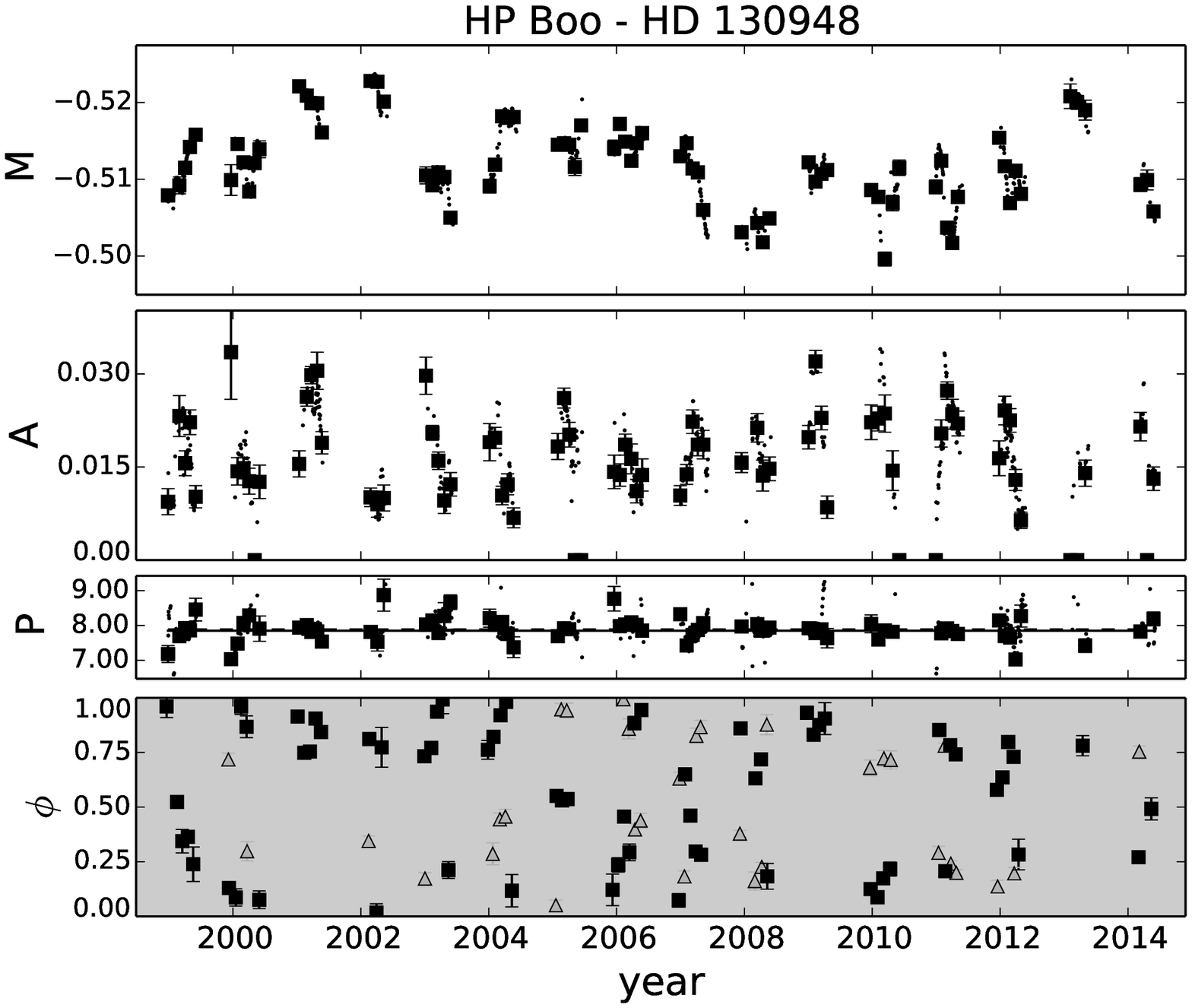} &
\includegraphics[width=0.485\linewidth]{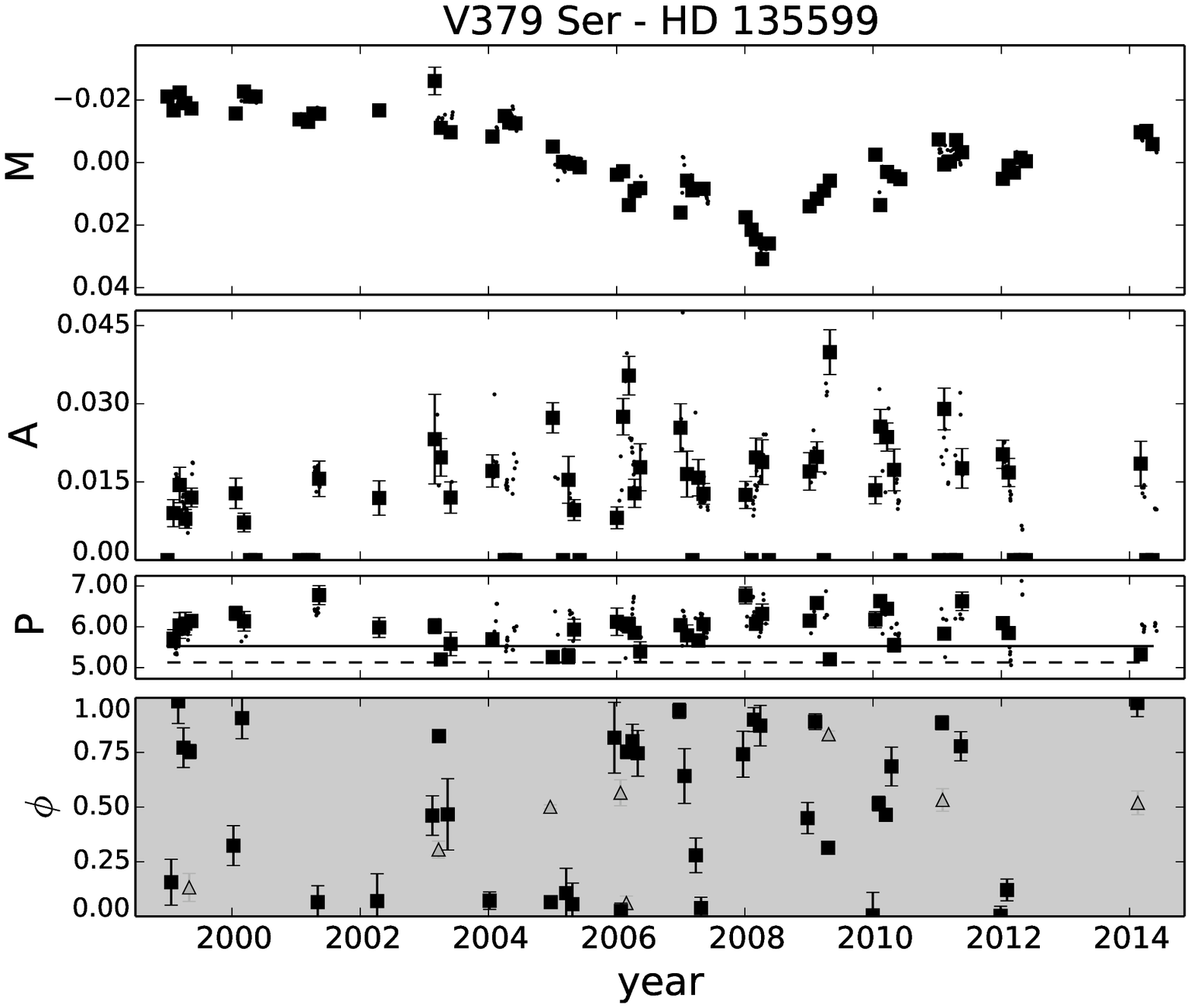}
\end{tabular}
\caption{Same as Fig. \ref{figres1} for HD 116956, HD 128987, 130948, and HD 135599.}
\label{figres4}
\end{figure*}

\subsection{NQ UMa -- HD 116956}

\object{HD 116956} (\object{NQ UMa}, \object{HIP 65515}) is an ``active'' G9V star identified as a TWA or LA member. \cite{lehtinen2011continuous} suggested the existence of a 3.3~yr cycle in a subset of the same V-band photometry as in the present study. We confirmed this here after finding a cycle of 2.9~yr in our $M$ and $M-A/2$ results. We also found an additional ``long'' cycle of 14.7~yr in the $A$ results. The same two active longitudes with the period $P_{\rm al}=7.8420$~d (compare with $P_{\rm w}=7.86$~d) found by \cite{lehtinen2011continuous} are also clearly present in our results and span the whole observation record without interruption. The active longitudes have undergone two phase jumps smaller than $\Delta\phi=0.25$ and their order of strength has switched around a few times, but for the most part the active longitudes are very constant. We previously estimated the differential rotation of the star as $Z=0.11$ \citep{lehtinen2011continuous} which is identical to the value we found from our new data.

\subsection{KU Lib -- HD 128987}

\object{HD 128987} (\object{KU Lib}, \object{HIP 71743}) is a ``moderately active'' G8V star with a low light curve amplitude. Kinematically it is identified as an IC member. We found a 5.4~yr cycle from its $M$, $M-A/2$, and $M+A/2$ results but did not find evidence of active longitudes.

\subsection{HP Boo -- HD 130948}

\object{HD 130948} (\object{HP Boo}, \object{HIP 72567}) is a ``moderately active'' F9IV-V star. It is orbited by a binary system of two L-type brown dwarfs at a projected distance of \citep{potter2002hokupaa}. It has not been identified as a member of any 48~AU kinematic group and \cite{maldonado2010spectroscopy} simply labelled it a young disk object. Estimates based on chromospheric and coronal activity imply ages between 190~Myr and 870~Myr. The spectral classification of the star would allow it to be classified as a more evolved subgiant. We calculated its minimum predicted radius from the rotation period and the projected rotational velocity $v\sin{i}=8.54$~km~s$^{-1}$ \citep{martinezarnaiz2010chromospheric} as $R\sin{i}=1.3R_{\sun}$. Based on this value, the age estimates, and because it falls directly on the ZAMS in our colour magnitude diagram (Fig. \ref{fighr}), it is most probable that this star is a main-sequence object.

We found a ``poor'' 3.9~yr cycle from the $M$ and $M+A/2$ results of the star. There is no evidence of any presence of active longitudes.

\subsection{V379 Ser -- HD 135599}

\object{HD 135599} (\object{V379 Ser}, \object{HIP 74702}) is an ``active'' K0V star. It was identified as an UMa member by \cite{gaidos2000spectroscopy} but this was refuted by \cite{maldonado2010spectroscopy} who described it as a ``probable non-member''. Age estimates based on chromospheric and coronal activity range between 200~Myr and 1340~Myr suggesting an older age for the star. We found a long 14.6~yr cycle from our $M$ and $M+A/2$ results. There is a suggestive correlation between the $M$ and $A$ results since the light curve amplitude reached its maximum values around the same time as the mean brightness dipped to its minimum. Still, we did not find any cycle from the $A$ results, nor did we find any evidence of active longitudes.

\begin{figure*}
\centering
\begin{tabular}{cc}
\includegraphics[width=0.485\linewidth]{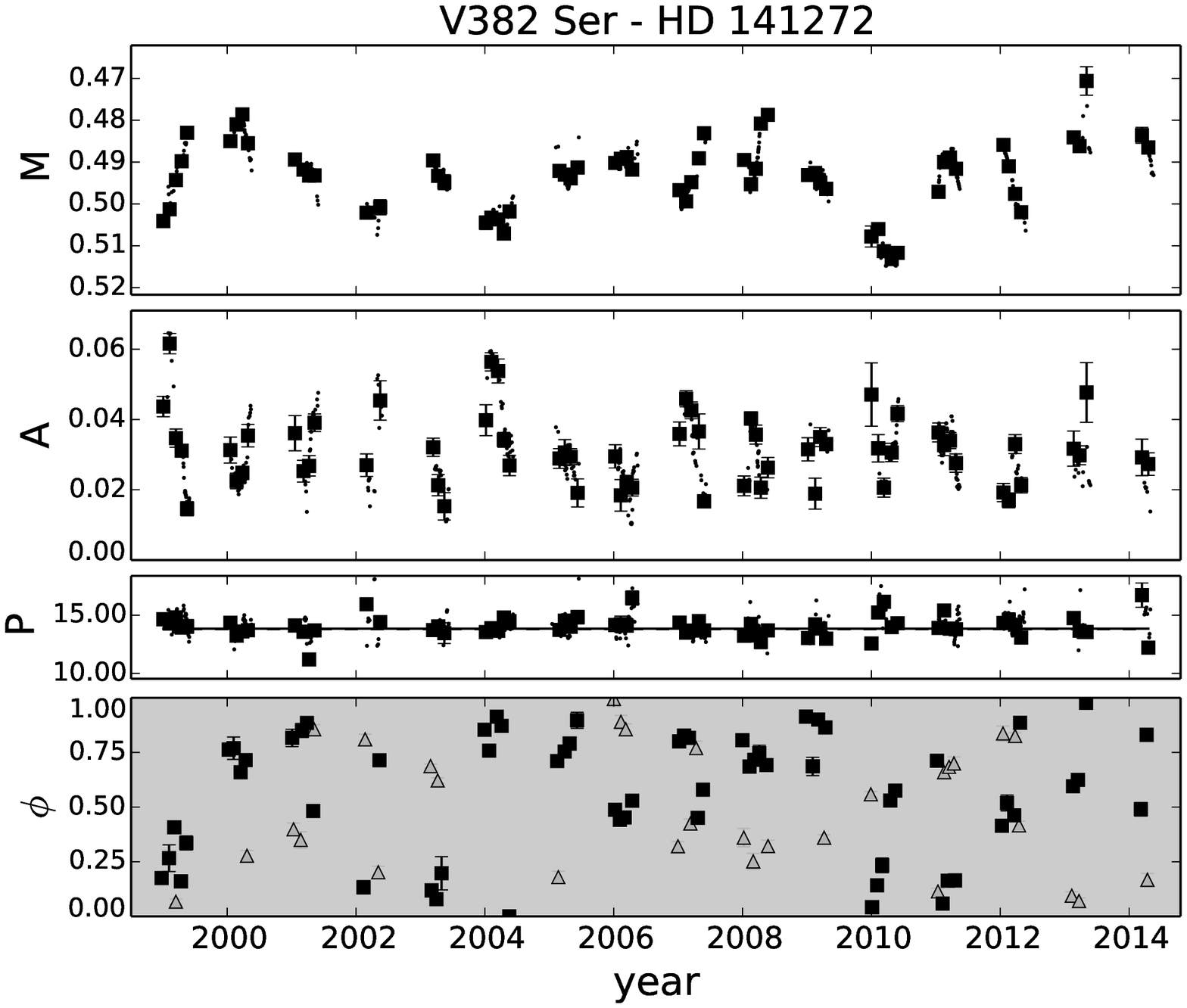} &
\includegraphics[width=0.485\linewidth]{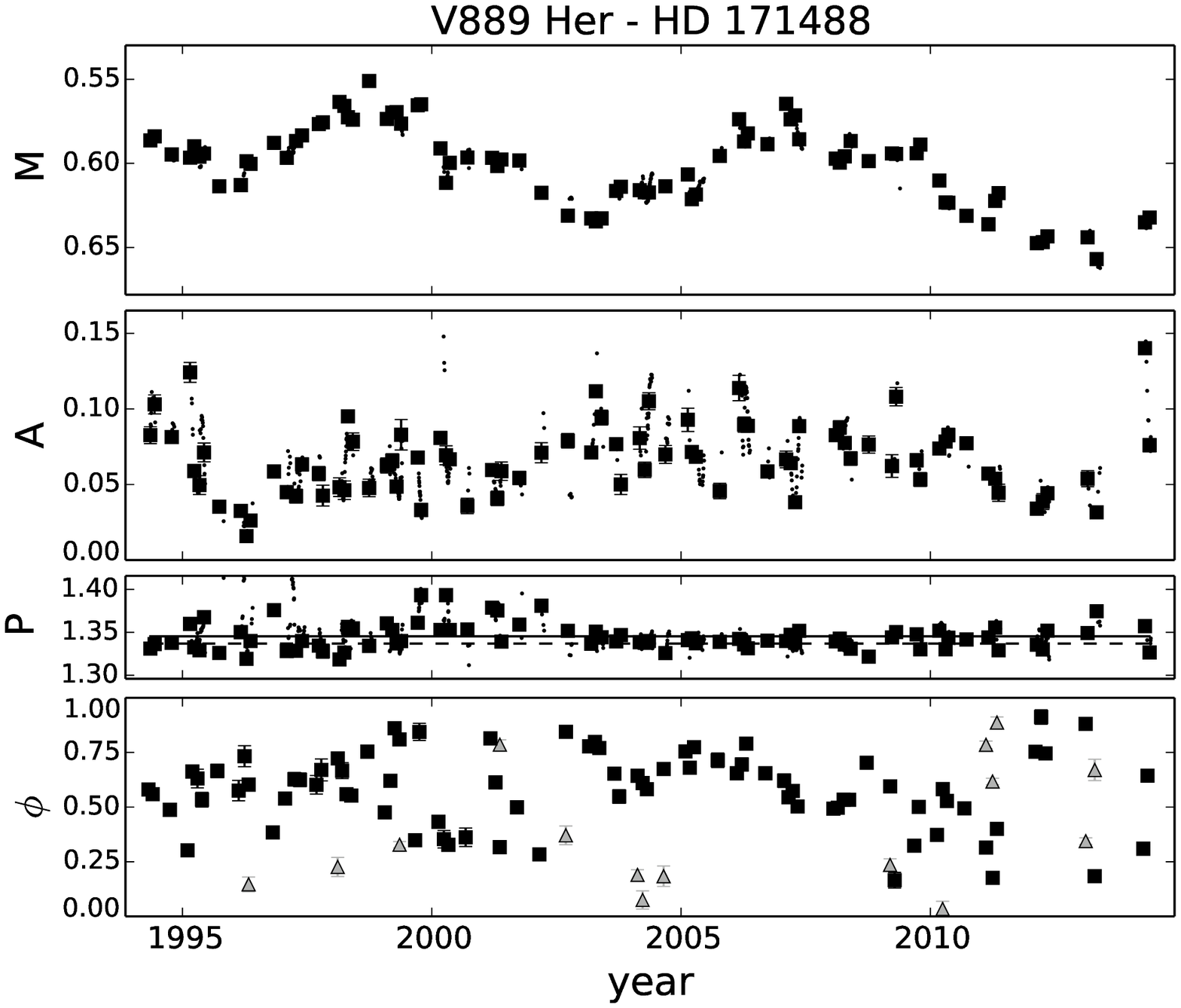} \\
\includegraphics[width=0.485\linewidth]{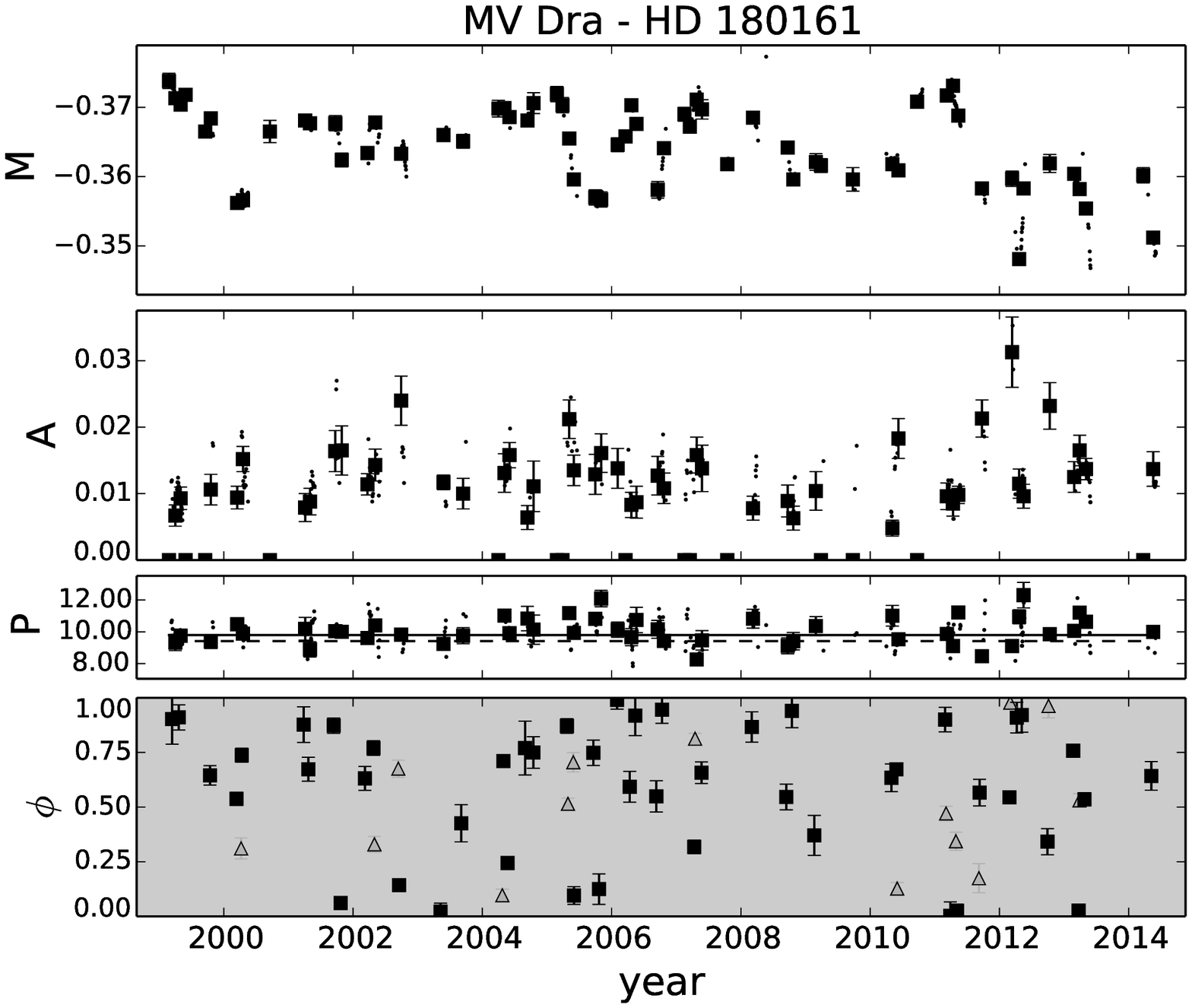} &
\includegraphics[width=0.485\linewidth]{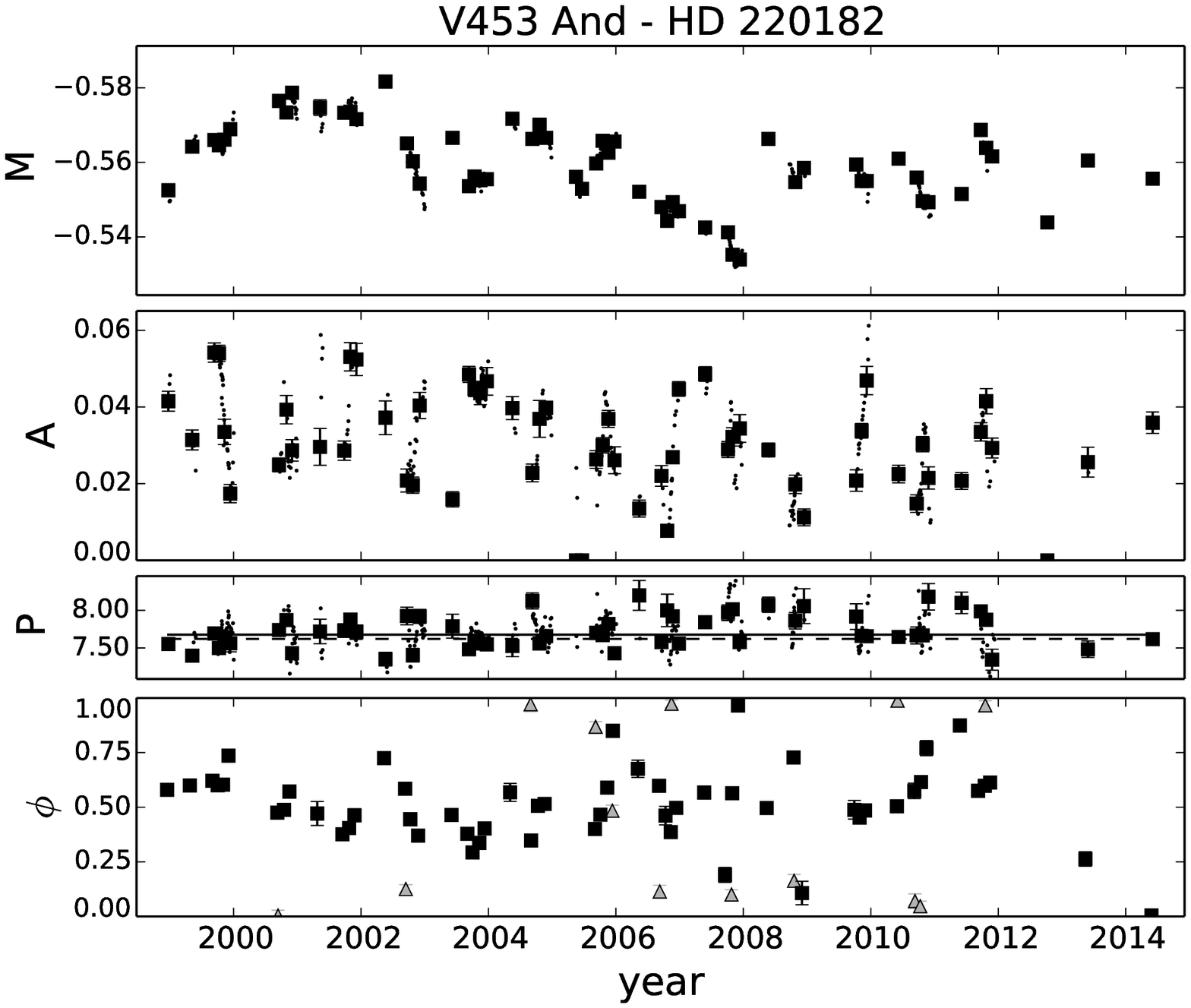}
\end{tabular}
\caption{Same as Fig. \ref{figres1} for HD 141272, HD 171488, HD 180161, and HD 220182.}
\label{figres5}
\end{figure*}

\subsection{V382 Ser -- HD 141272}

\object{HD 141272} (\object{V382 Ser}, \object{HIP 77408}) is the ``moderately active'' G9V primary component of a loose binary star. The secondary component of the system is an M dwarf at a projected distance of 350~AU from the primary \citep{eisenbeiss2007low}. The system is identified as a member of HLA or LA. We were able to find a ``poor'' 6.4~yr cycle from our $M$ results but did not find evidence of active longitudes.

\subsection{V889 Her -- HD 171488}

\object{HD 171488} (\object{V889 Her}, \object{HIP 91043}) is a ``very active'' G2V star with a rich reference history. It is among the youngest stars in our sample. \cite{strassmeier2003doppler} estimated an age of 30~Myr by comparing it with isochrones and 50~Myr from the lithium abundance, while \cite{frasca2010photometric} estimated an age of 50~Myr from isochrones. Kinematically the star is identified as a LA member.

The chromospheric emission of HD~171488 has been observed to be concentrated on the same phases as the dark spots \citep{mulliss1994search, frasca2010photometric}. Doppler images have been published by a number of authors and they all indicate that the spot activity is dominated by one or a few polar spots \citep{strassmeier2003doppler, marsden2006surface, jarvinen2008magnetic, jeffers2008high, huber2009long, frasca2010photometric}.

There is a clearly defined 9.5~yr cycle present in the $M$, $M-A/2$ and $M+A/2$ results of the star. This had previously been identified by \cite{jarvinen2008magnetic} as a 9~yr cycle. Two full cycles of roughly the same length can be seen in the $M$ results superimposed on a slight dimming trend. Variations can also be seen in the $A$ results, including two fast changes in 1995 and 2014, but no cycle could be found from them. There is also a pattern seen in the long-term evolution of the $P$ results but we are not able to claim anything about its nature.

An active longitude with the period $P_{\rm al}=1.33692$~d (compare with $P_{\rm w}=1.345$ d) has been present on the star between 1994 and 1998 and later between 2003 and 2008 with minor migration back and forth. For the rest of the time the phase distribution of the light curve minima has been less coherent. Tentatively, the active longitude appears to have been most clearly visible during the periods of increasing mean brightness or decreasing spottedness, but it is still too early to claim this with certainty.

The absolute value of surface shear on HD~171488 due to differential rotation has been estimated from Zeeman Doppler imaging by using the differential rotation rate as an optimized parameter. \cite{marsden2006surface} found a value of $\Delta\Omega=0.40$~rad~d$^{-1}$ from their Stokes I inversion and \cite{jeffers2008high} the values $\Delta\Omega=0.52$~rad~d$^{-1}$ from their Stokes I inversion and $\Delta\Omega=0.47$~rad~d$^{-1}$ from their Stokes V inversion, which correspond to differential rotation coefficients $k=0.084$, $k=0.11$, and $k=0.10$, respectively. They are somewhat larger than what our value of $Z=0.054$ suggests, possibly implying a narrow latitude range for the spots on the star.

\subsection{MV Dra -- HD 180161}

\object{HD 180161} (\object{MV Dra}, \object{HIP 94346}) is a ``moderately active'' G8V star identified as a Hya member. Its rotation period was estimated as $P_{\rm rot}=9.7$~d by \cite{gaidos2000spectroscopy}, but $P_{\rm rot}=5.49$~d by \cite{strassmeier2000vienna}. Using the projected rotational velocity $v\sin{i}=2.20$~km~s$^{-1}$ \citep{mishenina2012activity}, we can calculate the minimum expected radius of the star as $R\sin{i}=0.42R_{\sun}$ from the longer period and as $R\sin{i}=0.24R_{\sun}$ from the shorter period. This strongly favours the longer period, which was confirmed by our TSPA analysis. The final mean photometric rotation period from the CPS analysis was $P_{\rm w}=9.9$~d.

Some erratic short-term variation can be seen in the $M$ results of HD~180161 but no cycle could be found from the star. Neither did we find any evidence of active longitudes.

\subsection{V453 And -- HD 220182}

\object{HD 220182} (\object{V453 And}, \object{HIP 115331}) is an ``active'' G9V star. It has not been assigned to any kinematic group, but gyrochronology and the chromospheric and coronal activity point to an age of more than 100~Myr. The full range of published age estimates is from 40~Myr to 550~Myr.

We found a ``long'' cycle of 13.7~yr from the $M$ and $M+A/2$ results. By visual inspection a roughly 6~yr cycle might also be present in the $M$ results but we did not find any trace of it with the HB method. A persistent active longitude has been present on the star for the full span of the observing record, possibly excluding the last few observing seasons where the data was sparser. There have been two major migration patterns around the average active longitude period $P_{\rm al}=7.6200$~d. From the migration rates we estimated that between 2000 and 2002 the active longitude followed a shorter rotation period of approximately $P_{\rm migr}=7.610$~d and both before and after that a longer period of approximately $P_{\rm migr}=7.625$~d, which have all been shorter than the mean photometric period $P_{\rm w}=7.68$~d. A possible secondary active longitude separated by $\Delta\phi=0.5$ from the primary might also be present, but it is not particularly strong.

\subsection{V383 Lac -- SAO 51891}

\begin{figure}
\centering
\includegraphics[width=\linewidth]{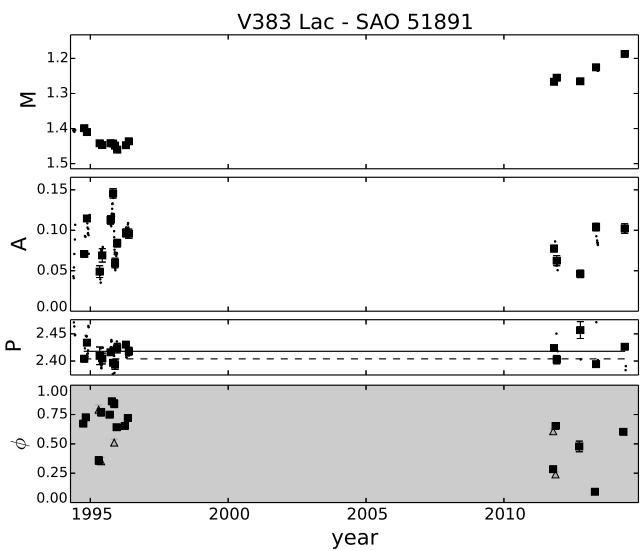}
\caption{Same as Fig. \ref{figres1} for SAO 51891.}
\label{figres6}
\end{figure}

\object{SAO 51891} (\object{V383 Lac}) is an ``active'' K1V star, although according to the previously estimated chromospheric emission index $\log{R'_{\rm HK}}=-4.05$ \citep{white2007high} it could also be classified as a ``very active'' star. \cite{biazzo2009young} observed that the \ion{Ca}{ii} H\&K, \ion{Ca}{ii} IRT and H$\epsilon$ emission is concentrated on the same rotational phases as the photospheric spots. They also observed significant variation in the H$\alpha$ emission, but did not see any dependence between it and the rotational phase. They reasoned that the H$\alpha$ emission might be dominated by processes like microflaring causing intrinsic variation and masking away the rotational modulation. The star is identified as a LA member and \cite{mulliss1994search} concluded that its lithium abundance is consistent with an age younger than the Pleiades.

Our photometric time series of SAO~51891 is dominated by the long gap separating the observations gathered between 1994 and 1996 from the observations gathered from 2011 onwards. Consequently, the CPS results consist of only 15 independent datasets with reliable model fits and we were not able to find activity cycles or active longitudes from them. Nevertheless, there is evidence of a possibly quite long cycle seen in the $M$ results. The mean V-band magnitude observed around 1995 was approximately 0.2 magnitudes dimmer than that observed during the recent years and its trends were opposite during the two parts of the photometry. The amplitude of the $M$ variations is exceptionally large compared to the rest of our sample stars. It is thus expected that with continued observations a distinct large amplitude cycle will become evident on SAO~51891. Likewise, the best period $P=2.40$~d found by the Kuiper method reveals a tightly confined feature from the light curve minima during the first three years of observations. It is thus also possible that well-defined active longitudes will surface from the star as more observations are gathered.

\section{Discussion}

\subsection{Differential rotation}
\label{diffsect}

\begin{figure*}
\centering
\includegraphics[width=\linewidth]{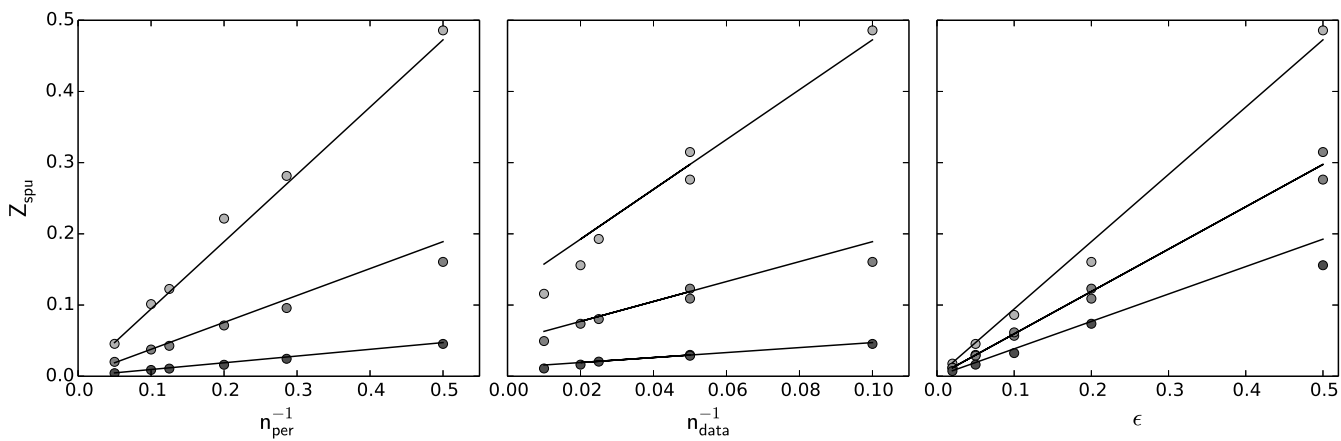}
\caption{Example slices of the fit of Eq. \ref{zspueq} onto a selection of $Z_{\rm spu}$ values. \textit{Left:} $Z_{\rm spu}$ computed for $n_{\rm data}=10$ and $\epsilon\in\{0.05,0.2,0.5\}$ ranging from the darkest to the lightest grey. \textit{Centre:} $Z_{\rm spu}$ for $n_{\rm rot}=2$ and $\epsilon\in\{0.05,0.2,0.5\}$ from the darkest to the lightest grey. \textit{Right:} $Z_{\rm spu}$ for $n_{\rm rot}=2$ and $n_{\rm data}\in\{50,20,10\}$ from the darkest to the lightest grey. The scale in $Z_{\rm spu}$ is identical for all the panels.}
\label{figzspu}
\end{figure*}

In Sect. \ref{sectrot} we touched on the problems of interpreting the $Z$ values as estimates of differential rotation. To quantify the effect of spurious fluctuations in the rotation period estimates, we simulated a set of periodic test time series with constant periods and using a range of values for their length, the number of data points contained in them, and the level of noise in the data. We then performed our period search for these simulated time series and calculated the resulting values of $Z$. It is evident that the three varied parameters (the number of data points per dataset $n_{\rm data}$, the number of complete rotations within each dataset $n_{\rm rot}$, and the ratio of noise to the signal amplitude $\epsilon$) contribute to the instability of the period determination. We chose their test ranges to be similar to what is expected for our ground-based automated photometry: $10 \le n_{\rm data} \le 100$, $2 \le n_{\rm rot} \le 20$, and $0.02 \le \epsilon \le 0.5$.

We found that the spurious fluctuations $Z_{\rm spu}$ quite expectedly increased as the data points became scarcer, the noise level increased, or the datasets included fewer complete rotations. There is a linear relation of $Z_{\rm spu}$ to each value of $n_{\rm data}^{-1}$, $n_{\rm rot}^{-1}$ and $\epsilon$. We found that the best fit to the simulation results was obtained by
\begin{equation}
Z_{\rm spu} = 16.0 \thinspace \epsilon n_{\rm rot}^{-1}(n_{\rm data}^{-1} + 0.023)
\label{zspueq}
\end{equation}
The fit is demonstrated in Fig. \ref{figzspu} for a selection of computed $Z_{\rm spu}$ values.

If we calculate the mean values of $n_{\rm data}$ and $n_{\rm rot}$ per dataset and the mean ratio $\epsilon$ of observational errors to the light curve amplitude for the CPS results of our stars, we find that the predicted values of $Z_{\rm spu}$ are always in the range of a few percent of the $Z$ values calculated from observations. The $Z$ estimates seem thus fairly rigid against numerical instability and correcting them for such small effects is not reasonable. There is also reason to believe that within the time scale of a few rotations, as used for our datasets, active region growth and decay also produce a relatively minor effect on the period fluctuations \citep{dobson1990variance}. We can thus say that the uncertainty in relating the $Z$ from our analysis to the relative differential rotation coefficient $k$ is dominated by the unknown latitude extent of the spot areas on the stars. The exact values of $k$ may remain unknown, but we can use $Z$ as a proxy measurement by using the proportionality $Z \propto k$.

\begin{figure}
\centering
\includegraphics[width=\linewidth]{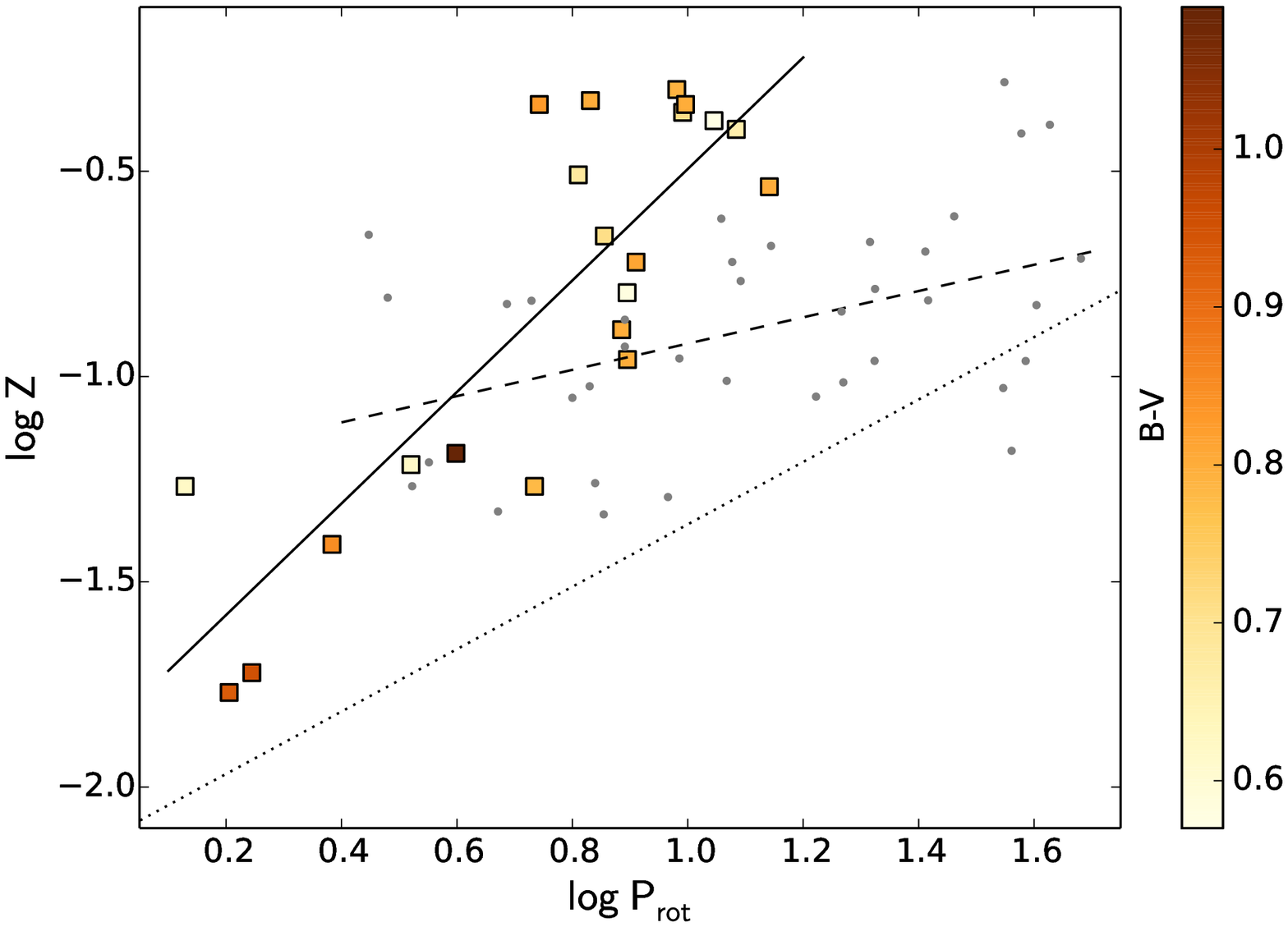}
\caption{$\log{Z}$ vs. $\log{P_{\rm rot}}$ for our sample stars shown with the squares. The colour coding denotes the $B-V$ colour of the stars. The fit to our results (Eq. \ref{zeq}) is shown with the solid line. For comparison we plot the stars from \cite{donahue1996relationship} as grey points and the fit to these stars with the dashed line. We also show the fit to the stars of \cite{henry1995starspot} with the dotted line for the $\log{P_{\rm rot}}$ range covered by our work and that of \cite{donahue1996relationship}.}
\label{figz}
\end{figure}

Despite the uncertainties associated with measuring the differential rotation, there is a clear dependence between $Z$ and the rotation period. Stars that have shorter rotation periods have smaller values of $Z$ and thus appear to have smaller differential rotation. For our stars we find a linear relation in the logarithmic scale as
\begin{equation}
\log{Z} = -1.85 + 1.36 \thinspace \log{P_{\rm rot}}.
\label{zeq}
\end{equation}
The stars and the fit are shown in Fig. \ref{figz} as the squares and the solid line.

This relation can be compared to previous results obtained by other authors. \cite{henry1995starspot} performed a study where they traced the rotation of light curve features identified as distinct spots. They estimated $k$ using the difference $\Delta P$ between the largest and smallest detected periods. For 87 stars taken from their study and that of \cite{hall1991learning} they found the relation
\begin{equation}
\log{k} = -2.12 + 0.76 \thinspace \log{P_{\rm rot}} - 0.57 \thinspace F,
\label{zheq}
\end{equation}
where $F$ is the Roche lobe filling factor. \cite{donahue1996relationship} investigated rotational variation seen in the \ion{Ca}{ii} H\&K line emission on 36 active stars and the Sun and estimated differential rotation by finding $\Delta P$ from seasonally computed periodograms. They found a scaling law $\Delta P \propto P_{\rm rot}^{1.3}$, corresponding to scaling the differential rotation coefficient as $k \propto \Delta P/P \propto P_{\rm rot}^{0.3}$.

The relations found by \cite{henry1995starspot} and \cite{donahue1996relationship} are both shown in Fig. \ref{figz} in comparison to our results. The dashed line and grey points denote the relation by \cite{donahue1996relationship} and the stars they used. The dotted line marks the relation by \cite{henry1995starspot} within the shown period range. Since all our stars lack close binary companions, we have set $F=0$ for Eq. \ref{zheq}. The relation predicts smaller values of differential rotation compared to the other results perhaps because \cite{henry1995starspot} traced the migration patterns of individual longer-lived spots instead of looking for the full period variation, thus potentially downplaying the effect of short-lived spots.

A further study was made by \cite{barnes2005dependence} based on Doppler imaging who found the scaling of the absolute equator to pole shear $\Delta\Omega$ to the rotation rate $\Omega$ to be $\Delta\Omega \propto \Omega^{0.15}$. This corresponds to $k \propto P_{\rm rot}^{0.85}$, which is again qualitatively similar to results from the other studies in that it shows increasing differential rotation towards slower rotators. In general we find that if we have $k \propto P_{\rm rot}^{\mu}$ for the differential rotation coefficient, we get $\Delta\Omega \propto \Omega^{\nu} = \Omega^{1-\mu}$ for the absolute shear. The power law indices $\mu$ and $\nu$ are listed for all the mentioned studies in Table \ref{diffpowtab}.

The values of $\mu$ and $\nu$ show considerable scatter between the different studies. The differences between the exact obtained values may originate from different choices of data analysis strategies and the physical rotation proxies being tracked, and from simple random scatter due to limited sample sizes. In the case of our sample and that of \cite{donahue1996relationship}, even the power law nature of the underlying scaling cannot be fully verified since they both cover only approximately one dex in $P_{\rm rot}$.

Nevertheless, at least in the case of $\mu$, the results can be interpreted to qualitatively agree with each other since they show slower rotating stars to have stronger differential rotation. In the case of the dependence of $\Delta\Omega$ on $\Omega$ the results are less clear, but as they are scattered closer to $\nu=0$ they may still suggest an agreement with theoretical results such as those of \cite{kuker2011difference} who found $\Delta\Omega$ to have little dependence on the rotation rate in their mean field models of lower main-sequence stars. Recently \cite{reinhold2013rotation} and \cite{reinhold2015rotation} performed a massive study of 24\,124 spotted Kepler field stars and measured differential rotation from two simultaneously observed rotation periods. Their results agree quite well with the theoretically predicted independence of $\Delta\Omega$ from $\Omega$ and are very close to the observational results of \cite{henry1995starspot}.

According to \cite{barnes2005dependence} and \cite{colliercameron2007differential} there is a strong dependence of $\Delta\Omega$ on the temperature of the star. We were unable to find evidence for this in our results, only finding a Pearson correlation coefficient $r=-0.21$ between the $B-V$ colour and $Z/P_{\rm rot} \propto \Delta\Omega$. This lack of correlation can also be seen from the colour coding for $B-V$ in Fig. \ref{figz}. Our results agree better with those of \cite{kuker2011difference}, who found only weak temperature dependence of $\Delta\Omega$ for stars with temperatures within the range 3500--6000~K and a steeper dependence only for stars hotter than this. The two-staged temperature dependence was also observationally confirmed by \cite{reinhold2013rotation} whose sample included stars from both sides of the 6000~K divide.

\begin{table}
\caption{Power law indices $\mu$ and $\nu$ for differential rotation fits $k \propto P_{\rm rot}^{\mu}$ and $\Delta\Omega \propto \Omega^{\nu}$ from the current study and the literature.}
\center
\begin{tabular}{r@{.}lr@{.}ll}
\hline
\hline
\multicolumn{2}{c}{$\mu$} & \multicolumn{2}{c}{$\nu$} & Reference \\
\hline
$1$&$36$ & $-0$&$36$ & this work \\
$0$&$76$ &  $0$&$24$ & \cite{henry1995starspot} \\
$0$&$3$  &  $0$&$7$  & \cite{donahue1996relationship} \\
$0$&$85$ &  $0$&$15$ & \cite{barnes2005dependence} \\
$0$&$71$ &  $0$&$29$ & \cite{reinhold2015rotation} \\
\hline
\end{tabular}
\label{diffpowtab}
\end{table}

\subsection{Active longitudes}

\begin{figure}
\centering
\includegraphics[width=\linewidth]{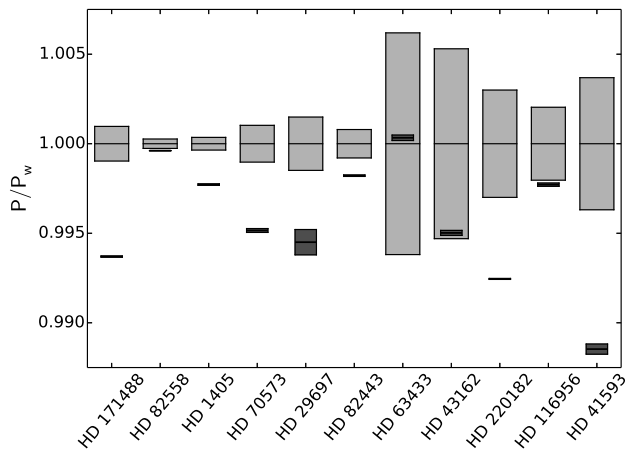}
\caption{Comparison of the active longitude periods $P_{\rm al}$ (dark boxes) to the photometric mean rotation periods $P_{\rm w}$ (light boxes), both normalized to the values of $P_{\rm w}$. The heights of the boxes show the estimated $\pm1\sigma$ error bars of the periods.}
\label{figprotpal}
\end{figure}

We were able to identify reliable active longitudes for roughly half of our stars. The values found for $P_{\rm al}$ are shown in Fig. \ref{figprotpal} in comparison to the weighted mean photometric periods $P_{\rm w}$. All the periods are normalized in the plot by dividing them by $P_{\rm w}$. The heights of the boxes corresponding to the period values denote the $\pm1\sigma$ error limits of the period estimates.

The $P_{\rm al}$ values are smaller than the $P_{\rm w}$ values nearly without exception. In the case of HD~41593 the relative difference between the two periods is as large as 1.1\% which corresponds to a lap time of 1.9 yr between $P_{\rm al}$ and $P_{\rm w}$. This raises the question of whether there is an underlying difference between the coherence period of the active longitudes and the photometric rotation period of the stars. This situation is possible if the active longitudes represent an activity generating structure rotating with a different angular velocity compared to the stellar photosphere. If individual spots or spot groups are formed at the active longitudes but start to follow the photospheric rotation once formed, we would expect there to be separate signals in the photometry for the periods $P_{\rm al}$ and $P_{\rm w}$.

To get an idea whether there is any significant difference between $P_{\rm al}$ and $P_{\rm w}$, we modelled the error distributions of $P_{\rm w}$ as Gaussians $\mathcal{N}(P_{\rm w},\sigma^2_{\rm P, w})$ with their standard deviations as the $P_{\rm w}$ standard errors of the mean. We then calculated the two sided tail-area probabilities $p={\rm Pr}(|P-P_{\rm w}|>|P_{\rm al}-P_{\rm w}|)$ that a random sample $P$ drawn from such a distribution has a larger deviation from $P_{\rm w}$ than $P_{\rm al}$ has. This gives us a measure analogous to the $Q_{\rm K}$ and $FAP$ used in the Kuiper test and the HB method. The process ignores the fact that the $P_{\rm al}$ values also have error distributions associated with them, but since they are always narrower than $\sigma_{\rm P, w}$, the simple $p$ values still provide a useful heuristic to assess the distinctiveness of $P_{\rm al}$ from $P_{\rm w}$.

The relative differences of $P_{\rm al}$ from $P_{\rm w}$, $\delta P=P_{\rm al}/P_{\rm w}-1$, and the tail area probabilities $p$ for each found active longitude period are listed in Table \ref{altab}. In the case of HD~43162, HD~63433, HD~82558, and HD~116956 there is little evidence of observed difference between $P_{\rm al}$ and $P_{\rm w}$. On the other hand, in the case of HD~1405, HD~29697, HD~41593, HD~70573, and HD~171488 the directly detected values of $P_{\rm al}$ have $p<0.01$ and for two more stars (HD~82443 and HD~220182) $p<0.05$. For these stars, a single underlying period appears inadequate to describe both $P_{\rm al}$ and $P_{\rm w}$ at the same time.

The observed differences between $P_{\rm al}$ and $P_{\rm w}$ can have several different interpretations. It might be that the magnetic structures that sustain the active longitudes reside at deeper levels in the stellar interior than the formed starspots and trace the rotation at those depths. In this case the difference between $P_{\rm al}$ and $P_{\rm w}$ would reflect radial shear in the outer layers of the stellar interior. Helioseismological results have shown that in the Sun there is a negative gradient of angular velocity on all latitudes just below the photosphere \citep{howe2000dynamic}. This behaviour of internal rotation would predict shorter rotation periods for structures rooted at deeper levels, just as observed for the active longitudes on our stars.

On the other hand, there is theoretical evidence from direct numerical simulations that a stellar dynamo may operate with a non-axisymmetric mode that propagates longitudinally with respect to the rotational reference frame of the star \citep{cole2014azimuthal}. If this is the case for our sample stars, what we are seeing as the migrating active longitudes are signs of azimuthal dynamo waves. Such a dynamo wave would feed a certain longitude in the rotational frame of $P_{\rm al}$ with spot structures, which could then start to drift with the local surface rotation. These are the azimuthal counterparts to the latitudinal dynamo waves seen on the Sun and are responsible for the well-known butterfly diagram \citep{radler1986investigations}.

Similar drift patterns have been observed before from active giants stars. \cite{hackman2011spot} observed that the active longitude on \object{II Peg} has rotated with a shorter period than the stellar surface, as is the case for our stars, while \object{FK~Com} \citep{hackman2013flipflops} seems to have followed an opposite pattern of the active longitude rotating more slowly than the stellar surface. What still need to be properly explained are, on the one hand, the observed values of the relative period differences $\delta P$ between $P_{\rm al}$ and $P_{\rm w}$ and, on the other, the variable migration trends of the active longitudes seen on some stars. In Fig. \ref{figprotpal} we have sorted the stars in the order of increasing $P_{\rm w}$ and no immediate relation between rotation period and the $\delta P$ value is apparent.

We note that it does not appear justified to interpret the differences seen between the periods $P_{\rm al}$ and $P_{\rm w}$ as measures of differential rotation, similarly to how \cite{reinhold2015rotation} used two simultaneous photometric periods detected in single quadrants of Kepler data for their differential rotation estimation. If such an interpretation were true for our much longer time series, often showing very stable active longitudes, we should expect a corresponding bimodality to be visible in the distribution of the local $P$ estimates produced by the CPS analysis. A look at the estimated $P$ values in Figs. \ref{figres1}--\ref{figres6} reveals instead only unimodal distributions centred at the $P_{\rm w}$ values.

A second significant result concerning the active longitudes is their presence on only the more active stars. This is shown in the right-hand panel in Fig. \ref{figcycro}, where we have circled those stars with activity cycles that also have reliably detected active longitudes. There is a sharp divide at approximately $\log{R'_{\rm HK}}=-4.46$ between active stars showing active longitudes and less active stars with poor or no evidence for them. In our sample there are only two exceptions to this pattern. One of them is HD~70573, which has a chromospheric emission level slightly below the above-mentioned divide and for which we have detected active longitudes. The other exception is SAO~51891, which shows high chromospheric activity at $\log{R'_{\rm HK}}=-4.327$ but for which we could not determine any active longitudes or definite activity cycles. We note that the lack of these detections is likely due to the long time gap in the observations from the star.

These results suggest that there are two well-defined domains of magnetic field geometry present on the late-type active stars that are defined by their activity level. On the less active stars the large-scale magnetic field appears to prefer an axisymmetric configuration, whereas on the more active stars the field configuration is dominated by a non-axisymmetric mode. This observation can be related to the numerical results of \cite{tuominen1999nonaxisymmetric} who found their mean field dynamo solutions to switch from axisymmetric to non-axisymmetric with increasing Taylor numbers, i.e. higher rotation rates.

\begin{table}
\caption{Relative differences $\delta P$ between the active longitude and photometric periods and the tail-area probabilities $p$ that the underlying photospheric rotation periods have a larger deviation from the estimated $P_{\rm w}$ than $P_{\rm al}$. Values of $\delta P<0$ indicate $P_{\rm al}<P_{\rm w}$.}
\center
\begin{tabular}{lccc}
\hline
\hline
Star      & $P_{\rm al}$ [d] & $\delta P$ & $p$ \\
\hline
HD 1405   & 1.752212  & -0.23\%   & $1.0\cdot10^{-10}$ \\
HD 29697  & 3.9433    & -0.55\%   & $2.2\cdot10^{-4}$ \\
HD 41593  & 8.0417    & -1.1\%    & $1.9\cdot10^{-3}$ \\
HD 43162  & 7.1323    & -0.50\%   & 0.35 \\
HD 63433  & 6.46414   &  0.033\%  & 0.96 \\
HD 70573  & 3.29824   & -0.49\%   & $2.3\cdot10^{-6}$ \\
HD 82443  & 5.41471   & -0.18\%   & $2.4\cdot10^{-2}$ \\
HD 82558  & 1.6037330 & -0.038\%  & 0.14 \\
HD 116956 & 7.84203   & -0.23\%   & 0.26 \\
HD 171488 & 1.336923  & -0.63\%   & $7.0\cdot10^{-11}$ \\
HD 220182 & 7.62002   & -0.76\%   & $1.2\cdot10^{-2}$ \\
\hline
\end{tabular}
\label{altab}
\end{table}

\subsection{Activity cycle lengths}

\begin{table}
\caption{Semi-empirical convective turnover times and inverse Rossby numbers.}
\center
\begin{tabular}{lcc}
\hline
\hline
Star      & $\tau_c$ [d] & $\log{{\rm Ro}^{-1}}$ \\
\hline
HD 1405   & 22.53        & 2.21 \\
HD 10008  & 19.27        & 1.55 \\
HD 26923  &  7.45        & 0.93 \\
HD 29697  & 23.71        & 1.88 \\
HD 41593  & 19.85        & 1.49 \\
HD 43162  & 15.51        & 1.43 \\
HD 63433  & 13.82        & 1.43 \\
HD 70573  & 10.48        & 1.60 \\
HD 72760  & 19.05        & 1.40 \\
HD 73350  & 12.28        & 1.10 \\
HD 82443  & 18.58        & 1.63 \\
HD 82558  & 22.36        & 2.24 \\
HD 116956 & 19.51        & 1.49 \\
HD 128987 & 15.35        & 1.31 \\
HD 130948 &  7.77        & 1.09 \\
HD 135599 & 20.34        & 1.66 \\
HD 141272 & 19.41        & 1.25 \\
HD 171488 & 10.13        & 1.98 \\
HD 180161 & 19.51        & 1.39 \\
HD 220182 & 19.41        & 1.50 \\
SAO 51891 & 20.80        & 2.03 \\
\hline
\end{tabular}
\label{tauctab}
\end{table}

\begin{figure*}
\centering
\includegraphics[width=\linewidth]{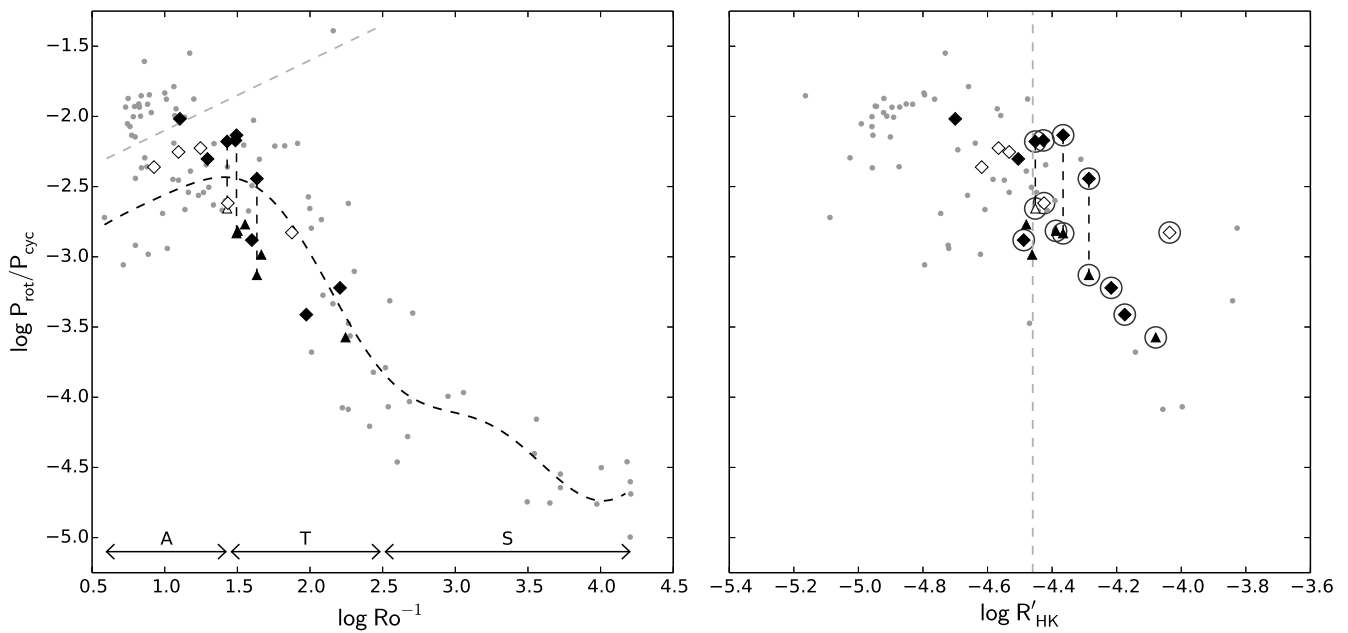}
\caption{\textit{Left:} $\log{P_{\rm rot}/P_{\rm cyc}}$ vs. $\log{{\rm Ro}^{-1}}$. Our stars are marked black if their cycles are graded ``fair'' or better and white if their cycles are ``poor''. Triangles denote stars with cycles labelled ``long'', while diamonds are used for the rest of the sample stars. Grey points denote reference data from \cite{saar1999time}. Vertical dashed lines connect the long and short cycles found on the stars HD 63433, HD 82443, and HD 116956. The curved dashed line is a Gaussian process fit to the activity branches apart from the inactive branch, approximately located above the light grey dashed line. Approximate ranges of the active, transitional and superactive branches are denoted by the horizontal arrows labelled ``A'', ``T'' and ``S''. \textit{Right:} Same as on the left but for $\log{P_{\rm rot}/P_{\rm cyc}}$ vs. $\log{R'_{\rm HK}}$. Stars in our sample that have detected active longitudes are circled in the plot. The grey vertical dashed line marks the approximate divide between the active stars with active longitudes and the less active ones without them. The scale for $\log{P_{\rm rot}/P_{\rm cyc}}$ is identical in both panels.}
\label{figcycro}
\end{figure*}

With a sample of estimated cycle periods it is possible to study the connection between the cycle lengths and other stellar parameters. This was done by \cite{brandenburg1998time} and later by \cite{saar1999time} who studied the relation of $P_{\rm rot}/P_{\rm cyc}$ to the inverse Rossby number ${\rm Ro}^{-1}$ and $\log{R'_{\rm HK}}$. They were able to group their stars into various activity branches characterized by the different behaviour of $P_{\rm rot}/P_{\rm cyc}$ and suggested that these branches are related to a dynamo evolution sequence. Slightly different types of studies were done by \cite{baliunas1996dynamo}, who compared $P_{\rm cyc}/P_{\rm rot}$ to $1/P_{\rm rot}$, and \cite{bohmvitense2007chromospheric} and \cite{olah2009multiple}, who also compared $P_{\rm cyc}$ directly to $P_{\rm rot}$. In our data we do not see any direct correlation between $P_{\rm cyc}$ and $P_{\rm rot}$ when setting $P_{\rm rot}=P_{\rm w}$. These two parameters show random scatter with a Pearson correlation coefficient $r=-0.27$. Thus, we decided to search for a connection of $P_{\rm rot}/P_{\rm cyc}$ to ${\rm Ro}^{-1}$ and $\log{R'_{\rm HK}}$.

We define the inverse Rossby number, equivalent to the Coriolis number, as ${\rm Ro}^{-1}={\rm Co}=2\Omega\tau_c=4\pi\tau_c/P_{\rm rot}$ \citep{brandenburg1998time}, where $\tau_c$ is the convective turnover time. The convective turnover time cannot be observed directly and needs to be determined in some other way. We used the semi-empirical formula of \cite{noyes1984magnetic} which gives $\tau_c$ as a function of the $B-V$ colour. The resulting $\tau_c$ and $\log{{\rm Ro}^{-1}}$ values are listed in Table \ref{tauctab}.

When the $\log{P_{\rm rot}/P_{\rm cyc}}$ values are plotted against $\log{{\rm Ro}^{-1}}$ (Fig. \ref{figcycro}, left panel) our results align nicely along the active and transitional branches defined by the data of \cite{saar1999time}. One of our stars (HD~73350) diverges from this pattern and lies closer to their inactive branch near the upper left corner of the plot. Our data shows that the superactive branch at high $\log{{\rm Ro}^{-1}}$ and low $\log{P_{\rm rot}/P_{\rm cyc}}$ has the transitional branch as its continuation. This in turn meets smoothly with the active branch at low $\log{{\rm Ro}^{-1}}$ and high $\log{P_{\rm rot}/P_{\rm cyc}}$ forming a long and meandering $\log{P_{\rm rot}/P_{\rm cyc}}$ sequence. The approximate $\log{{\rm Ro}^{-1}}$ ranges of the active, transitional and superactive branches on this sequence are denoted by the arrows labelled ``A'', ``T'' and ``S'' at the bottom of the left panel in Fig. \ref{figcycro}.

Our stars occupy the region around the turning point between the active and transitional branches. This provides new data for locating the $\log{{\rm Ro}^{-1}}$ value at which the apparent change in the behaviour of the activity cycle lengths occurs. To determine this value we chose our cycles graded ``fair'' or better and the cycle data from \cite{saar1999time} excluding the inactive branch. As a working criterion we defined all cycles with $\log{P_{\rm rot}/P_{\rm cyc}}>0.5\log{{\rm Ro}^{-1}}-2.6$ as belonging to the inactive branch. This limit is marked by the light grey dashed line in the left panel of Fig.\ref{figcycro}.  We model this cycle data with a simple Gaussian process applying a Gaussian covariance function over $\log{{\rm Ro}^{-1}}$ with a standard deviation of $\sigma_{\rm cov}=0.95$ \citep{haran2011gaussian}. This produces a predictive mean for $\log{P_{\rm rot}/P_{\rm cyc}}$ as a function of $\log{{\rm Ro}^{-1}}$, plotted as the black dashed curve in Fig. \ref{figcycro}. The curve peaks at $\log{{\rm Ro}^{-1}}=1.42$, locating the turnoff point between the active and the transitional branches near this value. As discussed by \cite{saar1999time}, this point may mark a transition from antiquenching of the $\alpha$-effect on low values of $\log{{\rm Ro}^{-1}}$ and magnetic field strength to the more traditional quenching on the higher values.

One problem with comparing $P_{\rm rot}/P_{\rm cyc}$ to ${\rm Ro}^{-1}$ is that the Rossby number cannot be observed directly and is itself a function of $P_{\rm rot}$. Some level of spurious correlation between the two values is thus expected. A resolution to this problem is to use the relation between ${\rm Ro}^{-1}$ and $\log{R'_{\rm HK}}$ \citep{noyes1984magnetic}. The chromospheric emission index $\log{R'_{\rm HK}}$ can be observed independently of $P_{\rm rot}$ and $P_{\rm cyc}$ and be used to estimate the spurious correlation. The right panel of Fig. \ref{figcycro} shows our results plotted as $\log{P_{\rm rot}/P_{\rm cyc}}$ vs. $\log{R'_{\rm HK}}$ over the reference data from \cite{saar1999time}. A $\log{R'_{\rm HK}}$ value is not available for all of the stars in the reference data so the plot appears sparser than the $\log{P_{\rm rot}/P_{\rm cyc}}$ vs. $\log{{\rm Ro}^{-1}}$ plot. Nevertheless, the same active and inactive branches and the turnoff to the falling transitional branch can be recognized from it. Now the Vaughan-Preston gap at $\log{R'_{\rm HK}}\approx-4.75$ \citep{vaughan1980survey,henry1996survey} approximately divides the active and inactive branches from each other. The similarity in the structures seen in the two plots gives a reassuring picture that the activity branches seen in the $\log{P_{\rm rot}/P_{\rm cyc}}$ vs. $\log{{\rm Ro}^{-1}}$ plot are not dominated by the suspected spurious correlation.

A feature that is not seen in the cycle data of \cite{saar1999time} but is evident in our data, especially in the $\log{P_{\rm rot}/P_{\rm cyc}}$ vs. $\log{R'_{\rm HK}}$ plot, is the split of the activity cycles into two parallel sub-branches both within the active and the transitional branches. The sub-branches are in fact defined so well in our data that even the uncertain ``poor'' cycles follow them closely. We find that none of our stars falls between the sub-branches. It is also significant that for the three stars with double cycles the two cycles fall on the separate sub-branches, the shorter on the upper and the longer on the lower sub-branch. Towards higher $\log{{\rm Ro}^{-1}}$ or $\log{R'_{\rm HK}}$ the behaviour of the sub-branches is less certain. However, if the ``poor'' cycle of HD~29697 at high $\log{R'_{\rm HK}}$ can be seen to fall on the upper sub-branch, this sub-branch continues as far down the transitional branch as the lower sub-branch does.

It is unclear how the sub-branches should be related to the previous results by other authors. \cite{bohmvitense2007chromospheric} suggested a division of the active branch into two sub-branches but these cannot be easily interpreted as our sub-branches.

\section{Conclusions}

In this paper we have analysed between 16 and 27 years of differential photometry from 21 young solar-type stars and characterized the nature of their activity with the aid of time series analysis. Furthermore, we have obtained high resolution spectra of the stars and calculated new chromospheric emission indices for them. The results show that the activity related phenomena of these stars follow a number of different trends.

We estimated the differential rotation of the stars by determining the range of the observed photometric period variations following the methodology of \cite{jetsu1993decade}. There are certain problems in interpreting the $Z$ parameter value directly as a measure of the differential rotation since both active region growth and decay and computational instabilities can increase the uncertainties in the measured period values. We conclude, however, that these effects do not dominate our results. The remaining uncertainty concerns the unknown latitude extent of the spot activity on the observed stars and how the period variation ranges should be scaled into the differential rotation coefficient $k$. Nevertheless, the measured period variations correlate with the surface differential rotation and reveal on the one hand a steep trend of increasing $k$ towards longer $P_{\rm rot}$, as $k \propto P_{\rm rot}^{1.36}$, and on the other a much flatter dependence of $\Delta\Omega$ on $\Omega$, as $\Delta\Omega \propto \Omega^{-0.36}$. Both of these results agree qualitatively with previous studies \citep{henry1995starspot, donahue1996relationship, barnes2005dependence,reinhold2013rotation}. However, it has to be noted that the published results for the surface shear are widely scattered around constant $\Delta\Omega$, and it is not immediately clear how its underlying rotation dependence should be interpreted. We did not find any temperature dependence of the differential rotation, which again fits the recent results of \cite{kuker2011difference} and \cite{reinhold2013rotation} for stars with effective temperatures between 3500 K and 6000 K.

It should be borne in mind that there are still considerable uncertainties in measuring differential rotation with the $Z$ parameter. It is possible to obtain information on the rotation and temperature dependence of $k$ and $\Delta\Omega$ using $Z$ for a population of stars, but the differential rotation estimates of individual stars can have substantial errors and should be taken with a grain of salt.

Activity cycles were found from the photometry of nearly all of our stars with some level of significance (see Table \ref{pertab}). When the ratio of the rotation and cycle periods was plotted against $\log{{\rm Ro}^{-1}}$ or $\log{R'_{\rm HK}}$, our stars fell neatly on the active and transitional activity branches defined by \cite{saar1999time} with one star being closer to their inactive branch. Our stars show that the transitional branch meets smoothly with the active branch at around $\log{{\rm Ro}^{-1}}=1.42$. Thus, the superactive and transitional branches form a continuous trend of decreasing scaled cycle periods with decreasing ${\rm Ro}^{-1}$ down to the turnoff point of meeting with the active branch and turning smoothly into an opposite trend. Something not clearly seen until now is the split of the active and transitional branches into two parallel sub-branches, as is visible in Fig. \ref{figcycro}. This finding suggests that there are multiple cycle modes available for the dynamos across a wide range of rotation rates.

Active longitudes were also frequently found, although they do not appear to be quite as common as the activity cycles (see Table \ref{pertab}). We found temporary or persistent active longitudes from 11 of our stars, all of them belonging to the more active part of our sample. There appears to be a divide at approximately $\log{R'_{\rm HK}}=-4.46$ so that nearly all of the stars on the less active side of the limit have no active longitudes, while nearly all on the more active side have them. This suggests that the large scale magnetic fields are dominated in the less active stars by axisymmetric dynamo modes up to a certain limit, after which the more active stars develop strong non-axisymmetric modes. In Fig. \ref{figcycro} the limit appears close to the break point between the active and transitional activity branches, suggesting a deeper connection between the opposite trends seen in the activity cycle lengths and the dominant magnetic field geometry.

In all but one case the estimated active longitude periods $P_{\rm al}$ were shorter than the mean photometric rotation periods $P_{\rm w}$. In this single deviating case of opposite order of $P_{\rm al}$ and $P_{\rm w}$ the two estimated periods were practically identical. We compared the two periods on each star and found that for seven of them a single underlying rotation period does not seem adequate to simultaneously describe them. Our interpretation is either that the magnetic structures supporting the active longitudes in these stars reside on deeper levels in the interiors, where the rotation is faster or that these stars have prograde propagating azimuthal dynamo waves resulting in a trend of systematically shortened active longitude periods. Similar migrating active longitudes have been observed previously from active giant stars \citep{hackman2011spot, hackman2013flipflops} and propagating azimuthal dynamo waves have been seen in direct numerical dynamo simulations \citep{cole2014azimuthal}. The picture is somewhat complicated by the observation that the active longitudes commonly show some level of migration around the average active longitude period. Both the migration patterns and the disappearance and reappearance of the shorter-lived active longitudes show little regularity and still need to be fully explained.

\begin{acknowledgements}
This work has made extensive use of the SIMBAD data base at CDS, Strasbourg, France and NASA's Astrophysics Data System (ADS) bibliographic services. The work of J.L. was supported by Vilho, Yrj\"{o} and Kalle V\"{a}is\"{a}l\"{a} Foundation. The automated astronomy program at Tennessee State University has been supported by NASA, NSF, TSU and the State of Tennessee through the Centers of Excellence program. We thank Dr. Maarit K\"{a}pyl\"{a} and Prof. Axel Brandenburg for valuable comments on the manuscript.
\end{acknowledgements}

\bibliographystyle{aa}
\bibliography{solan}

\end{document}